\documentclass[10pt,journal,twoside]{IEEEtran}
\usepackage{amsmath}
\usepackage[mathscr]{eucal}
\usepackage{amsfonts}
\usepackage{amssymb}
\usepackage{psfrag}
\usepackage{cite}
\usepackage{color}
\usepackage{graphicx}
\usepackage{multirow}
\usepackage{stfloats}
\usepackage{lettrine}
\usepackage{amsfonts}
\usepackage{amssymb}
\usepackage{subfigure}
\usepackage{booktabs}

\usepackage{threeparttable}

\input{TNDs_IEEEtran_v16_mods}
\newcommand{\paperstatus}{Submitted to \textit{IEEE Transactions on Information Theory}}

\newtheorem{theorem}{Theorem}
\newtheorem{lemma}{Lemma}

\newtheorem{problem}{Problem}
\newtheorem{design problem}{Design Problem}

\newtheorem{example}{Example}

\newtheorem{proposition}{Proposition}
\newtheorem{definition}{Definition}
\newtheorem{property}{Property}
\newtheorem{algorithm}{Algorithm}

\newtheorem{subproblem}{Subproblem}

\newcommand{\smatlabaxislabel}[1]{\fontsize{12}{\f@baselineskip}%
\textsf{#1}}
\newcommand{\matlabaxislabel}[1]{\fontsize{14.4}{\f@baselineskip}%
\textsf{#1}}
\newcommand{\mmatlabaxislabel}[1]{\fontsize{17.28}{\f@baselineskip}%
\textsf{#1}}
\newcommand{\bmatlabaxislabel}[1]{\fontsize{20.74}{\f@baselineskip}%
\textsf{#1}}
\newcommand{\bbmatlabaxislabel}[1]{\fontsize{24.88}{\f@baselineskip}%
\textsf{#1}} \makeatother

\allowdisplaybreaks

\centerfigcaptionstrue

\newcommand{\half}%
{\raisebox{2.5pt}{\scriptsize 1}{\small
/}\raisebox{-1pt}{\scriptsize 2}}

\newcommand{\beq}{\begin{equation}}
\newcommand{\eeq}{\end{equation}}

\title{Reliable MIMO  Optical Wireless Communications Through Super-Rectangular Cover}
\author{Yan-Yu Zhang, Hong-Yi Yu, Jian-Kang Zhang and  Jin-Long Wang 
\thanks{This work was supported by NHTRDP (863 Program) of China (Grant No.2013AA013603). The work of Jian-Kang Zhang was partially supported by NSERC. }
\thanks{
Yan-Yu Zhang, Hong-Yi Yu and Jin-Long Wang
are with National Digital Switching System Engineering and Technology Center, Henan Province (450000), China. Emails:
yyzhang.xinda@gmail.com, maxyucn@sohu.com and wjl543@sina.com}
\thanks{Jian-Kang Zhang is with the Department of Electrical and Computer Engineering,
McMaster University, 1280 Main Street West, L8S 4K1, Hamilton,
Ontario, Canada. Email: jkzhang@mail.ece.mcmaster.ca.
}}

\begin{document}
\maketitle

\begin{abstract}
In this paper, we consider an intensity modulated direct detection multiple-input-multiple-output optical wireless communication (IM/DD MIMO-OWC) system, where the channel coefficients are assumed to be completely known at the receiver.  For such a system, since both the transmitted signals and the channel coefficients are nonnegative, the full rank condition on the unique identification of the signals for a noise-free channel and fully reliable (diversity) estimation of the signals for a noisy channel for MIMO radio frequency (MIMO-RF) communications is sufficient but not necessary. For this reason, from the viewpoint of detection theory, a novel so-called super-rectangular cover theory  is developed  to characterize both the unique identifiability and full reliability for IM/DD MIMO-OWC. In this theory, two important concepts, cover order and cover length are introduced. Among all the super-rectangles covering the feasible domain in the first quadrate determined by a semidefinite positive quadratic form smaller than any given positive constant, the largest super-rectangular cover is the one with the maximum number of finitely lengthy sides, while the smallest super-rectangular cover is the largest super-rectangular cover with the length of each finitely lengthy side being the minimum. Cover order and cover length are defined, respectively, as the number of the finitely lengthy sides of the largest super-rectangle cover and the side length of the smallest super-rectangular cover. This theory states that a transmitted matrix signal can be uniquely identified if and only if the cover order is equal to the transmitter aperture number (maximum cover order), i.e., full cover. In particular, when this theory is applied to the diversity analysis for space-time block coded IM/DD MIMO-OWC over commonly used log-normal fading channels, we prove that full reliability is guaranteed  with a maximum likelihood (ML) detector if and only if the space-time block code (STBC) enables full cover. In addition, for this system, the diversity gain can be geometrically interpreted as the cover order of the super-rectangle, which should be maximized, and the volume of this super-rectangle, as the diversity loss, should be minimized. Using this established error performance criterion, the optimal linear STBC for block fading channels is proved to be spatial repetition code with an optimal power allocation. The design of the optimal non-linear STBC is shown to be equivalent to constructing the optimal multi-dimensional constellation, which is a classic and long-standing topic. Specifically, a multi-dimensional constellation from Diophantine equations is proposed and then, shown to be more energy-efficient than the commonly used nonnegative pulse amplitude modulation (PAM) constellation. Furthermore, for fast fading channels, a linear STBC is constructed by collaborating the signals of two successive channel uses and proved to be related to the Golden number $\frac{\sqrt{5}+1}{2}$, say, Golden code. This Golden code has an important property of \textit{non-increasing} diversity loss as constellation size increases and thus, presents encouraging error performances. These observations obtained in this paper reveal useful insights into how the space-time block coded IM/DD MIMO-OWC  systems are remarkably different from,  rather than ``mimic'', the   conventional  space-time coded MIMO-RF systems (Tarokh, Seshadri, and Calderbank, 1998) from the perspectives of  detection theory, error performance criterion and code design techniques.
\end{abstract}

\begin{keywords}
 Cover length, cover order, Diophantine equation, full diversity, Golden codes, intensity modulation and direct detection (IM/DD),  maximum-likelihood (ML) detection, multidimensional constellation, multiple input multiple output (MIMO), optical wireless communications (OWC), repetition coding (RC), rectangular cover, space-time block code (STBC).
\end{keywords}

\section{Introduction}\label{sec:intro}
\subsection{Motivation}
\lettrine[lines=2]{M}{odern} society has witnessed an explosively increasing demand of information. This demand has ever been triggering off an enormous expansion of wireless communications. As an extensively investigated area, radio frequency (RF) wireless communication has played an important role in our daily life. However, the continually increasing demand of data transmission almost makes the radio spectrum approaching saturated. As an adjunct or alternative to RF communications, intensity modulation direct detection optical wireless communication (IM/DD OWC), due to its potential for bandwidth-hungry applications,  has become a very important area of  research~\cite{li2003optical,o2005optical,boucouvalas2005challenges,chan2006free,das2008requirements,Brien2008challenges,Lubin2009MIMO,Kumar2010Led-based,Elgala2011review,Borah2012review, Gancarz13,khalighi2014survey,chaudhary2014role}. The importance of IM/DD OWC lies in the advantages of low cost, high security, freedom from spectral licensing issues etc. Therefore, IM/DD OWC is considered to be the next frontier for net-centric connectivity for  bandwidth, spectrum and security issues in numerous scenarios. For the road map of optical communications recently developed in~\cite{Ioannis2016Roadmap}, the authors even predicted that ``\textit{the next decade will undoubtedly see the widespread deployment of OWC systems}.''

Among the diverse environments using IM/DD OWCs, the three most broadest incarnations~\cite{Ioannis2016Roadmap} are visible light communications (VLC), terrestrial free space optical (FSO) communications and deep space systems.
\begin{enumerate}
  \item IM/DD OWC operating in the visible band (390$\sim$750 nm) is commonly referred to as VLC~\cite{langer2007recent,Brien2008challenges,Lubin2009MIMO,wang2011high,Kumar2010Led-based,Elgala2011review,Borah2012review}, which utilizes ubiquitous light emitting diodes as data transmitters,  e.g., indoor lights, street lights, car brake lights, remote control units and countless other applications. This novel IM/DD OWC has attracted extensive attention and has had great breakthrough of transmission rate up to several Gbit/s~\cite{Cossu201234,chi2014,Tsonev3g}. It should be noted that for indoor IM/DD OWC, when the receiver and transmitter  are in fixed locations, the optical intensity channel is considered to be  deterministic. However, some challenges remain, especially in the mobile environments,

  \item For FSO over mobile or atmospheric channels, \textit{reliability} is a key consideration~\cite{farid2007outage,yang2014free,JinYuanWang2014,borah2009pointing}. In the mobile link, there will be inevitable impairments such as terminal-sway, aerosol scattering and pointing errors, etc. In addition, for  the atmospheric environments, atmospheric effects, such as rain, snow, fog and temperature variation, will significantly affect the  link performance by resulting in the fading of the received intensity signal. Therefore, in the design of IM/DD OWC links, we need to consider these impairments-induced fading for reliability issues.
  \item For the near-Earth and interplanetary deep-space communications~\cite{hemmati2006deep,Hemmati11}, OWC systems have the ability to provide high data rate communication and enjoy compelling advantage over currently well-established RF communications. For these very long distance deep-space applications, such as a 400,000km OWC link between a satellite in lunar orbit and a ground station demonstrated by NASA in 2013~\cite{Khatri2014Overview}, there is no viable alternative to OWC~\cite{Ioannis2016Roadmap}.
\end{enumerate}

 To assure reliable IM/DD OWC links, IM/DD multiple-input-multiple-output (MIMO) OWC (IM/DD MIMO-OWC) systems use multiple transmitter and receiver apertures with sufficient separation between each such that multiple receiver apertures receive diverse replicas of the transmitted signals. Reliable detection can be achieved  by  introducing a design for the transmitted symbols distributed over transmitter apertures (space) and (or) symbol periods (time). Such proper design is an important and challenging task. Therefore, in this paper, we focus on the space-time signal processing theory and design techniques for IM/DD MIMO-OWC systems.

\subsection{Previous Work and Current Challenge}
Historical perspective on the space-time block code (STBC) is mainly resulted from the landmark work in~\cite{tarokh98}. Unfortunately, compared with the well-developed theory and techniques for MIMO-RF communications~\cite{tarokh98,yuen2005quasi,jkz06,jkz-it06,dao2008four,lei2011quasi,dongxia2013energy} and coherent MIMO-OWC~\cite{aghajanzadeh2010diversity,bayaki2012performance,niu2012performance,song2014high,niu2014alamouti}, there are the following \textit{three} main features in IM/DD MIMO-OWC communications.

The \textit{first} feature is a nonnegative constraint on the design of transmission for IM/DD MIMO-OWC and the channel coefficient with direct detection of  optical intensity. In a typical optical system with IM/DD, the received current from the optical detector is proportional to the power of light wave. However, in the MIMO-RF and coherent MIMO-OWC systems, the received current is proportional to the amplitude of the carrier~\cite{Barry1994Ifrd,wilson2005ijsac}. In other words, both the channel and the signals of IM/DD MIMO-OWC are located in the \textit{positive orthants of a real space}, whereas the channel and signals of MIMO-RF and coherent are generally complex-valued and bipolar. Because of this feature of IM/DD MIMO-OWC, the full rank condition for the reliable signal identification of MIMO-RF~\cite{tarokh98,yuen2005quasi,jkz06,jkz-it06,dao2008four,lei2011quasi,dongxia2013energy} is sufficient but not necessary for IM/DD MIMO-OWC. Hence, the currently well-developed techniques for MIMO-RF and coherent MIMO-OWC cannot be applicable to IM/DD MIMO-OWC completely. To characterize the detection reliability of matrix signals for IM/DD MIMO-OWC, new mathematical theory indeed needs to be developed.

The \textit{second} feature is that there does not exist any available mathematical tool that could be directly applied to the analysis of the average pair-wise probability (PEP) for IM/DD MIMO-OWC when the commonly used log-normal~\cite{haas2002space,Liu2004itct,Filho2005el,navidpour2007itwc,Beaulieu2008itct,giggenbach2008fading}, Gamma-Gamma~\cite{vetelino2007fade,andrews1999theory,al2001mathematical}, K-distributed~\cite{andrews2001laser,sandalidis2008outage} are involved. In these scenarios, since the resulting PEP relating diversity gain analysis to signal-to-noise ratio (SNR) becomes very complicated, it is indeed a challenge to extract a dominant term of the average PEP. Let alone say what the best diversity gain is. Here, it should be mentioned that there really exist mathematical formulae in literature for numerically and accurately computing the integral involving channel fading~\cite{haas2002space,Liu2004itct,Filho2005el,navidpour2007itwc,Beaulieu2008itct}. However, it still seems difficult to provide a general guideline on STBC design. To overcome this difficulty, novel techniques are indeed in demand.

The \textit{third} feature is that the power constraint of IM/DD MIMO-OWC is on the amplitude average of the nonnegative signals, whereas the power constraint for MIMO-RF and coherent MIMO-OWC is on the squared amplitude average.  Despite the fact that the nonnegative constraint can be satisfied by properly adding some direct-current components (DC) into transmitter designs so that the existing advanced MIMO  techniques~\cite{tarokh98,jkz-it06,shang08} for the RF communications such as orthogonal STBC (OSTBC)~\cite{geramita79,alamouti98,tarokh99,su01,ganesan01,tirkkonen02,liang03,liu-icassp04,cui07} could be  used in IM/DD MIMO-OWC, the power loss arising from DC is considered to incur the fact that this modified OSTBC~\cite{simon2005alamouti,wang2009mimo,tianhigh} in IM/DD channels has worse error performance than the  repetition coding (RC), which is based on the spatial-repetitional transmission of equally spaced nonnegative pulse amplitude modulation (PAM) and is shown to be the best code available in the literature for this application~\cite{navidpour2007itwc,majid2008twc,bayaki2010space,abaza2014diversity}. In addition, it is known that for a single transmitter transmission, the nonnegative PAM is indeed the most energy-efficient in terms of the maximization of the minimum Euclidean distance under an optical power budget. However, our recent results in~\cite{zhuspace} have revealed that this is no longer true even for a two-transmitter transmission IM/DD MIMO-OWC system. Unfortunately, attaining a general optimal multi-dimensional constellation is as hard as solving a parallel and long-standing  well-known optimization problem in modern RF digital communications, which, to the best knowledge of the authors, still remains unsolved thus far~\cite{gallager2008principles,forney1989multidimensional1,forney1989multidimensional2,conway1993sphere}. Since the feasible signal domain and power constraint of OWC are significantly different from those of RF communications, the design of energy-efficient space-time constellations is indeed needed to further improve its system performance.

\subsection{Our Main Work}
 Indeed, all the aforementioned factors motivate us to propose a novel theory for the signal design of a general IM/DD MIMO-OWC system and then, in particular, to develop new techniques for the corresponding energy-efficient signal designs. Therefore, our main tasks in this paper are summarized as follows.
\begin{enumerate}
\item \textit{To develop a theory to characterize the properties of STBC of IM/DD MIMO-OWC}. With this aim, we will develop a fundamentally important \textit{super-rectangular cover} theory in Section~\ref{sub:cover_theory} to characterize the identification of signals for nonnegative matrix channels from the viewpoint of detection theory. In addition, full cover will be proved to a necessary and sufficient condition for the unique identification of signals in any noise-free nonnegative channel.

  \item \textit{To establish a general criterion for the STBC design of IM/DD MIMO-OWC over log-normal fading channels}. To this end, the super-rectangular cover theory will be specifically applied to the diversity analysis of IM/DD MIMO-OWC in presence of the log-normal fading in Section~\ref{sec:performance_analysis}. Our main idea here is that by fragmenting the integral region of the average PEP involving log-normal into two sub-domains, the dominant term will be successfully extracted. With this, we will prove that full cover is also a  necessary and sufficient condition to assure a full diversity achievement in a noisy IM/DD MIMO-OWC log-normal channel. 
 \item\textit{To systematically design STBC of IM/DD MIMO-OWC over log-normal fading channels}. For this purpose, we will, in Section~\ref{sec:code_costruction}, use the newly developed criterion for IM/DD MIMO-OWC over log-normal fading channels to attain the following results.
     \begin{enumerate}
       \item \textit{Optimal linear STBC structure}. By fully utilizing the nonnegativity of signals and linear power constraint, it will be proved that for block fading channels, RC with an optimal power allocation  is the optimal linear STBC.
       \item \textit{Optimal non-linear STBC structure}. For block fading channels, the design of optimal STBC will be shown to be equivalent to constructing the optimal multi-dimensional constellation with an optimized spatial power allocation.
       \item \textit{Multidimensional Diophantine constellations}. A specific energy-efficient multi-dimensional constellation from Diophantine equations will be proposed to construct an energy-efficient collaborative STBC and provide a useful insight into how to design multidimensional energy-efficient constellation for IM/DD MIMO-OWC.
       \item \textit{Golden codes}. For fast fading channels, by linearly collaborating the signals of two successive independent channel uses, an optimal linear STBC will be designed via analytical optimization and shown to be related to the \textit{Golden number} $\frac{\sqrt{5}+1}{2}$.
     \end{enumerate}

\end{enumerate}

\textit{Notation}: Most notations used throughout this paper are standard: column vectors and matrices are boldface lowercase and uppercase letters, respectively; the matrix transpose is denoted by $\left(\cdot \right)^T$; $\mathbf{I}_{N\times N}$ denotes an $N\times N$ identity matrix; Notation  $\|\mathbf{v}\|_2$ denotes the $\ell$-2 norm of $\mathbf{v}$; and $\emptyset$ denotes an empty set. $\mathbb{R}_{+}^{L\times N}$ is the set of all $L\times N$ matrices with all entries being nonnegative and $\mathbb{R}_{++}^N$ is the set of all the $N\times1$ vectors with all $N$ entries being positive.

\section{System  Model}\label{sec:model}

Let us consider an $N\times M$ IM/DD MIMO-OWC system having $M$ receiver apertures and $N$ transmitter apertures. During the $L$ time slots (channel use), the $N$ transmitter apertures transmit the codeword matrices $\mathbf{X}_k,k=0,1,~\cdots,~2^K-1$, which are randomly, independently and equally likely selected from a given nonnegative space-time constellation $\mathcal{X}$ satisfying the unipolarity requirement of intensity modulation, i.e., $\mathcal{X}\subseteq\mathbb{R}_+^{L\times N}$. These matrix codewords are then transmitted to the receivers through flat-fading path coefficients, which form the elements of the $N\times M$ channel matrix $\mathbf{H}$. The received space-time signal, denoted by the $L\times M$ matrix $\mathbf{Y}$, can be written as
 \begin{eqnarray}\label{eqn:system_model}
\mathbf{Y}=\mathbf{X}\mathbf{H}+\mathbf{N},
\end{eqnarray}
  where the entries of  channel matrix $\mathbf{H}$ are nonnegative, i.e., $\mathbf{H}\in\mathbb{R}_+^{N\times M}$.
In addition, regarding noise matrix $\mathbf{N}$ in~\eqref{eqn:system_model}, the two primary sources at the receiver front end are due to noise from the receiver electronics and shot noise from the received DC photocurrent induced by background  radiation~\cite{Karp1988,Barry1994Ifrd}. Firstly, the exact number of the arriving photons at the receiver during a given duration is random and modelled as Poisson  distribution~\cite{Karp1988,Barry1994Ifrd} with a rate proportional to the input. This model reflects the physical nature of the transmitted signal consisting of many photons and thus, this noise component is signal-dependent. Secondly, the signal is also corrupted by background radiation (called dark current) that is modelled by an additional constant rate added to the rate of the Poisson distribution.  In addition to the above two components, the received signal is also impaired by the  thermal noise  from the receiving device. In this paper, we assume that the shot noise caused by background radiation is dominant compared with the other noise components~\cite{zhu2002free,Lee2004}. By the central limit theorem, this high-intensity shot noise matrix $\mathbf{N}$ is closely approximated as additive, signal-independent, white, Gaussian noise~\cite{Barry1994Ifrd,zhu2002free} with zero mean and covariance matrix $\frac{\sigma_{\mathbf{N}}^{2}}{M}\mathbf{I}_{LM\times LM}$.

It can be seen that the channel matrix of IM/DD MIMO-OWC lies in the positive orthants of a real space. In addition, the channel fading is normally modelled as log-normal~\cite{haas2002space,Liu2004itct,Filho2005el,navidpour2007itwc,Beaulieu2008itct,giggenbach2008fading}, Gamma-Gamma~\cite{vetelino2007fade,andrews1999theory,al2001mathematical}, K-distributed~\cite{andrews2001laser,sandalidis2008outage}  not Rayleigh anymore. Therefore, we cannot straightforwardly resort to the well-developed techniques for MIMO-RF wireless communications~\cite{tarokh99} to the design of STBC for IM/DD MIMO-OWC. For this reason, we will develop a new theory called cover theory for this application in the following section.

\section{Establishment of Super-Rectangular Cover Theory}\label{sub:cover_theory}
 In this section, our primary purpose is to develop a novel  super-rectangular cover theory to characterize the properties of the coding structure from the viewpoint of signal identification.

\subsection{Unique Identification and Full Super-Rectangular Cover}

This subsection aims at presenting the motivation of our super-rectangular cover theory from the perspective of detection theory and introducing the definition of full cover. Then, we will show that the unique identification of signals and the full cover are equivalent.
\subsubsection{Unique Identification of Space-Time Constellations}
 Let us first consider the noise-free version of  \eqref{eqn:system_model},  given by
\begin{eqnarray}\label{eqn:noise_free_system_model}
\mathbf{Y}=\mathbf{X}\mathbf{H}
\end{eqnarray}
From signal detection theory, it is of interest to determine  a condition that guarantees the existence of a unique codeword matrix $\mathbf{X}$ in the constellation $\mathcal{X}$ for any given non-zero received signal $\mathbf{Y}$ and nonzero channel matrix $\mathbf{H}\in\mathbb{R}_+^{N\times M}$, that being said, the unique identification of codewords. Naturally, one may ask: \textit{Why is the study of the noise-free case so important when all realistic communication systems are noisy?} The main reason for this can be explained as follows: The reliability (specifically referred to as diversity gain for fading channels) is to measure how quickly the error probability decreases against increasing SNR~\cite{tarokh98,brehler01}, and thus, quantitatively characterizes how accurately the transmitted signal is estimated in a noisy environment, particularly in a high-SNR scenario. Therefore, a full-reliability (or full diversity) code must have the ability to allow the transmitted signal to be uniquely identified in a noise-free case. That is to say, if a constellation design is not able to provide the unique identification of the transmitted signals in the noise-free case, then, the reliable detection of the signal will not be guaranteed, even in a sufficiently high SNR region. In fact, our results in Section \ref{sec:performance_analysis} will show that it is the unique identification of  the transmitted signal in the noise-free case that guarantees a full diversity achievement in the noisy case with log-normal fading.

What follows gives a formal definition of the   unique identification in \eqref{eqn:noise_free_system_model}:
\begin{definition}
 For an arbitrarily given nonzero channel vector $\mathbf{h}\in \mathbb{R}_+^N$, a space-time constellation $\mathcal{X}$ is said to be uniquely identified if for any two distinct codeword matrices $\mathbf{X},\tilde{\mathbf{X}}\in\mathcal{X}$, then,  it holds that
$(\mathbf{X}-\tilde{\mathbf{X}})\mathbf{h}\neq\mathbf{0}$.~\hfill\QED
\end{definition}

In the following, we are interested in the condition under which the unique identification is assured.
\subsubsection{Super-Rectangular Cover}
In this subsection, we formally give the definition of super-rectangular cover and then, prove the equivalence between unique identification and full cover.

To start with, let us introduce some related definitions.
\begin{definition}\label{definition:cover_order}
A positive semi-definite (PSD) matrix $\mathbf{A}^T\mathbf{A}\in\mathbb{R}^{N\times N}$ is  said to be  $r$-covered ($r\le N$) by a super-rectangle if for any given positive real-valued number  $\tau>0$, the domain determined by $\{\mathbf{h}:\mathbf{h}\in\mathbb{R}_{+}^N, \mathbf{h}^T\mathbf{A}^T\mathbf{A}\mathbf{h}\le\tau^2 \}$ lies in a super-rectangle $\{\mathbf{h}:0\le h_{n_k}\le\bar{c}_{n_k}\tau,k=1, 2,~\cdots,~r\}$, i.e.,
\begin{eqnarray*}
&&\{\mathbf{h}:\mathbf{h}\in\mathbb{R}_{+}^N, \mathbf{h}^T\mathbf{A}^T\mathbf{A}\mathbf{h}\le\tau^2 \}\nonumber\\
&&\subseteq\{\mathbf{h}:0\le h_{n_k}\le\bar{c}_{n_k}\tau,k=1, 2,~\cdots,~r\}
\end{eqnarray*}
where all $\bar{c}_{n_k}$ are absolutely constants independent of $\tau$. Then, $\mathbf{A}^T\mathbf{A}$ or ${\mathbf A}$ is said to have a cover link $n_1,~\cdots~,n_r$. In addition,
\begin{enumerate}
  \item The maximum $r$ is named the cover order of $\mathbf{A}^T\mathbf{A}$ and denoted by $R_{c}$.
  \item  When $R_c=N$, $\mathbf{A}^T\mathbf{A}$ is said to have full-cover.
  \item  When $R_c=0$, $\mathbf{A}^T\mathbf{A}$ has zero-cover.
  \item For a fixed $n_k$ satisfying $1\le k\le R_c$, the minimum constant $\bar{c}_{n_k}$, which is specifically denoted by $c_{n_k}$, is named the $n_k$-th cover length
  and the product of all the cover lengths is called cover volume.
  \item   The index sequence $n_1,~\cdots~,n_{R_c}$ is called one of the longest cover links of $\mathbf{A}^T\mathbf{A}$.
\end{enumerate}~\hfill\QED
\end{definition}
To make our presentation as clear as possible, we provide the following non-trivial examples to illustrate the definitions of cover order and cover length.
\begin{example} \label{example:non_degenerate} Consider
 $\mathbf{A}^T\mathbf{A}=\left({
  \begin{array}{cc}
  1& 1\\
  1& 1\\
  \end{array}
  }\right)$.
In this case, we have $\mathbf{h}^T\mathbf{A}^T\mathbf{A}\mathbf{h}=\left(h_1+h_2\right)^2$.  We show the domain $\{\mathbf{h}\in\mathbb{R}_+^2:\mathbf{h}^T\mathbf{A}^T\mathbf{A}\mathbf{h}\le \tau^2\}$ in Fig.~\ref{fig:exampe_full_cover}. As illustrated by Fig.~\ref{fig:exampe_full_cover}, it can be seen that the domain $\{\mathbf{h}\in\mathbb{R}_+^2:\mathbf{h}^T\mathbf{A}^T\mathbf{A}\mathbf{h}\le \tau^2\}$ is triangle-shaped with three vertexes being $(0,0), (1,0)$ and $(0,1)$ and can be covered by a square with four vertexes determined by $(0,0),(1,0),(0,1)$, and $(1,1)$. Hence, for any $\tau>0$,
\begin{eqnarray*}
&&\{\mathbf{h}\in\mathbb{R}_+^2:\mathbf{h}^T\mathbf{A}^T\mathbf{A}\mathbf{h}\le \tau^2\}\nonumber\\
&&\subseteq\{\mathbf{h}\in\mathbb{R}_+^2: 0\le h_1, h_2\le \tau\}
\end{eqnarray*}
Therefore, $R_c=2$, i.e., $\mathbf{A}^T\mathbf{A}$ has full-cover.
~\hfill\QED

\begin{figure}[t]
    \centering
    \resizebox{8cm}{!}{\includegraphics{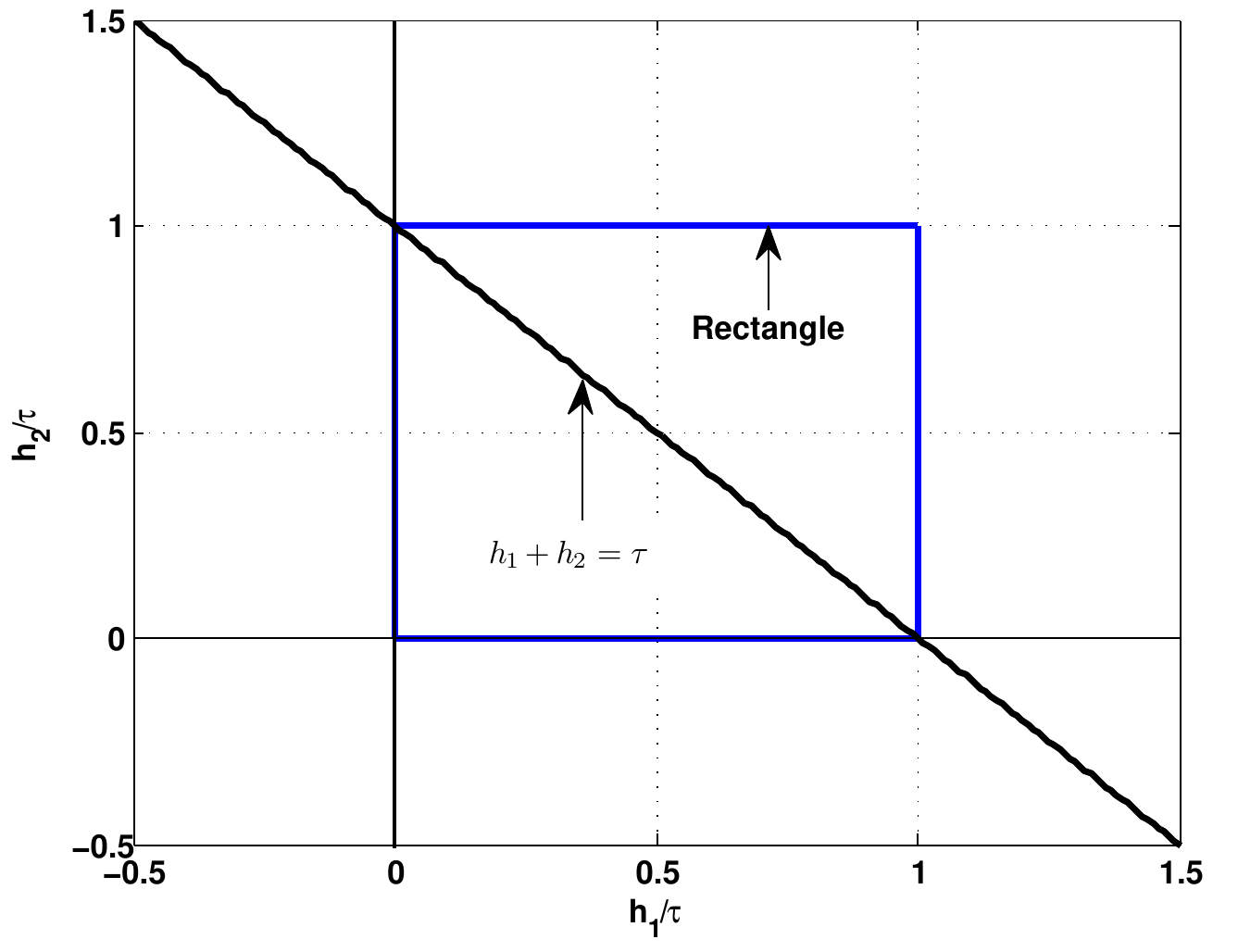}}
    \centering \caption{Example of a full-cover $2\times 2$ PSD matrix.}
    \label{fig:exampe_full_cover}
\end{figure}

\end{example}

\begin{example} \label{example:non_full_cover}
\begin{figure}[t]
    \centering
    \resizebox{8cm}{!}{\includegraphics{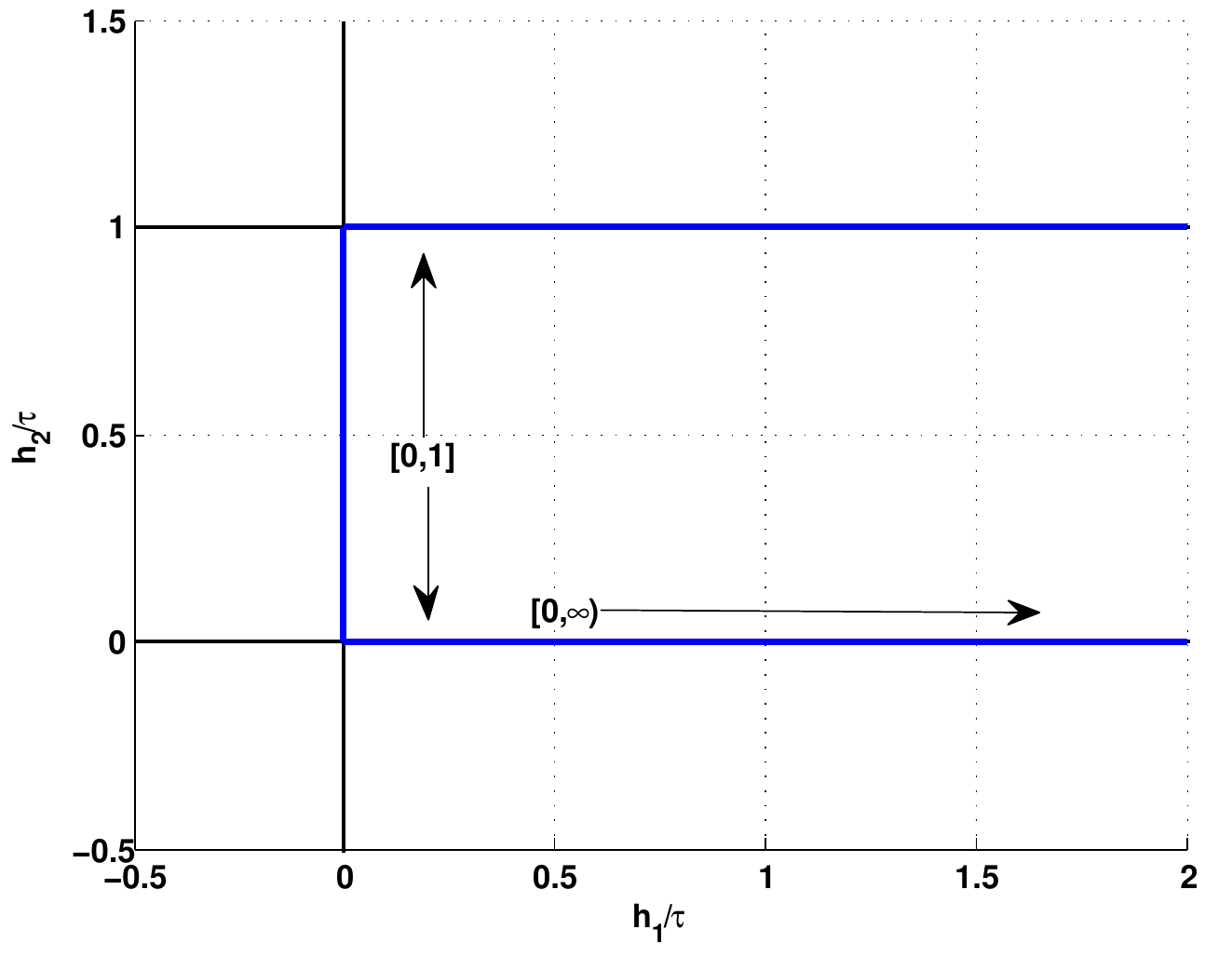}}
    \centering \caption{Example of a one-cover $2\times 2$ PSD matrix.}
    \label{fig:exampe_one_cover}
\end{figure}
Consider
 $\mathbf{A}^T\mathbf{A}=\left({
  \begin{array}{cc}
  1& 0\\
  0& 0\\
  \end{array}
  }\right)$. For this matrix, $\{\mathbf{h}\in\mathbb{R}_+^2:\mathbf{h}^T\mathbf{A}^T\mathbf{A}\mathbf{h}\le \tau^2\}=\{\mathbf{h}\in\mathbb{R}_+^2:0\le h_2 \le \tau,h_1\ge0\}$, which is shown in Fig.~\ref{fig:exampe_one_cover} and open in the x-axis direction. In other words, only one dimension is covered. Thus, we attain that $R_c=1$ and $c_1=1$.
~\hfill\QED
\end{example}

\begin{example} \label{example:full_degenerate}
\begin{figure}[t]
    \centering
    \resizebox{8cm}{!}{\includegraphics{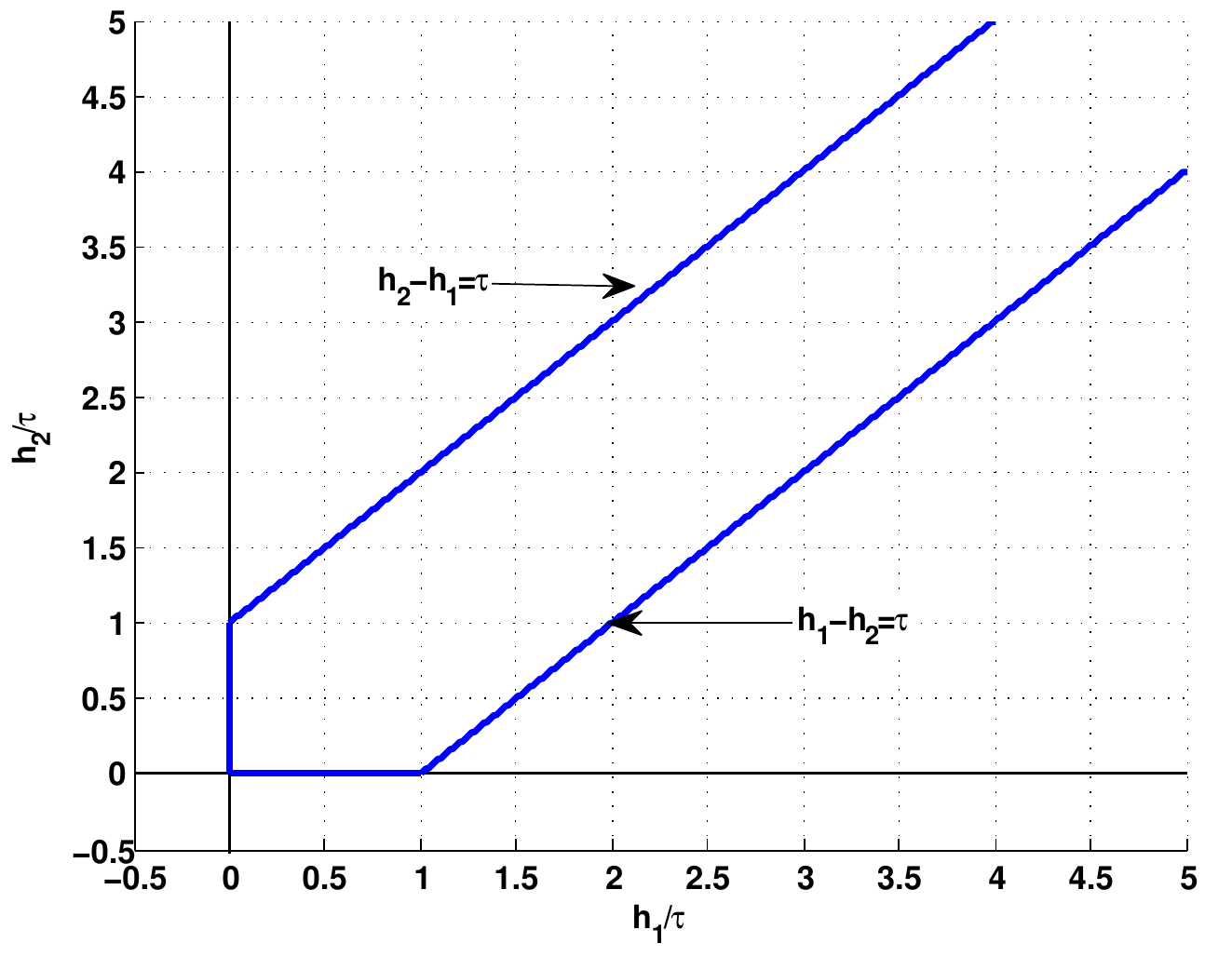}}
    \centering \caption{Example of a zero-cover $2\times 2$ PSD matrix.}
    \label{fig:exampe_zero_cover}
\end{figure}
 $\mathbf{A}^T\mathbf{A}=\left({
  \begin{array}{cc}
  1& -1\\
  -1& 1\\
  \end{array}
  }\right)$ has zero-cover\footnote{In Subsection~\ref{subsec:an_example_degenerate_code}, a zero-cover STBC is designed specifically for IM/DD MIMO-OWC over log-normal fading channels}. In this case, the feasible set supported by $\mathbf{h}^T\mathbf{A}^T\mathbf{A}\mathbf{h}=\left(h_1-h_2\right)^2\le \tau^2$ is shown in Fig.~\ref{fig:exampe_zero_cover} and is open and unbounded with respect to both $h_1$ and $h_2$. Hence, $R_c=0$.
~\hfill\QED
\end{example}
From the above three examples, we can find that 1) essentially, the cover order $R_{c}$ and the cover length $c_i$, respectively, represent the maximal dimension and minimal side lengths of the super-rectangle that covers $\{\mathbf{h}:\mathbf{h}\in\mathbb{R}_{+}^N, \mathbf{h}^T\mathbf{A}^T\mathbf{A}\mathbf{h}\le\tau^2 \}$ and, 2) these two identities are robust to the arbitrarily given positive number $\tau$.

 \subsubsection{Unique Identification and Full Cover}\label{subsec:full_cover}

Now, we formally present the first main result in this paper on the equivalence between full cover and unique identification.
\begin{theorem} \label{theorem:full_cover_uniquely}
 An $N\times N$ PSD matrix $\mathbf{A}^T\mathbf{A}$ has full-cover if and only if for any non-zero $\mathbf{h}\in\mathbb{R}_+^N$, we have $\mathbf{A}\mathbf{h}\neq\mathbf{0}$.
~\hfill\QED
\end{theorem}

On Theorem~\ref{theorem:full_cover_uniquely}, whose proof is postponed into Appendix~\ref{app:full_cover_identification}, we have the following remarks:
\begin{enumerate}
  \item Now, we can conclude that  full cover of any nonzero $\mathbf{X}-\tilde{\mathbf{X}}$ is indeed a necessary and sufficient condition on the unique identification of space-time coded constellation $\mathbb{X}$. Very interestingly, our full cover condition is parallel to the full rank condition for RF STBC to assure the unique identification given in~\cite{tarokh99,jkz-it06,shang08} in the sense of the existence of the unique solution to \eqref{eqn:noise_free_system_model} conditioned on non-zero $\mathbf{Y}$ and $\mathbf{H}\in\mathbb{R}^{N\times M}$. The difference between IM/DD MIMO-OWC and MIMO-RF as well as coherent MIMO-OWC lies in the sets of channel coefficients: the former is only the positive orthants of  the latter. This difference of the channel vector sets implies that full rank is sufficient, but not necessary for signal unique identification. This observation is also the main \textit{theoretical motivation} leading us to establishing this so-called super-rectangular cover theory.
  \item Although our super-rectangular cover theory is developed under the assumption that the channel coefficients are nonnegative, this idea can be generalized. For example, in the MIMO-RF case.  If the channel coefficients are bipolar valued, then, the full cover condition is exactly full column rank condition on  any nonzero $\mathbf{X}-\tilde{\mathbf{X}}$, which is the well-known full diversity condition developed in~\cite{tarokh99}. For this reason, our super-rectangular cover theory can be used for a generic applications, where the channel coefficients can be located at any subset of the whole space.
\end{enumerate}

Using Theorem~\ref{theorem:full_cover_uniquely}, we can immediately obtain the following properties.
\begin{property} \label{property:rank_one}
  For an $N\times N$ rank-one matrix $\mathbf{A}^T\mathbf{A}$, $R_{c}=N$ if and only if all the entries of $\mathbf{A}^T\mathbf{A}$ are positive.
~\hfill\QED
\end{property}
\begin{property} \label{property:rank_n}
  Full-rank matrix $\mathbf{A}^T\mathbf{A}$ has  full-cover.
~\hfill\QED
\end{property}

In the following, we will give some algebraic conditions on full cover. For presentation convenience, we introduce some notations. For any PSD matrix $\mathbf{P}\in\mathbb{R}^{N\times N}$ and $\forall\mathbf{h}\in\mathbb{R}_+^{N}$, notation $\bar{\mathbf{h}}_i$ denotes an $(N-1)\times1$ vector attained by deleting $i$-th entry from $\mathbf{h}$, $\bar{\mathbf{P}}_{ii}$ is the $(N-1)\times(N-1)$ sub-matrix  formed by deleting $i$-th row and $i$-th column from $\mathbf{P}$, and $\bar{\mathbf{p}}_{i}$ is the $(N-1)\times1$ vector generated by deleting $i$-th entry from the $i$-th row of $\mathbf{P}$. If there exist $n$ negative entries in $\bar{\mathbf{p}}_i$, then, for presentation simplicity, we denote the indexes of the negative entries of $\bar{\mathbf{p}}_i$  by $i_1^{(-)},~\cdots,~i_n^{(-)}$, where $0\le n\le N-1$. The discriminant of the equation $p_{ii}h_i^2+2h_i\bar{\mathbf{p}}_i^T\bar{\mathbf{h}}_i+
\bar{\mathbf{h}}_i^T\bar{\mathbf{P}}_{ii}\bar{\mathbf{h}}_i=0$ with respect to $h_i$ is given by
\begin{eqnarray}
\Delta_i\triangleq-4 \bar{\mathbf{h}}_i^T(p_{ii}\bar{\mathbf{P}}_{ii}-\bar{\mathbf{p}}_i\bar{\mathbf{p}}_i^T)\bar{\mathbf{h}}_i
\end{eqnarray}
Notice that $\bar{\mathbf{P}}_{ii}-\frac{\bar{\mathbf{p}}_i\bar{\mathbf{p}}_i^T}{p_{ii}}$ is the Schur complement of $\bar{\mathbf{P}}_{ii}$, which tells us that $p_{ii}\bar{\mathbf{P}}_{ii}-\bar{\mathbf{p}}_i\bar{\mathbf{p}}_i^T$ is a PSD matrix if $p_{ii}\neq0$, since $\mathbf{P}$ is PSD. Thus, we have the following
\begin{eqnarray}\label{eqn:discreminat}
  \forall i=1,2,~\cdots,~N, \Delta_i\le0
\end{eqnarray}
Then, we form an $n\times n$ sub-matrix, denoted by $\bar{\mathbf{C}}_{i_1 i_2\cdots i_n}^{(-)}$, using the rows and  columns of $p_{ii}\bar{\mathbf{P}}_{ii}-\bar{\mathbf{p}}_i\bar{\mathbf{p}}_i^T$ indexed by $i_1^{(-)},~\cdots,~i_n^{(-)}$.

\begin{theorem} \label{theorem:full_cover_algebraic}
For nonzero PSD matrix $\mathbf{P}\in\mathbb{R}^{N\times N}$, the following statements are true.
\begin{enumerate}
\item Suppose that there exists at least one $i$ such that $\bar{\mathbf{p}}_i$ is nonnegative, i.e., $\{\bar{\mathbf{p}}_i:\bar{\mathbf{p}}_i\in\mathbb{R}_+^{N-1},i=1,2,~\cdots,~N\}\neq\emptyset$. Then, $\mathbf{P}$ has full-cover if and only if  all the diagonal entries of $\mathbf{P}$ are nonzero and  $\bar{\mathbf{P}}_{ii}$ has full-cover for any $i$ satisfying $\bar{\mathbf{p}}_i\in\mathbb{R}_+^{N-1}$.
\item If there exists $i$ such that $p_{ii}>0$ and $(p_{ii}\bar{\mathbf{P}}_{ii}-\bar{\mathbf{p}}_i\bar{\mathbf{p}}_i^T)$ has full-cover, then, $\mathbf{P}$ has full-cover.\footnote{The necessity does not holds. An counterexample is given by $\mathbf{P}=\left(\begin{array}{lll}
1&1&1\\
1&1&1\\
1&1&1    \end{array}\right)$.}
\item  $\mathbf{P}$ has full-cover if and only if all the principal sub-matrices of $\mathbf{P}$ have full-cover.
\item Suppose that there exists $\bar{\mathbf{p}}_i$ having at least one negative entry, say, $\{\bar{\mathbf{p}}_i:\bar{\mathbf{p}}_i\notin\mathbb{R}_+^{N-1},i=1,2,~\cdots,~N\}\neq\emptyset$. If $\mathbf{P}$ has full-cover, $\bar{\mathbf{C}}_{i_1 i_2\cdots i_n}^{(-)}$ has full-cover for any $i$ satisfying $\bar{\mathbf{p}}_i\notin\mathbb{R}_+^{N-1}$.\footnote{The sufficiency  is not true since there exits an counterexample given by $\mathbf{P}=\left(\begin{array}{lll}
1&-1&0\\
-1&1&0\\
0&0&1    \end{array}\right)$.}~\hfill\QED
\end{enumerate}
 \end{theorem}

The detailed proof of Theorem~\ref{theorem:full_cover_algebraic} is provided in Appendix~\ref{app:sufficient_necessary}.

In addition, for the block-diagonal PSD matrix, we have the following property on full cover.
\begin{property}\label{property:block_diagonal}
Let each $\mathbf{P}_\ell$ for $\ell=1, 2, \cdots, L$ denote $N_\ell\times N_\ell$ PSD matrix. Then, the block diagonal PSD matrix $\mathbf{P}=\textrm{diag}\left(\mathbf{P}_1,~\cdots,~\mathbf{P}_L\right) $ has full-cover if and only if all $\mathbf{P}_\ell$ have full-cover.~\hfill\QED
\end{property}

The proof of Property~\ref{property:block_diagonal} is postponed into Appendix~\ref{app:block_diagonal}.

\subsection{Super-Rectangular Cover Order}
Our primary target in this subsection is twofold. On one hand, we will present cover properties to effectively and efficiently determine the cover order. On the other hand, the properties related to the zero cover case will be investigated.

\subsubsection{Cover Order Determination}
For completeness of exposition, we begin by introducing some related results extracted from \cite{roman1992advanced}.
\begin{proposition}\label{proposition:non_full_cover}
 Let $\mathbb{S}$ be a subspace of $\mathbb{R}^N$ and $\mathbb{S}^{\bot}$ be the orthogonal complementary subspace of $\mathcal{S}$. Then,
 \begin{enumerate}
   \item $\mathbb{S}\cap\mathbb{R}_+^N=\emptyset$ if and only if $\mathbb{S}^{\bot}\cap\mathbb{R}_{++}^N\neq\emptyset$.
   \item $\mathbb{S}\cap\mathbb{R}_{++}^N=\emptyset$ if and only if $\mathbb{S}^{\bot}\cap\mathbb{R}_{+}^N\neq\emptyset$.~\hfill\QED
 \end{enumerate}
\end{proposition}

Denote the row space of $\mathbf{A}$ by $\mathbb{S}_{\mathbf{A}}$ and the kernel space of $\mathbf{A}$ by $\mathbb{S}_{\mathbf{A}}^{\bot}$. Using Proposition~\ref{proposition:non_full_cover}, we now develop a necessary and sufficient condition to determine the cover order of a matrix, which is stated as the following theorem.

\begin{theorem}\label{theorem:positve_vector}
Let notation $\mathbb{\bar{R}}_{++}^K$ denote the set of all the nonnegative vectors with $K$ positive entries and specifically, $\mathbb{\bar{R}}_{++}^0$ be $\{\mathbf{0}_{N\times1}\}$. Then, the cover order of $\mathbf{A}^T\mathbf{A}$ is equal to  $\max_{\mathbb{S}_{\mathbf{A}}\cap\mathbb{\bar{R}}_{++}^{K}\neq\emptyset}K$.~\hfill\QED
\end{theorem}

The proof of Theorem~\ref{theorem:positve_vector} is given in Appendix~\ref{app:theorem_positive_vector}. Although it does not provide us with an explicit condition in terms of the matrix entries, Theorem~\ref{theorem:positve_vector}  indeed implicitly suggests us that we can perform a linear row transformation to determine the cover order of $\mathbf{A}^T\mathbf{A}$. In fact, by Theorem~\ref{theorem:positve_vector}, we know that if we can find some nonnegative vectors in $\mathbb{S}_{\mathbf{A}}$, then, the cover order of $\mathbf{A}^T\mathbf{A}$ is equal to the largest number of the positive entries of these vectors. To this end, we first attain the \textit{echelon form} of $\mathbf{A}$ by using row transformation and column permutation if necessary. Then, we constrain ourselves to \textit{positive elementary transformation} to assure that the first $R_r$ (matrix rank) columns of the transformed matrix have constant signs, and to construct a positive vector with as many positive entries as possible. These above procedures can be summarized as the following algorithm.

\begin{algorithm} \label{algor:cover_order}(\textbf{Cover Order Determination})
This algorithm consists of the following six progressive steps.
\begin{enumerate}
  \item \textit{Echelon form}. Find the elementary transformation matrix $\mathbf{E}$ such that
     \begin{eqnarray}\label{eqn:echelon}
      \mathbf{E}\mathbf{A}=
      \left(
          \begin{array}{ccc}
            \mathbf{I}_{R_r\times R_r}& \mathbf{\bar{A}}_0\\
             \mathbf{0}_{\left(N-R_r\right)\times R_r }&\mathbf{0}_{\left(N-R_r\right)\times\left(N-R_r\right) }
         \end{array}
       \right)
    \end{eqnarray}
\item \textit{Initialization}. Initialize $k=1$ and $R_c=R_r$.
\item\label{item:fail} \textit{Sign check: negative entries}. Check the signs of entries. Let $\mathbf{A}_0=\mathbf{\bar{A}}_0$. Denote the $i$-th entry of the $k$-th column of $\mathbf{A}_0$ by $a_{ik}$. If $a_{ik}\le 0$ for $i=1,~\cdots,~R_r$ and $\sum_{i=1}^{R_r}a_{ik}^2\neq0$, then, $R_c=0$ and jump to \ref{item:end}). In addition, if $k=N-R_r+1$, then, jump to~\ref{item:end}). Otherwise, go to~\ref{item:positive_transform}).
\item\label{item:positive_transform}\textit{Sign check: zero entries}. If $\forall i=1,\cdots,R_r, a_{ik}=0$, then, increase $k$ by 1 and go to \ref{item:fail}). Otherwise, go to \ref{item:bipolar}).
\item\label{item:bipolar}\textit{Sign check: bipolar entries}.  If there are $a_{ik}>0$ and $a_{jk}<0$ for $i\ne j$, then, multiply the $i$-th row of $\mathbf{A}_0$ by $-\frac{a_{jk}}{a_{ik}}$ and add the multiplied $i$ row to the $j$ row. Repeat this process  until the $k$-th column of transformed $\mathbf{A}_0$, $\mathbf{\bar{A}}_0$, is nonnegative.  Then, set $R_c=\min\{R_c+1,N\}$ and increase $k$ by 1 and go to \ref{item:fail}).
\item \label{item:end} \textit{Output}. Output the cover order $R_c$.~\hfill\QED
\end{enumerate}
\end{algorithm}

It should be noted that the cover order can be computed by employing the same operations on $\mathbf{A}^T\mathbf{A}$ or the eigen-matrix formed by the eigenvectors corresponding to the non-zero eigenvalues of $\mathbf{A}^T\mathbf{A}$, since these matrices have the same row space. Therefore, the method to determine the cover order in Algorithm~\ref{algor:cover_order} is not  unique. In addition, by Algorithm~\ref{algor:cover_order}, we can attain the following relationship between the rank and cover order of a non block-diagonal matrix.
\begin{property}\label{property:rank_cover}
$\forall\mathbf{A}\in\mathbb{R}^{M\times N}$, if $\mathbf{A}^T\mathbf{A}$ is not block-diagonal, then, the cover order and the rank of $\mathbf{A}^T\mathbf{A}$ satisfy $R_{c}=0$ or $R_c\ge R_r$.~\hfill\QED
\end{property}

Proof: By Algorithm~\ref{algor:cover_order}, we know that if performing positive row transformation to $\bar{\mathbf{A}}_0$ in~\eqref{eqn:echelon} gives no non-negative row vector, then, the corresponding cover order is zero, i.e., $R_c=0$. If $R_c\neq0$ and $\mathbf{A}^T\mathbf{A}$ is not block-diagonal, then, transforming $\bar{\mathbf{A}}_0$ with positive row operations will give us a non-negative vector with at least $R_r$ positive entries. Hence, we have $R_c\ge R_r$. This completes the proof of Property~\ref{property:rank_cover}.~\hfill$\Box$

Here, it should be pointed out that when $\mathbf{A}^T\mathbf{A}$ is block-diagonal, $R_c< R_r$ may happen. The following example serves to show this phenomenon.
\begin{example}For the following $4\times 4$ block-diagonal PSD matrix
 \begin{eqnarray*}
 \mathbf{A}^T\mathbf{A}=\left({
  \begin{array}{cccc}
 2&1&0&0\\
 1&2&0&0\\
  0&0&1& -1\\
  0&0&-1& 1\\
  \end{array}
  }\right)
 \end{eqnarray*}
 its cover order is given by $R_c=2$. However, the rank of this matrix is equal to $R_r=3$, giving thus, $R_c<R_r$.
~\hfill\QED
\end{example}

\begin{figure}[t]
    \centering
    \resizebox{8cm}{!}{\includegraphics{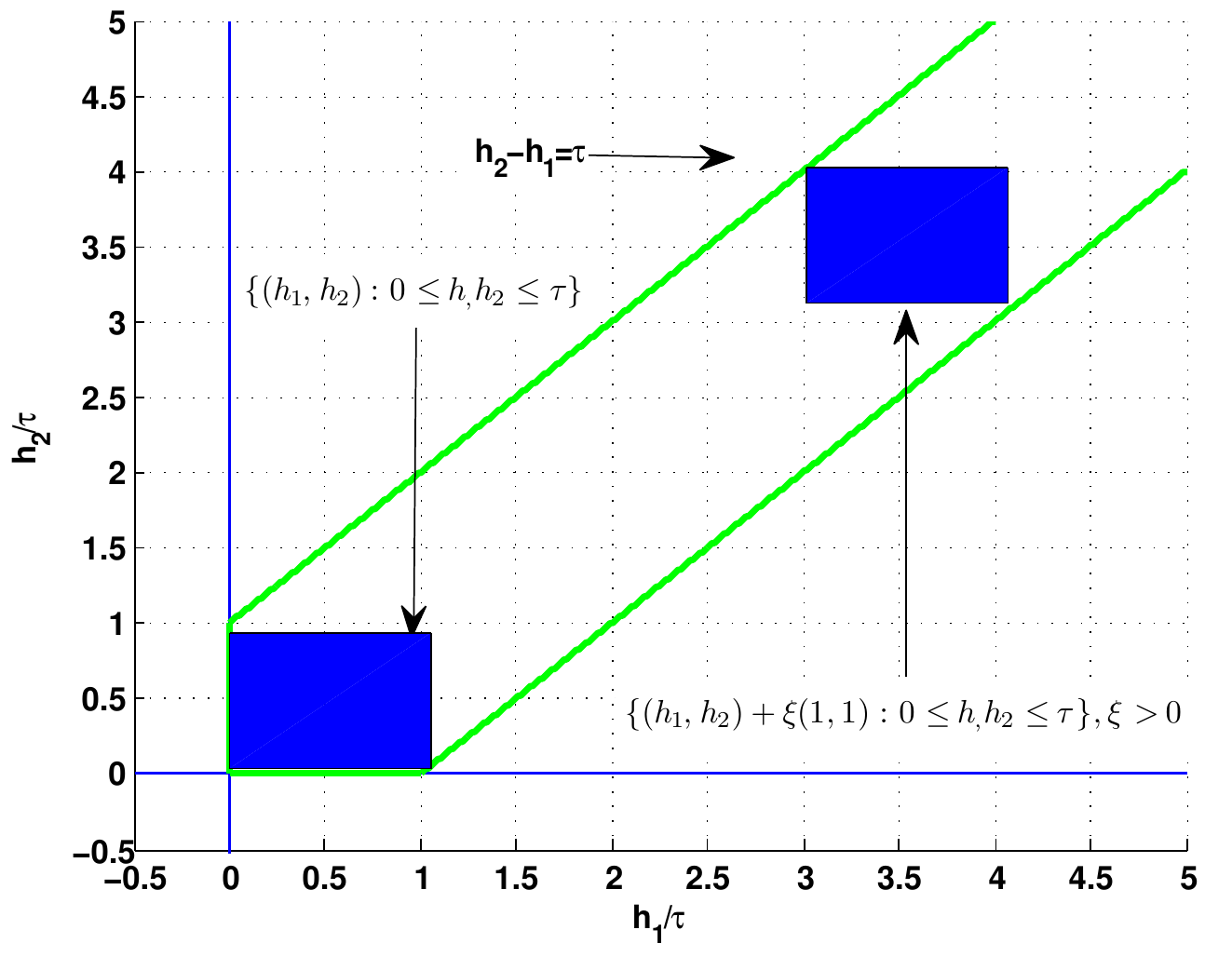}}
    \centering \caption{Example of a zero-cover $2\times 2$ PSD matrix.}
    \label{fig:property_example}
\end{figure}
\subsubsection{Cover Order Properties}
In the following, our main task is to reveal the properties of inner super-rectangle in the set $\left\{\mathbf{h}:0\le \mathbf{h}^T\mathbf{A}^T\mathbf{A}\mathbf{h}\le\tau^2,\mathbf{h}\in\mathbb{R}_+^{N}\right\}$. These properties will be shown to be closely related to zero-cover. To make our idea more understandable, we revisit our Example~\ref{example:full_degenerate} by providing new insight into Fig.~\ref{fig:exampe_zero_cover}, which is illustrated in Fig.~\ref{fig:property_example}. From Fig.~\ref{fig:property_example}, it is noticed that inside domain determined by $\{(h_1,h_2):\left(h_1-h_2\right)^2\le \tau^2, h_1,h_2\ge0\}$, there is a largest inner square with four vertexes being $(0,0),(0,1), (1,0)$ and $(1,1)$. More importantly, we also notice that shifting this square in the direction $\xi(1,1)$ for any positive number $\xi$ will gives us a new square which is still inside $\{(h_1,h_2):\left(h_1-h_2\right)^2\le \tau^2, h_1,h_2\ge0\}$. Now, we formally state this important observation in a generic case as the following theorem, whose detailed proof is provided in Appendix~\ref{app:zero_cover}.
\begin{theorem}\label{theorem:zero_cover}
If the cover order of $N\times N$ nonzero PSD matrix $\mathbf{A}^T\mathbf{A}$ is $R_c$, $0\le R_c<N$, then, the following two statements are true.
\begin{enumerate}
  \item There exists a nonnegative vector $\mathbf{v}\in\bar{\mathbb{R}}_{++}^{N-R_c}$ such that\
   \begin{eqnarray*}
   &&\left\{\mathbf{h}+\xi\mathbf{v}:0\le h_i\le\frac{\tau}{\sqrt{N\lambda_{\max}}},1\le i\le N,\xi>0\right\}\nonumber\\
   &&\subseteq\left\{\mathbf{h}:0\le \mathbf{h}^T\mathbf{A}^T\mathbf{A}\mathbf{h}\le\tau^2,\mathbf{h}\in\mathbb{R}_+^{N}\right\}
   \end{eqnarray*}
   holds for any given positive constant $\tau$, where $\lambda_{\max}$ denotes the maximum eigenvalue of $\mathbf{A}^T\mathbf{A}$.
  \item The subprincipal matrix of $\mathbf{A}^T\mathbf{A}$ which is formed by the rows and columns  having the  same indexes as the positive entries of $\mathbf{v}$ has zero-cover.~\hfill\QED
\end{enumerate}
\end{theorem}
From Theorem~\ref{theorem:zero_cover}, we arrive at the following property without much difficulty and thus, its proof is omitted.
\begin{property}\label{property:zero_cover}
If an $N\times N$ PSD matrix $\mathbf{P}$ has zero-cover, then, there exists an $N\times1$ vector $\mathbf{v}$ with all entries being positive such that
\begin{eqnarray*}
&&\left\{\mathbf{h}+\xi\mathbf{v}:0\le h_i\le\frac{\tau}{N\sqrt{\lambda_{\max}}},1\le i \le N,\xi>0\right\}\nonumber\\
&&\subseteq\left\{\mathbf{h}:0\le \mathbf{h}^T\mathbf{P}\mathbf{h}\le\tau^2,\mathbf{h}\in\mathbb{R}_+^{N}\right\}
\end{eqnarray*}
holds for any given positive constant $\tau$, where $\lambda_{\max}$ is the maximum eigenvalue of $\mathbf{P}$.
~\hfill\QED
\end{property}

The following theorem reveals another important property for PSD matrices having $R_c$-cover.
\begin{theorem} \label{theorem:cover_link}
For any $N\times N$ PSD matrix $\mathbf{P}$ with cover order $R_c$ satisfying $1\le R_c<N$, its longest cover link is unique.~\hfill\QED
 \end{theorem}
 The proof of Theorem~\ref{theorem:cover_link} is postponed into Appendix~\ref{app:cover_link}.

\subsection{Super-Rectangular Cover Length}
In this subsection, we give the closed-form cover length for a two by two case and then, using this result leads us to a lower-bound of cover length.

For a $2\times2$ PSD matrix, we can arrive at the following  more precise result.
\begin{theorem}\label{theorem:2_by_2_full_diversity}
 A $2\times2$ PSD $\mathbf{P}$ has full-cover if and only if either $\mathbf{P}$ has full-rank or all its entries are positive.
 Furthermore,
 \begin{enumerate}
   \item If all the entries of $\mathbf{P}$ are positive, then, we have
           $c_i=\frac{1}{\sqrt{\left[\mathbf{P}\right]_{ii}}},i=1,2$.
   \item If $\mathbf{P}$ has full-rank and $[\mathbf{P}]_{12}\le 0$, then,
       $c_i=\sqrt{\left[\mathbf{P}^{-1}\right]_{ii}}$ for $i=1,2$.~\hfill\QED
 \end{enumerate}
\end{theorem}

The proof of Theorem~\ref{theorem:2_by_2_full_diversity} is provided in Appendix~\ref{app:two_by_two}.

\begin{theorem} \label{theorem:rank_one_cover_length}
For any $N\times N$ PSD matrix $\mathbf{P}\in\mathbb{R}^{N\times N}$, if  $\mathbf{P}$ has full-cover, then, all the $N$ cover lengths of $\mathbf{P}$ are lower-bounded by
\begin{eqnarray*}
  c_i\ge\frac{1}{\sqrt{\left[\mathbf{P}\right]_{ii}}},i=1,2,~\cdots,~N
\end{eqnarray*}
where all the equalities hold simultaneously if and only if all the entries of $\mathbf{P}$ are nonnegative. ~\hfill\QED
 \end{theorem}

Up to now, the discussions on the super-rectangular cover theory have been complete. Basically, this theory deals with the identification of matrix signals over nonnegative matrix channels by investigating the non-existence of the nonnegative solution to a linear equation $\mathbf{A}\mathbf{h}=\mathbf{0}$ with respect to variable vector $\mathbf{h}$. One of important elements in cover theory is cover order. As we will see in ensuing section, it will play the same significant role in IM/DD MIMO-OWC as the rank does in  MIMO-RF communications. Essentially, cover theory reveals the fact that any transmitted matrix signal can be uniquely recovered from a noise-free channel if and only if each error coding matrix has full cover.
In addition, just as full rank is necessary and sufficient condition for assuring full diversity over Rayleigh fading channels for MIMO-RF communications, we will prove that full cover is also a necessary and sufficient condition for assuring full diversity over log-normal fading channels for IM/DD MIMO-OWC. Therefore, full cover can be reasonably regarded a universal algebraic definition for charactering full reliability over any independent fading channel regardless of the probability density function of the channel coefficients. At this point,  our cover theory is very general for the signal unique identification of IM/DD MIMO-OWC with direct detection and thus, applicable to IM/DD MIMO-OWC over any fading channels such as log-normal~\cite{haas2002space,Liu2004itct,Filho2005el,navidpour2007itwc,Beaulieu2008itct,giggenbach2008fading}, Gamma-Gamma~\cite{vetelino2007fade,andrews1999theory,al2001mathematical}, K-distributed~\cite{andrews2001laser,sandalidis2008outage} and so on. Since the probability density functions of these distributions for IM/DD MIMO-OWC channels are usually complicated, a useful signal design criterion remains unavailable in literature. As an initial exploration, we will apply this cover theory to specifically analyze the diversity gain of space-time block coded IM/DD MIMO-OWC over the commonly used log-normal fading channels.

\section{Applications of Cover Theory to Log-Normal Fading Channels: Performance Criterion }\label{sec:performance_analysis}

In this section, we fully take advantage of  the cover theory developed in Section \ref{sub:cover_theory}  to  establish an error performance criterion for  the STBC design of IM/DD MIMO-OWC with an ML detector. Then, we prove that full cover is a necessary and sufficient condition for assuring full diversity.
\subsection{Diversity Analysis Through Super-Rectangular Cover}

For IM/DD MIMO-OWC, we specifically consider the log-normal fading channels, i.e.,  the entries of the channel matrix $\mathbf{H}$ are assumed to be independent and log-normal distributed, i.e., $h_{ij}=e^{z_{ij}}$, where $z_{ij}\sim\mathcal{N}\left(\mu_{ij},\sigma_{ij}^2\right)$ for $i=1,2,~\cdots,~N, j=1,2,~\cdots,~M$. Then, the probability density function (PDF) of $h_{ij}$ is
\begin{eqnarray}
f_{H_{ij}}\left(h_{ij}\right)=\frac{1}{\sqrt{2\pi}h_{ij}\sigma_{ij} }\exp\left(-\frac{\left(\ln h_{ij}-\mu_{ij}\right)^2}{2\sigma _{ij}^{2}}\right)
\end{eqnarray}
The PDF of $\mathbf{H}$ is $f_{\mathbf{H}}\left(\mathbf{H}\right)=\prod_{i=1}^{N}\prod_{j=1}^{M}f_{H_{ij}}\left(h_{ij}\right)$. We know from~\cite{forney98} that given a channel realization $\mathbf{H}\in\mathbb{R}_{+}^{N\times M}$ and a transmitted signal vector $\mathbf{X}\in\mathcal{X}$, the pair wise probability of transmitting $\mathbf{X}$ and deciding in favor of $\hat{\mathbf X}$ with the ML receiver is given by~\cite{forney98}
\begin{eqnarray}\label{eqn:ml_detection_pep_conditional}
P\left(\mathbf{X}\rightarrow\hat{\mathbf{X}}|\mathbf{H}\right)=Q\left(\frac{d(\mathbf{X}\rightarrow\hat{\mathbf{X}})}{2}\right) \end{eqnarray}
where $d^2(\mathbf{X}\rightarrow\hat{\mathbf{X}})=\frac{\rho}{N}\sum_{j=1}^{M}\mathbf{h}_j^T\Delta\mathbf{X}^{T}\Delta\mathbf{X}\mathbf{h}_j$, ${\mathbf h}_j$ denotes the $j$-th column of $\mathbf{H}$ for $j=1,\cdots,M$, $\rho$ is the SNR and  $\mathbf{X}\neq\hat{\mathbf{X}}$ and $\mathbf{X},\hat{\mathbf{X}}\in\mathcal{X}$.  Taking an average of~\eqref{eqn:ml_detection_pep_conditional}  over $\mathbf{H}$ yields
\begin{eqnarray} \label{eqn:ml_detection_pep2}
P(\mathbf{X}\rightarrow\hat{\mathbf{X}})
=\int_{{\mathbf H}\in {\mathbb R}_+^{N\times M}}P\left(\mathbf{X}\rightarrow\hat{\mathbf{X}}|\mathbf{H}\right)f_{\mathbf{H}}\left(\mathbf{H}\right)d\mathbf{H}
\end{eqnarray}
Just as we have mentioned in Section~\ref{sec:intro}, due to the log-normal PDF, dealing with integral~\eqref{eqn:ml_detection_pep2} for the extraction of its asymptotic dominant term is much more difficult than dealing with the one for MIMO-RF communications, where the PDF is Gaussian distributed. To the best knowledge of authors, there are some mathematical formulae available in literature only for the purpose of numerically computing the integral~\cite{haas2002space,Liu2004itct,Filho2005el,navidpour2007itwc,Beaulieu2008itct}. Recently, the authors in~\cite{zyyisit15} have proposed a new approach to handle the integral~\eqref{eqn:ml_detection_pep2} for the design of space codes in which the rank of $\Delta\mathbf{X}$ is one.
However, their method still seems difficult to be extended to a general guideline on the design of STBC. Now, by fully taking advantage of the super-rectangular cover theory established in Section~\ref{sub:cover_theory}, we successfully arrive at the following results.
\begin{theorem} \label{theorem:full_diversity_condition}  For any $\mathbf{X},\hat{\mathbf{X}}\in\mathcal{X}$ with $\mathbf{X}\neq\hat{\mathbf{X}}$, define quantity $\mathcal{D}_{l}\left(\Delta\mathbf{X}\right)$ as
\begin{eqnarray}\label{def:larg}
\mathcal{D}_{l}\left(\Delta\mathbf{X}\right)\triangleq -\lim_{\rho\rightarrow\infty}\frac{8\times \ln P\big(\mathbf{X}\rightarrow\hat{\mathbf{X}}\big)}{\ln^2 \rho}
\end{eqnarray}
Then,  $\mathcal{D}_{l}\left(\Delta\mathbf{X}\right)=\sum_{j=1}^M \sum_{k=1}^{R_c}\sigma_{i_k, j}^{-2}$ if and only if the cover order of $\Delta\mathbf{X}^{T}\Delta\mathbf{X}$ is $R_c$.
 ~\hfill\QED
\end{theorem}
The proof of Theorem~\ref{theorem:full_diversity_condition} is given in Appendix~\ref{app:cover_diversity}. We would like to make the following observations on Theorem~\ref{theorem:full_diversity_condition}.
\begin{enumerate}
\item It is known that for MIMO-RF communications through complex Gaussian fading channels, the corresponding quantity $\mathcal{D}\left(\Delta\mathbf{X}\right)$ for PEP $P\left(\mathbf{X}\rightarrow\hat{\mathbf{X}}\right)$ is defined as $\mathcal{D}\left(\Delta\mathbf{X}\right)\triangleq-\lim_{\rho\rightarrow\infty}\frac{\log P\left(\mathbf{X}\rightarrow\hat{\mathbf{X}}\right)}{\log \rho}$ and that $\mathcal{D}\left(\Delta\mathbf{X}\right)=M R_r$ if and only if the rank of $\Delta\mathbf{X}^H\Delta\mathbf{X}$ is $R_r$.
Essentially, this quantity characterizes that the error curve in this scenario decays polynomially. However, for MIMO-OWC, we cannot follow this definition, since some experiment evidences~\cite{haas2002space,Liu2004itct,Filho2005el,navidpour2007itwc,Beaulieu2008itct} and analysis~\cite{zyyisit15} have already indicated that the error curve decays exponentially not polynomially. This is a significant difference between MIMO-RF communications and MIMO-OWC. In fact, our definition~\eqref{def:larg} quantitatively characterizes the exponential decaying speed of the error curve (see the proof of Theorem~\ref{theorem:full_diversity_condition} and the following Theorem~\ref{theorem:mimo_ovlc_pep}).
\item Particularly when $\sigma_{i, j}=1$ for $i=1, 2, \cdots, N, j=1, 2, \cdots, M$,  Theorem~\ref{theorem:full_diversity_condition} reveals the fact that  $\mathcal{D}_{l}\left(\Delta\mathbf{X}\right)=M R_c$ if and only if the cover order of $\Delta\mathbf{X}^{T}\Delta\mathbf{X}$ is $R_c$. At this point, therefore,  the cover order plays the same important role in MIMO-OWC for the design of STBC as the rank does in MIMO-RF communications.
\item In general, a matrix of rank $R_r$ has several $R_r\times R_r$ submatrices with each having rank $R_r$. However, according to our Theorem~\ref{theorem:cover_link},  a matrix of cover order $R_c$ has only one submatrix which has cover order $R_c$, i.e., the longest cover link is unique.  It exactly tells us that all the positions of entries of the matrix link the specific random channel variables which are not only covered by the super-rectangle but contribute to $\mathcal{D}_{l}\left(\Delta\mathbf{X}\right)$ as well.

\end{enumerate}

Particularly for full cover, we can attain the following sharper error performance bounds than Theorem~\ref{theorem:full_diversity_condition} does.
\begin{theorem} \label{theorem:mimo_ovlc_pep}
If $\forall \mathbf{X},\hat{\mathbf{X}}\in\mathcal{X}$ with $\mathbf{X}\neq\hat{\mathbf{X}}$, $\Delta\mathbf{X}^{T}\Delta\mathbf{X}$ has full-cover, then, the average PEP of STBC for $N\times M$ IM/DD MIMO-OWC is asymptotically bounded by
   \begin{eqnarray} \label{eqn:general_stbc_design_MIMO}
    &&C_{L}\prod_{j=1}^{M}\prod_{i=1}^{N}\frac{1-\left(\frac{\ln\rho +2\ln \lambda_{\max}+2\mu_{ij}}{2\sigma_{ij}}\right)^{-2}}{\frac{\ln\rho +2\ln \lambda_{\max}+2\mu_{ij}}{2\sigma_{ij}}}\nonumber\\
&&\times
\exp\left(-\sum_{j=1}^{M}\sum_{i=1}^{N}\frac{\left(\ln\rho +2\ln \lambda_{\max}+2\mu_{ij}\right)^2}{8\sigma_{ij}^2}\right)
\nonumber\\
&&\le P(\mathbf{X}\rightarrow\hat{\mathbf{X}})
\le
P_U(\mathbf{X}\rightarrow\hat{\mathbf{X}})
\nonumber\\
&&+ o\left(\left(\frac{\rho}{\ln^2\rho}\right)^{-\frac{MN}{2}}
\exp\left(-\frac{\Omega\ln^2\rho}{8}\right)\right)
  \end{eqnarray}
 where
 \begin{eqnarray*}
 \left\{
 \begin{array}{lll}
 \Omega_i=\sum_{j=1}^M\sigma_{ij}^{-2},
 \Omega=\sum_{i=1}^N\Omega_i, \\ \tilde{\Omega}_i=\sum_{j=1}^M\mu_{ij}\sigma_{ij}^{-2}, \tilde{\Omega}=\sum_{i=1}^N\tilde{\Omega}_i, \\ C_{\min}=\min_{\mathbf{z}\in\mathbb{R}_+^N\cap\left\{\mathbf{z}:\|\mathbf{z}\|_2^2=1\right\}}\mathbf{z}^T\Delta\mathbf{X}^T\Delta\mathbf{X}\mathbf{z}, \\ C_{L}=\frac{\sqrt{\prod_{j=1}^M\prod_{i=1}^N\sigma_{ij}^{2}}Q\left(\sqrt{\frac{M}{4N}}\right)}{\sqrt{2\pi}\prod_{i=1}^N\lambda_{\max}^{\tilde{\Omega}_i}} . \end{array}
 \right.
 \end{eqnarray*}
  and
\begin{eqnarray*}
 {P}_U(\mathbf{X}\rightarrow\hat{\mathbf{X}})&=&C_{U}\left(\frac{\rho}{\ln^2\rho}\right)^{\frac{1}{2}\ln\prod_{i=1}^Nc_i^{\Omega_i}-\frac{1}{2}\Omega\ln\left(N\Omega\right)-\frac{1}{4}\tilde{\Omega}}
 \nonumber\\
&\times&\left(\ln\rho\right)^{-MN}\exp\left(-\frac{\Omega\ln^2\frac{\rho}{\ln^2\rho}}{8}\right)
 \end{eqnarray*}
where
\begin{eqnarray*}
C_{U}=\frac{\left({\Omega C_{\min}}/{N^2}\right)^{-{MN}/{2}} }{2\sqrt{\prod_{j=1}^M\prod_{i=1}^N\sigma_{ij}^{2}}} \prod_{i=1}^Nc_i^{\tilde{\Omega}_i-\Omega_i\ln\left(N\Omega\right)}
\end{eqnarray*}
~\hfill\QED
\end{theorem}

The detailed proof of Theorem~\ref{theorem:mimo_ovlc_pep} is postponed into Appendix~\ref{app:mimo_ovlc_pep}.

\subsection{Super-Rectangular Cover Criteria for STBC Designs}
For given $M$, $N$ and the channel statistical parameters, the following \textit{three} factors dictate the minimization of $P_{U}(\mathbf{X}\rightarrow\hat{\mathbf{X}})$, which are discussed in detail as follows.
\subsubsection{\textbf{Cover Order: Large-Scale Diversity Gain}}
The exponent $\Omega$ governs the behavior of the upper bound with respect to $\ln\frac{\rho}{\ln^2{\rho}}$. To keep the upper bound as low as possible, we should make  $\min \mathcal{D}_{l}\left(\Delta\mathbf{X}\right)$ as large as possible. In addition, the cover order of $\Delta\mathbf{X}^{T}\Delta\mathbf{X}$ also dictates the polynomial decaying in terms of $\ln\rho$ and thus, two scales of decaying  can be kept as low as possible  at the same time. Because of the above-mentioned reasons, we name $\mathcal{D}_{l}\left(\Delta\mathbf{X}\right)$ the \textit{large-scale diversity gain} for space-time coded IM/DD MIMO-OWC systems.

To address the remarkable difference between the error performance behaviour of space-time block coded IM/DD MIMO-OWC and MIMO-RF, we make the following remarks.
\begin{enumerate}
  \item [(i)] By Property~\ref{property:rank_cover}, if $R_c\neq0$ for a non block-diagonal coding matrix, then, $MR_c$ is larger than or equal to $M R_r$, where $R_r$ is the rank of the error coding matrix. This observation tells us that the diversity gain defined for MIMO-RF is upper-bounded by our defined large-scale diversity order, showing the difference between IM/DD MIMO-OWC and MIMO-RF in terms of the definition of diversity gain.
  \item [(ii)]The diversity order for MIMO-RF over complex-valued Gaussian channels is the negative power deciding the power-law decaying of the error curves, $\rho^{-M R_r}$. In addition, from the proof of Theorem~\ref{theorem:full_diversity_condition} in Appendix~\ref{app:cover_diversity}, for non full-cover STBC of IM/DD MIMO-OWC, the resulting power-law decaying is lower-bounded by $\rho^{-\frac{M(N-R_c)}{2}}$.
  \item [(iii)]Moreover, when $R_c=0$, the decaying speed of the error curves for IM/DD MIMO-OWC in high SNRs is polynomial, which will be further investigated in Subsection~\ref{subsec:an_example_degenerate_code}.
  \item [(iv)]When $R_c=N$, the power-law decaying of $\rho^{-\frac{M(N-R_c)}{2}}$ disappears and the decaying speed of the error curves is maximized, implying a full diversity achievement.
\end{enumerate}

Full large-scale diversity is achieved if and only if $\forall \mathbf{X}\ne \tilde{\mathbf{X}}\in \mathcal{X}, R_c=N$, i.e., $\Delta\mathbf{X}^{T}\Delta\mathbf{X}$ has full cover. This implies that the full large-scale diversity gain achieved by an ML detector depends on $\Delta\mathbf{X}$, which is decided by the type of signalling. Now, we can see clearly that the condition to guarantee the unique identification of signals in the noise-free case is equivalent to ensuring full large-scale diversity gain in the noisy case. Thus, when we design STBC, full large-scale diversity must be assured \textit{in the first place}.

 \subsubsection{\textbf{Geometrical Cover Volume: Small-Scale Diversity Loss}}
  $\prod_{i=1}^Nc_i^{\Omega_i}$ is defined as the \textit{small-scale diversity loss}~\footnote{The reason for the term ¡®loss¡¯ will be explained in the remarks on Theorem~\ref{theorem:worst_case_pep} in Subsection~\ref{subsec:optimal_linear}}, which affects the polynomial decaying in terms of $\frac{\rho}{\ln^2 \rho}$. Note that when $\Omega_1=~\cdots~=\Omega_N=1$,  $\prod_{i=1}^Nc_i$ is equal to the volume of the $N$-dimensional super-rectangle covering $\{\mathbf{H}:\Delta\mathbf{X}^{T}\Delta\mathbf{X}\le\tau^2,\mathbf{h}_j\in\mathbb{R}^{N}_{+},\tau>0\}$. Thus, when we design STBC, under the condition that full large-scale diversity is assured, $\max_{\mathbf{X},\tilde{\mathbf{X}}\in \mathcal{X},\mathbf{X}\neq\tilde{\mathbf{X}}}\prod_{i=1}^Nc_i^{\Omega_i}$ should be minimized to optimize the error performance  in terms of the small-scale diversity loss.
  In addition, by taking the logarithm of $\prod_{i=1}^Nc_i^{\Omega_i}$, we attain $\sum_{i=1}^N\Omega_i\ln c_i$, leading to a \textit{log-sum} form. This equivalent transformation leads us to the well-known log-sum inequality, which, in fact, is of significance to our optimal design of STBC. Thus, we extract this useful inequality from~\cite{cover91} as follows.
\begin{proposition}\label{pro:log-sum}
For any nonnegative real numbers, $x_1,x_2,~\cdots,~x_N$ and $y_1,y_2,~\cdots,~y_N$, it holds that
\begin{eqnarray*}
  \sum_{i=1}^Nx_i\log\frac{x_i}{y_i} \ge\left(\sum_{i=1}^Nx_i\right)\log\frac{\sum_{i=1}^Nx_i}{\sum_{i=1}^Ny_i}
\end{eqnarray*}
with equality if and only if $\frac{x_i}{y_i}=\frac{\sum_{i=1}^Nx_i}{\sum_{i=1}^Ny_i}$.~\hfill\QED
\end{proposition}
For our purpose, the log-sum inequality in Proposition~\ref{pro:log-sum} can be restated as the following lemma.
\begin{lemma}\label{lemma:product_power}
For any positive real numbers, $x_1,x_2,~\cdots,~x_N$ and $y_1,y_2,~\cdots,~y_N$, it holds that
\begin{eqnarray}
  \prod_{i=1}^Nx_i^{y_i}\le\prod_{i=1}^Ny_i^{y_i}\left(\frac{\sum_{i=1}^Nx_i}{\sum_{i=1}^Ny_i}\right)^{\sum_{i=1}^Ny_i}
\end{eqnarray}
where the equality holds if and only if $\frac{x_i}{y_i}=\frac{\sum_{i=1}^Nx_i}{\sum_{i=1}^Ny_i}$ for $i=1,~\cdots,~N$.~\hfill\QED
\end{lemma}

 \subsubsection{\textbf{Coding Gain}}
$C_{\min}$ is determined by the structure of $\Delta\mathbf{X}^T\Delta\mathbf{X}$ and  $\min_{\mathbf{X},\tilde{\mathbf{X}}\in \mathcal{X},\mathbf{X}\neq\tilde{\mathbf{X}}}C_{\min}$ should be maximized. Accordingly, $\min_{\mathbf{X},\tilde{\mathbf{X}}\in \mathcal{X},\mathbf{X}\neq\tilde{\mathbf{X}}}C_{\min}$ is called the \textit{coding gain}, which decides the horizontal shift of the error curve. Under the conditions that the large-scale diversity gain is maximized and that the small-scale diversity loss is minimized, if there is still some freedom left for further optimization of the coding gain, $\min_{\mathbf{X},\tilde{\mathbf{X}}\in \mathcal{X},\mathbf{X}\neq\tilde{\mathbf{X}}}C_{\min}$ should be maximized.

Thus far, we have established a general criterion for the STBC design of IM/DD MIMO-OWC. With this, we can provide a guideline for full large-scale diversity design in the ensuing section.

\section{Code Constructions For Log-Normal Channels Using Super-Rectangular Cover Criteria}\label{sec:code_costruction}
In this section, our main task is to systematically design STBCs for IM/DD MIMO-OWC over log-normal fading channels by using the established  general performance criterions in Section~\ref{sec:performance_analysis}. In particular, our code constructions are based on the following two basic assumptions.
\begin{enumerate}
\item \textit{Power constraint}. In this paper, we  assume that the average optical power $\frac{1}{2^K}\sum_{\mathbf{X}\in\mathcal{S}}\mathbf{1}^T\mathbf{X}\mathbf{1}$ is constrained, which is commonly used in practical modulated optical sources~\cite{hranilovic2003optical}.
\item\textit{Channel state information}. For the log-normal fading channels, we assume that the transmitter apertures have the knowledge of $[\Omega_1,\Omega_2,~\cdots,~\Omega_N]$, where $\Omega_i=\sum_{j=1}^M\sigma_{ij}^{-2},i=1,2,~\cdots,~N$.
\end{enumerate}

Under the aforementioned assumptions, we will design several classes of STBCs for IM/DD MIMO-OWC systems as follows.
 \begin{enumerate}
   \item Zero-cover codes (ZCCs) will be investigated in Subsections~\ref{subsec:an_example_degenerate_code} for block fading channels.  For ZCCs, a common lower-bound of PEP in the high SNR is derived. In particular, a specific example of ZCCs is provided to exemplify the existence of ZCCs and verify the necessity of our established error performance criterion through super-rectangular cover theory.
   \item Optimal linear STBCs are designed  in Subsection~\ref{subsec:optimal_linear} by maximizing the large-scale diversity gain and minimizing the small-scale diversity loss. The optimal linear STBCs are shown to be RC with an optimal power allocation. In addition, this optimal design admits a fast ML decoding algorithm.

   \item In Subsection~\ref{sec:non_linear}, the optimal non-linear space-time block coding structure is proposed. It will be proved that the design of the optimal non-linear STBC is reduced to the design of the optimal multi-dimensional constellation. Then, a specific energy-efficient multi-dimensional constellation from Diophantine equations is constructed to form a collaborative STBC in Subsection~\ref{sec:collaborative_code}.
   \item In Subsection~\ref{sec:golden_code}, an optimal linear STBC is constructed for fast fading channels by collaborating the transmitted signals of two successive channel uses and proved to be related to the Golden number $\frac{\sqrt{5}+1}{2}$. This specific linear STBC design has an important property that the resulting small-scale diversity gain is non-increasing against increasing constellation size, which mimics the nonvanishing STBC for MIMO-RF communications.
 \end{enumerate}

More details of code constructions are described in the ensuing subsections.

\subsection{Zero-Cover Codes} \label{subsec:an_example_degenerate_code}
The main purpose of this subsection is to examine the significance of our proposed super-rectangular cover criterion by giving a ``bad'' STBC,  for which there exists a zero-cover coding matrix $\Delta\mathbf{X}^T\Delta\mathbf{X}$. Therefore, we name this code ZCC. For ZCCs, we have the following common lower-bound of PEP.

\begin{property}\label{property:zero_pep}
If $\Delta\mathbf{X}^T\Delta\mathbf{X}$ is zero-cover, then, in the high SNR regimes, $P(\mathbf{X}\rightarrow\hat{\mathbf{X}})$ is lower-bounded by
\begin{eqnarray*}
P(\mathbf{X}\rightarrow\hat{\mathbf{X}})\ge C_{zcc}\rho^{-\frac{MN}{2}}
\end{eqnarray*}
where $C_{zcc}$ is a constant independent of SNR.~\hfill\QED
\end{property}

Property~\ref{property:zero_pep} can be proved by using Property~\ref{property:zero_cover} and following the similar techniques in Subsection~\eqref{sec:lower_bound} of Appendix~\ref{app:cover_diversity}. From Property~\ref{property:zero_pep}, it can be seen that the decaying speed of the ZCCs' PEP is not faster than $\rho^{-\frac{MN}{2}}$ and the large-scale diversity gain is zero.
To further exemplify this result,  we construct a specific ZCC.

\begin{example}\label{example:zcc}
Let us consider a STBC with the codeword matrix given by
\begin{eqnarray}\label{eqn:zero_cover}
 \mathbf{X}\left(\mathbf{s}\right)=
  \left({
  \begin{array}{cc}
  x_1& x_2\\
  x_1 & x_2\\
  \end{array}
  }\right)
\end{eqnarray}
For this example, the modulation  format is fixed to be OOK with 1 bit per channel use (pcu) and we vary the variance of channel coefficients. The transmitted signal vector is $\mathbf{x}=[x_1,x_2]^T $, where $x_1$ and $x_2$ are equally likely chosen  from $\{0,1\}$. The optical power is normalized by $E\left[x_1+x_2\right]=\frac{1}{2}+\frac{1}{2}=1$.
The autocorrelation error coding matrix is
\begin{eqnarray*}
  \mathbf{X}^T\left(\mathbf{e}\right)\mathbf{X}\left(\mathbf{e}\right) =
  \left({
  \begin{array}{cc}
  e_1^2& e_1 e_2\\
  e_1 e_2& e_2^2\\
  \end{array}
  }\right)
   \end{eqnarray*}
where $e_1,e_2\in\{0,\pm1\}$ and $e_1^2+e_2^2\neq0$.
For this coding structure, we have the following error events.
\begin{enumerate}
  \item $e_1e_2=0$. In this case,
    \begin{eqnarray*}
   \mathbf{X}^T\left(\mathbf{e}\right)\mathbf{X}\left(\mathbf{e}\right) =
  \left({
  \begin{array}{cc}
  1& 0\\
  0& 0\\
  \end{array}
  }\right)
  \end{eqnarray*}
  or
  \begin{eqnarray*}
   \mathbf{X}^T\left(\mathbf{e}\right)\mathbf{X}\left(\mathbf{e}\right) =
  \left({
  \begin{array}{cc}
  0& 0\\
  0& 1\\
  \end{array}
  }\right).
  \end{eqnarray*}
  For this typical event, $R_c=1$  (please see Example \ref{example:non_full_cover}) and as a result, $\min R_c\le 1$. Thus, full large-scale diversity  is not achieved.
  \item $e_1e_2=1$. In this instance,
   \begin{eqnarray*}
  \mathbf{X}^T\left(\mathbf{e}\right)\mathbf{X}\left(\mathbf{e}\right) =
  \left({
  \begin{array}{cc}
  1& 1\\
  1& 1\\
  \end{array}
  }\right)
   \end{eqnarray*}
$R_c=2$ by Example \ref{example:non_degenerate}. The full large-scale diversity is attained.

  \item $e_1e_2=-1$. Let us now consider some typical error events satisfying $e_1e_2=-1$ and the resulting autocorrelation error coding matrix is
\begin{eqnarray*}
  \mathbf{X}^T\left(\mathbf{e}\right)\mathbf{X}\left(\mathbf{e}\right) =
 \left({
  \begin{array}{cc}
  1& -1\\
  -1& 1\\
  \end{array}
  }\right)
   \end{eqnarray*}
It has been illustrated by Example~\ref{example:full_degenerate} that the cover order of $\mathbf{X}^T\left(\mathbf{e}\right)\mathbf{X}\left(\mathbf{e}\right)$ is zero. Hence, the large-scale diversity attained is zero.
\end{enumerate}
~\hfill\QED
\end{example}
In fact, for this specific ZCC given in Example~\ref{example:zcc},  we can obtain a tighter lower bound than that provided in Property~\ref{property:zero_pep}.
\begin{property}\label{theorem:degenerate_code}
In high SNR regimes, PEP of STBC in~\eqref{eqn:zero_cover} for $2\times 2$ IM/DD MIMO-OWC is lower bounded by
\begin{eqnarray*}
P\left(\mathbf{s}\rightarrow\mathbf{\hat{s}}\right)\ge
C_0 \rho^{-1}+o\left(\rho^{-1}\right)
\end{eqnarray*}
where $C_0$ is a constant independent of SNR.
  ~\hfill\QED
\end{property}
The detailed proof of Property~\ref{theorem:degenerate_code} is given in Appendix~\ref{app:mimo_degenerate}.

\begin{figure}[t]
    \centering
    \resizebox{8cm}{!}{\includegraphics{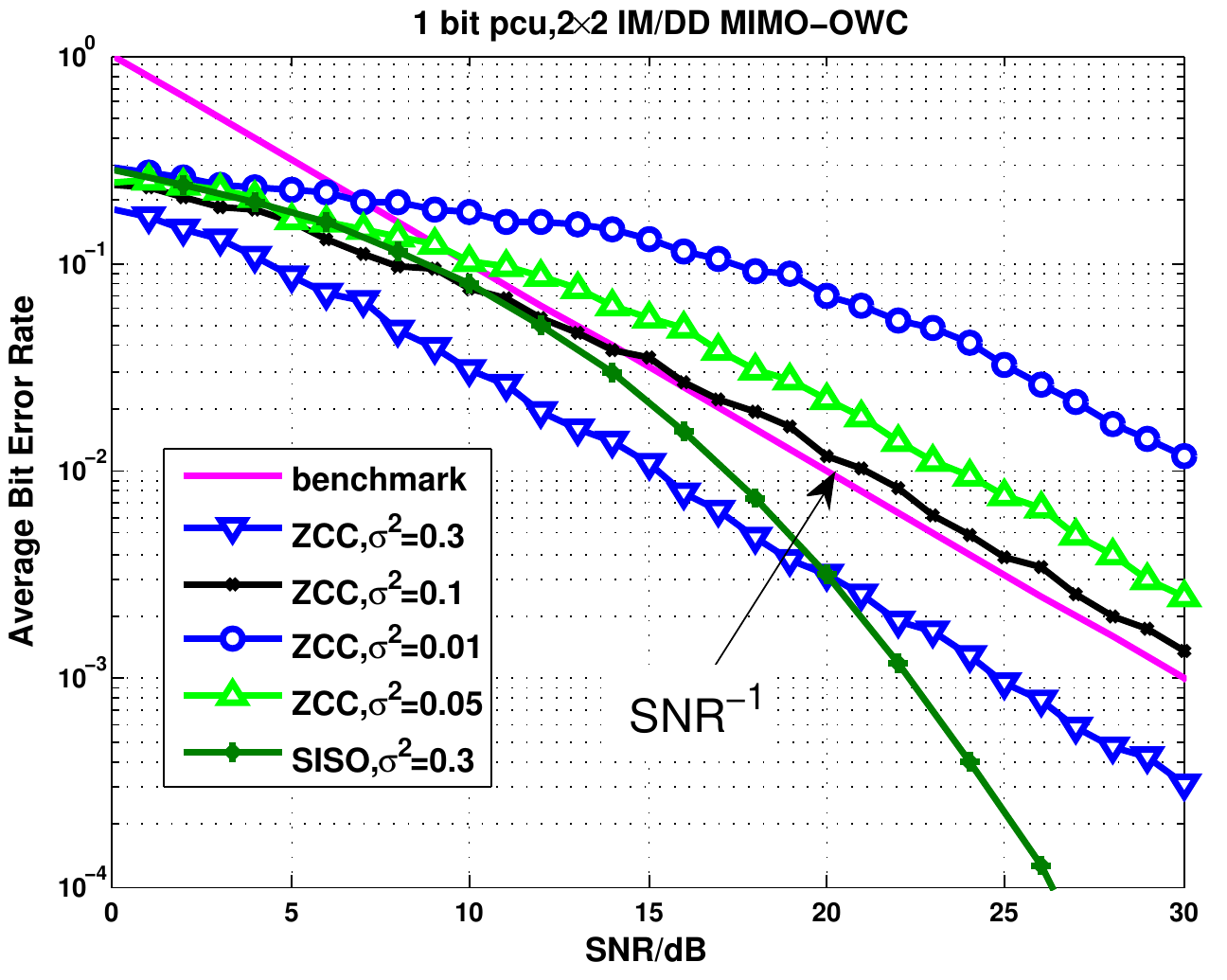}}
    \centering \caption{Error performance of zero-cover codes (ZCC) in~\eqref{eqn:zero_cover} for $2\times 2$ IM/DD MIMO-OWC.}
    \label{fig:zero_cover_code}
\end{figure}
Now, we show the simulation results of this ZCC in Fig.~\ref{fig:zero_cover_code}, which indicates that an improperly designed STBC may have error performance inferior to that of single-input single-output (SISO) schemes. Indeed, the STBC design of IM/DD MIMO-OWC differs significantly from its MIMO-RF counterpart. Generally speaking, the decaying rate of the error curves for OWC over log-normal fading channels is exponential. However, this never implies that all the transmission schemes can assure this rate. If the transmission scheme has zero-cover, then, the resulting decaying rate of error performance is not faster than power-law and thus, the attained large-scale diversity gain is zero. This observation gives us more insight into the design of STBC for the system. The existence  of ZCCs verifies the necessity  of  the full-cover condition, which is the \textit{main numerical motivation} of this paper. With this in mind, we proceed to construct ``good''  codes satisfying our newly developed super-rectangular cover criterion.

\subsection{Optimal Linear STBCs for Block Fading Channels}\label{subsec:optimal_linear}

This subsection aims at designing the optimal linear STBCs in terms of maximizing both the large-scale diversity gain and minimizing the small-scale diversity loss.

 For a linear STBC, the codeword matrix is given by
\begin{eqnarray}\label{eqn:codeword_stbc}
   \mathbf{X}\left(\mathbf{p}\right)= \sum_{\ell=1}^L\mathbf{A}_\ell p_\ell,p_\ell\in\mathcal{P}
 \end{eqnarray}
where $\mathbf{A}_\ell $ is an $L\times N $ nonnegative matrix, that is, $\mathbf{A}_\ell\in\mathbb{R}_+^{L\times N}$ and $\mathcal{P}$ is a nonnegative constellation satisfying $\mathcal{P}\subseteq\mathbb{R}_+$ and $|\mathcal{P}|=2^K$.
According to our established super-rectangular cover criterion, the optimal linear STBC design problem is formulated into what follows:
\begin{problem}\label{prob:optimal_linear} (Linear STBC Design).
 For fixed positive integers $L,K,M,N$, find $L$ matrices $\mathbf{A}_\ell\in\mathbb{R}_+^{L\times N}$ and a nonnegative constellation $\mathcal{P}\subseteq\mathbb{R}_+$
 such that 1) Any nonzero error coding matrix $\mathbf{X}\left(\mathbf{p}-\tilde{\mathbf{p}}\right)$ has full-cover;
 and 2) The worst-case small-scale diversity loss $\max_{\mathbf{p}\neq\tilde{\mathbf{p}}}\prod_{i=1}^N c_i^{\Omega_i}$
  is minimized, subject to a power constraint that~${\mathbb E}[{\mathbf 1}^T {\mathbf X}({\mathbf p}){\mathbf 1}]=\frac{\sum_{p\in\mathcal{P}}p}{2^K}\sum_{\ell=1}^L\mathbf{1}^T\mathbf{A}_\ell\mathbf{1}=L$.~\hfill\QED
\end{problem}

In order to attain a closed-form solution to this problem, we first establish a lower bound of the worst-case small-scale diversity loss  for any linear STBC and then, construct a specific linear STBC achieving this lower-bound. The final optimal solution is summarized as a theorem below:
\begin{theorem} \label{theorem:worst_case_pep}
An optimal solution to Problem~\ref{prob:optimal_linear} is given by
\begin{eqnarray}\label{eqn:linear_optimal}
  \mathbf{X}\left(\mathbf{p}\right)=\frac{2}{\Omega\left(2^K-1\right)}
  \left(\begin{array}{llll}
  \Omega_1p_1&\ldots&\Omega_Np_1\\
  \vdots&\ddots&\vdots\\
  \Omega_1p_L&\ldots&\Omega_Np_L
  \end{array}
  \right)
\end{eqnarray}
where $p_1,p_2,~\cdots,~p_L\in\{0,1,~\cdots,~2^K-1\}$. Furthermore, the minimum worst-case small-scale diversity loss obtained by an ML detector for $N\times M$ IM/DD MIMO-OWC systems is equal to the following
\begin{eqnarray}\label{eqn:union_bound_small_scale_diversity_gain}
\max_{\mathbf{p}\neq\tilde{\mathbf{p}}}\prod_{i=1}^N c_i^{\Omega_i}\ge\left(\frac{2^K-1}{2}\right)^{\Omega}\prod_{i=1}^N\left(\frac{\Omega}{\Omega_i}\right)^{\Omega_i}
\end{eqnarray}
~\hfill\QED
\end{theorem}
The detailed proof of Theorem~\ref{theorem:worst_case_pep} can be found in Appendix~\ref{app:optimal_linear_stbc}. In the following, we would like to make some comments on Theorem~\ref{theorem:worst_case_pep}.
\begin{enumerate}
  \item \textit{Optimal Space-Time Constellation}. Theorem~\ref{theorem:worst_case_pep} indicates that  the design of the optimal linear space-time constellation $\mathcal{X}$ is equivalent to constructing  the optimal one-dimensional unipolar constellation under the power-loaded spatial-repetitional coding structure. Intuitively speaking, for the nonnegative channels, the addition of multiple sub-channels in the propagation media is constructive and thus, the SNR-maximal at the receiver side.
  \item \textit{Optimality of RC}. When $\Omega_1=\Omega_2=~\cdots~=\Omega_N$, the optimal linear STBC can be specifically given in a form as follows.
      \begin{eqnarray}\label{eqn:repetition_code}
  \mathbf{X}\left(\mathbf{p}\right)=\frac{2}{N\left(2^K-1\right)}
  \left(\begin{array}{llll}
  p_1&p_1&\ldots&p_1\\
  p_2&p_2&\ldots&p_2\\
  \vdots& \vdots&\ddots&\vdots\\
  p_L&p_L&\ldots&p_L
  \end{array}
  \right)_{L\times N}
\end{eqnarray}
where $p_1,p_2,~\cdots,~p_L\in\{0,1,~\cdots,~2^K-1\}$. Thus, the optimal design in Theorem~\ref{theorem:worst_case_pep} also tells us that RC is the optimal linear STBC for IM/DD MIMO-OWC when $\Omega_1=\Omega_2=~\cdots~=\Omega_N$. This is the first formal proof of the optimality of RC as a linear STBC.

\item \textit{Small-Scale Diversity Loss}. From Theorem~\ref{theorem:worst_case_pep}, the lower-bound of $\max_{\mathbf{p}\neq\tilde{\mathbf{p}}}\prod_{i=1}^N c_i^{\Omega_i}$ is given by $\left(\frac{2^K-1}{2}\right)^{\Omega}\prod_{i=1}^N\left(\frac{\Omega}{\Omega_i}\right)^{\Omega_i}$, which is larger than one when $K>1$ and thus, whose logarithm is positive. For this reason, we  name $\prod_{i=1}^N c_i^{\Omega_i}$ the so-called \textit{small-scale diversity loss}.

\end{enumerate}

In the following, we provide a fast ML detection algorithm for the optimal design given by Theorem~\ref{theorem:worst_case_pep}. Notice that for the optimal linear STBC, the resulting MIMO channel model~\eqref{eqn:system_model} becomes
\begin{eqnarray}\label{eqn:opt_model}
\mathbf{y}_\ell=\mathbf{H}\mathbf x^{(\ell)}+\mathbf{n}_\ell,\ell=1,~\cdots,~L
\end{eqnarray}
where $\mathbf x^{(\ell)}=\left[\Omega_1,~\cdots,~\Omega_N\right]^T k$ with $k\in \{0, 1,~\cdots,~2^{K_\ell}-1\}$. For such channel model, a linear zero-forcing receiver is equivalent to the optimal ML receiver. Hence, the optimal estimate of the transmitted signal can be efficiently obtained below:

\begin{algorithm}\textbf{(Fast ML Detection)}:
 Given the received signal ${\mathbf y}_\ell$ and the non-zero channel matrix ${\mathbf H}$,  the output of  the ML detector for the optimal linear STBC is given by
 \begin{eqnarray}\label{eqn:fast_ml_decoding}
         \hat{\mathbf x}^{(\ell)}=
              \left\{\begin{array}{llll}
              \mathbf{0},& \frac{\mathbf{y}_\ell\mathbf{H}\mathbf{v}}{\|\mathbf{H} \mathbf{v}\|_2^2}<0,\\
              \left\lfloor\frac{\mathbf{y}_\ell\mathbf{H}\mathbf{v}}{\|\mathbf{H} \mathbf{v}\|_2^2}+\frac{1}{2}\right\rfloor\mathbf{v},& 0\le\frac{\mathbf{y}_\ell\mathbf{H}\mathbf{v}}{\|\mathbf{H} \mathbf{v}\|_2^2}\le 2^{K_\ell}-1,\\
              \left(2^{K_\ell}-1\right)\mathbf{v},&\frac{\mathbf{y}_\ell\mathbf{H}\mathbf{v}}{\|\mathbf{H} \mathbf{v}^T\|_2^2}> 2^{K_\ell}-1.
              \end{array}
              \right.
       \end{eqnarray}
       where $\mathbf{v}=\left(\Omega_1,~\cdots,~\Omega_N\right)^T$.
       ~\hfill\QED
       \end{algorithm}

On the other hand, we know from~\cite{simon98} that for any given channel realization $\mathbf{H}$ at the receiver, the exact conditional symbol error probability (SEP) for the channel model~\eqref{eqn:opt_model} with the PAM constellation using the ML detector is given by
 \begin{eqnarray}\label{eqn:conditional_codeword}
  {\rm P}_{{\rm s_\ell}|\mathbf{H}}=\frac{2\left(2^{K_\ell}-1\right)}{2^{K_\ell}}Q\left(\sqrt{\frac{\|\mathbf{H} \mathbf{v}\|_2^2}{2\sigma_{\mathbf{N}}^2}}\right)
 \end{eqnarray}
 This exact conditional SEP allows us to numerically and efficiently calculate the average codeword error probability by
\begin{eqnarray}\label{eqn:codeword_error}
{\rm P}_{\mathbf{s}|\mathbf{H}}=1-\prod_{\ell=1}^L\left(1-{\rm P}_{{\rm s_\ell}|\mathbf{H}}\right)
\end{eqnarray}
in the simulation section without running time-consuming computer simulations.

\subsection{Optimal Non-Linear STBCs for Block Fading Channels}\label{sec:non_linear}

In Subsection~\ref{subsec:optimal_linear}, Theorem~\ref{theorem:worst_case_pep} gives us an encouraging fact that an STBC design is indeed necessary and RC is not optimal in a general case. This result encourages us to further investigate the design of non-linear STBCs. To this end, we will give an optimal coding structure among all the full-cover coding structures.

\subsubsection{Optimal Space-Time Block Coding Structure}
Our main task in this subsection is to solve the following optimization problem:
\begin{problem} \label{prob:optimal_non_linear} (Non-linear STBC Design).
Let $\tilde{d}_{\min}(\mathcal{X})$ be determined by $\tilde{d}_{\min}(\mathcal{X})=\min_{\mathbf{X}\neq\tilde{\mathbf{X}},\mathbf{X},\tilde{\mathbf{X}}\in\mathcal{X} }\sum_{i=1}^N\|\mathbf{x}_i-\tilde{\mathbf{x}}_i\|_{2}$, where notation $\mathbf{x}_{i}$ denotes the $i$-th column of the matrix $\mathbf{X}$.
Under a constraint that $\tilde{d}_{\min}(\mathcal{X})=1$\footnote{This constraint is necessary for a full large-scale diversity gain achievement. Using this constraint for linear STBC design, we can arrive at the same result as Theorem~\ref{theorem:worst_case_pep}.}, for any given positive integers $J$, $L$ and $N$, find a constellation  $\mathcal{X}\subseteq\mathbb{R}_+^{L\times N}$ with $|\mathcal{X}|=2^J$ such that 1) the worse-case small-scale diversity loss $\max_{\mathbf{X}\neq\tilde{\mathbf{X}},\mathbf{X},\tilde{\mathbf{X}}\in\mathcal{X} }\prod_{i=1}^N c_i^{\Omega_i}$ is minimized , and 2) the average optical power $\frac{1}{2^J}\sum_{\mathbf{X}\in\mathcal{X}}\mathbf{1}^T\mathbf{X}\mathbf{1}$ is minimized.
~\hfill\QED
\end{problem}

\begin{theorem}\label{theorem:optimal_structure}
An optimal solution to Design Problem~\ref{prob:optimal_non_linear} is given by
\begin{eqnarray}\label{eqn:coding_structure}
\mathcal{X}=\left\{ \left(\begin{array}{llll}
  \Omega_1s_1&\Omega_2s_1&\ldots&\Omega_Ns_1\\
  \Omega_1s_2&\Omega_2s_2&\ldots&\Omega_Ns_2\\
  \vdots& \vdots&\ddots&\vdots\\
  \Omega_1s_L&\Omega_2s_L&\ldots&\Omega_Ns_L
  \end{array}
  \right),\mathbf{s}\in\mathcal{S}_{opt}\right\}
\end{eqnarray}
where
$\mathcal{S}_{opt}$ is the optimal solution to the following problem:
\begin{problem}\label{problem:multi_dimension_constellation} (Constellation Design).
 \textit{For any given positive integers $L$ and $K$, find a constellation $\mathcal{S}\subseteq\mathbb{R}_+^L$ with $|\mathcal{S}|=2^K$ such that the average optical power $\frac{1}{2^K}\sum_{\mathbf{s}\in\mathcal{S}}\mathbf{1}^T\mathbf{s}$ is minimized under a constraint that $\min_{\mathbf{s}\neq\tilde{\mathbf{s}}, \mathbf{s},\tilde{\mathbf{s}}\in\mathcal{S}}\|\mathbf{s}-\tilde{\mathbf{s}}\|_2=1$.}~\hfill\QED
\end{problem}
\end{theorem}
The proof of Theorem~\ref{theorem:optimal_structure} is provided in Appendix~\ref{app:optimal_coding_structure}.

\subsubsection{Orthogonal Equivalent Channel}

 For the spatial-repetitional coding structure defined in~\eqref{eqn:coding_structure}, by aligning the received matrix $\mathbf{Y}$, the channel noise matrix $\mathbf{N}$, and the signal code-channel product $\mathbf{X}\mathbf{H}$,  we can write
         \begin{eqnarray*}
         \left\{\begin{array}{lllll}
\mathbf{y}=\left[
    \begin{array}{llllllllll}
    y_{11}&\ldots&y_{1L}&\ldots&y_{M1}&\ldots&y_{ML}
    \end{array}
        \right]^T,\\
      \mathbf{n}=\left[
    \begin{array}{lllllllllllll}
    n_{11}&\ldots&n_{1L}&\ldots&n_{M1}&\ldots&n_{ML}
    \end{array}
        \right]^T,\\
       \mathcal{H}=\left(
    \begin{array}{lllllllll}
    \sum_{i=1}^N\sqrt{\Omega_{i}}h_{i1}\mathbf{I}_{L\times L}\\
        \sum_{i=1}^N\sqrt{\Omega_{i}}h_{i2}\mathbf{I}_{L\times L}\\
  ~~~~~~~~~~~~~\vdots\\
   \sum_{i=1}^N\sqrt{\Omega_{i}}h_{iM}\mathbf{I}_{L\times L}
    \end{array}
        \right),
         \mathbf{s}=\left(
    \begin{array}{ll}
    s_1\\
    \vdots\\
    s_L
    \end{array}
        \right)_{L\times 1}
        \end{array}
        \right.
  \end{eqnarray*}
  so that the original channel model $\mathbf{Y}=\mathbf{X}\mathbf{H}+\mathbf{N}$  can be rewritten in the following equivalent form
    \begin{eqnarray}\label{eqn:awgn}
    \mathbf{y}={\mathcal H}\mathbf{s}+\mathbf{n}
    \end{eqnarray}
Since ${\mathcal H}^H{\mathcal H}=\Big(\sum_{j=1}^M \big(\sum_{i=1}^N\sqrt{\Omega_i}h_{ij}\big)^2\Big){\mathbf I}_{L\times L}$, the channel model~\eqref{eqn:awgn} becomes an orthogonal matrix channel. Now, it becomes more clear that
      one of significant advantages of optimal STBC coding structure for the IM/DD MIMO-OWC system is like OSTBC for a MIMO-RF system, to transform the original MIMO channel into a scaled version of ideal MIMO channel~\eqref{eqn:awgn}, thereby significantly simplifying ML detection.
Therefore, a zero-forcing receiver is exactly equivalent to ML detection and thus, this coding structure has the potential to admitting fast ML detection if the time-collaborative constellation $\mathcal{S}$ is properly designed.

\subsubsection{Optimal Time-Collaborative Constellation}
From  Theorem~\ref{theorem:optimal_structure}, we know that design of the optimal non-linear $L\times N$ STBC is equivalent to constructing  the optimal $L$-dimensional constellations. Unfortunately, the optimal design of multi-dimensional constellations is a classic and long-standing problem in modern  wireless communications~\cite{forney1988coset1,forney1988coset2,forney1989multidimensional1,forney1989multidimensional2,conway1993sphere,forney98}.  Since the resulting discrete optimization problem for RF digital communications is extremely challenging to be formulated into a tractable optimization problem~\cite{gallager2008principles}, the systematic design of the optimal constellation, to the best knowledge of authors, still remains unsolved thus far. Nevertheless, even when $\Omega_1=\Omega_2=~\cdots~=\Omega_N$, the original RC based on PAM is not the optimal because of the optical power efficiency, which will be verified by the design examples in Section~\ref{sec:collaborative_code}.
 Theorem~\ref{theorem:optimal_structure} gives us  the optimal coding structure of the optimal non-linear STBC, which we name \textit{collaborative space-time block codes} (CSTBCs).

\subsection{Collaborative STBC Design From Diophantine Equations}\label{sec:collaborative_code}
In this subsection, we construct a novel multi-dimensional constellation  from Diophantine equations and then, examine its energy efficiency compared with the currently available schemes based on PAM.

\begin{theorem}\label{theorem:diophantine_conste}
For any given positive integers $L$ and $K$,
let
\begin{eqnarray*}
\mathcal{S}^{(L)}_{q}=\cup_{n=0}^{\lfloor\sqrt{L}\rfloor-1}\left\{\frac{n}{\lfloor\sqrt{L}\rfloor}\mathbf{1}_{L\times 1}+\mathbf{x}:\mathbf{1}^T\mathbf{x}=q,\mathbf{x}\in\mathbb{N}^L\right\}
\end{eqnarray*}
where notation $\lfloor x \rfloor$ denotes the largest integer number not larger than $x$. Then,  it holds that  $\min_{\mathbf{s}\neq\tilde{\mathbf{s}},\mathbf{s},\tilde{\mathbf{s}}\in\cup_{q=0}^{\infty}\mathcal{S}^{(L)}_{q}}\|\mathbf{s}-\tilde{\mathbf{s}}\|_2=1$.
~\hfill\QED
\end{theorem}
The proof of Theorem~\ref{theorem:diophantine_conste} is given in Appendix~\ref{app:diopantine_constellation}.
Since the constellation $\cup_{q=0}^{\infty}\mathcal{S}^{(L)}_{q}$ has infinite elements, we provide a strategy to construct an energy-efficient size-$2^K$ subset $\mathcal{S}^{(L,K)}$ of $\cup_{q=0}^{\infty}\mathcal{S}^{(L)}_{q}$ as follows.
\begin{itemize}
  \item \textbf{Rule 1:} Selecting the elements of $\mathcal{S}^{(L,K)}$ such that $\mathcal{S}^{(L,K)}\subseteq\cup_{q=0}^{\infty}\mathcal{S}^{(L)}_{q}$.
  \item \textbf{Rule 2:} Selecting the elements of $\mathcal{S}^{(L,K)}$ such that $\mathcal{S}^{(L,K)}$ is the set of the $2^K$ elements of $\cup_{q=0}^{\infty}\mathcal{S}^{(L)}_q$, which have the smallest total power, say, $\sum_{\mathbf{s}\in\mathcal{S}^{(L)}}\mathbf{1}^T\mathbf{s}\le \sum_{i=0}^{2^K}\mathbf{1}^T\mathbf{s}_i$ for any $\mathbf{s}_{0},\mathbf{s}_1,~\cdots,~\mathbf{s}_{2^K-1}\in\cup_{q=0}^{\infty}\mathcal{S}^{(L)}_q$.
  \item \textbf{Rule 3:} Selecting the elements of $\mathcal{S}^{(L,K)}$ such that the elements of $\mathcal{S}^{(L,K)}$ with the largest optical energy have the smallest neighbour points.
\end{itemize}

From our designed constellations, it can be seen that  for given nonnegative integers $q$ and $n$, the cardinality of $\{\frac{n}{\lfloor\sqrt{L}\rfloor}+\mathbf{x}:\mathbf{1}^T\mathbf{x}=q,\mathbf{x}\in\mathbb{N}^L\}$ is equal to the number of the nonnegative integer solution to the Diophantine equation $\mathbf{1}^T\mathbf{x}=q$, say, $\frac{(L+q)!}{L!q!}$. For this reason, we name our designed multi-dimensional constellation \textit{Diophantine constellation}. To make the constellation structures given by Theorem~\ref{theorem:diophantine_conste} more clear, some specific examples are presented below.
For notational simplicity, we denote the $L$-dimensional PAM-based constellation by
\begin{eqnarray*}
\mathcal{P}^{(L,K)}=\mathcal{P}^{(1,K_1)}\times \mathcal{P}^{(1,K_2)}\times~\cdots~\times\mathcal{P}^{(1,K_L)}
\end{eqnarray*}
where $\mathcal{P}^{(1,K_1)}=\{0,1,~\cdots,~2^{K_i}-1\}$ and $\sum_{i=1}^LK_i=K$ such that the optical power of $\mathcal{P}^{(L,K)}$ is minimized.  The following are some specific examples of $\mathcal{S}^{(L,K)}$.

\begin{example} When $L=1$, $\mathcal{S}^{(1,K)}=\{0,1,~\cdots,~2^K-1\}$, giving us a $2^K$-ary PAM constellation. In this example, $\mathcal{S}^{(1,K)}=\mathcal{P}^{(1,K)}$.~\hfill\QED
\end{example}

\begin{figure}[t]
    \centering
    \resizebox{9cm}{!}{\includegraphics{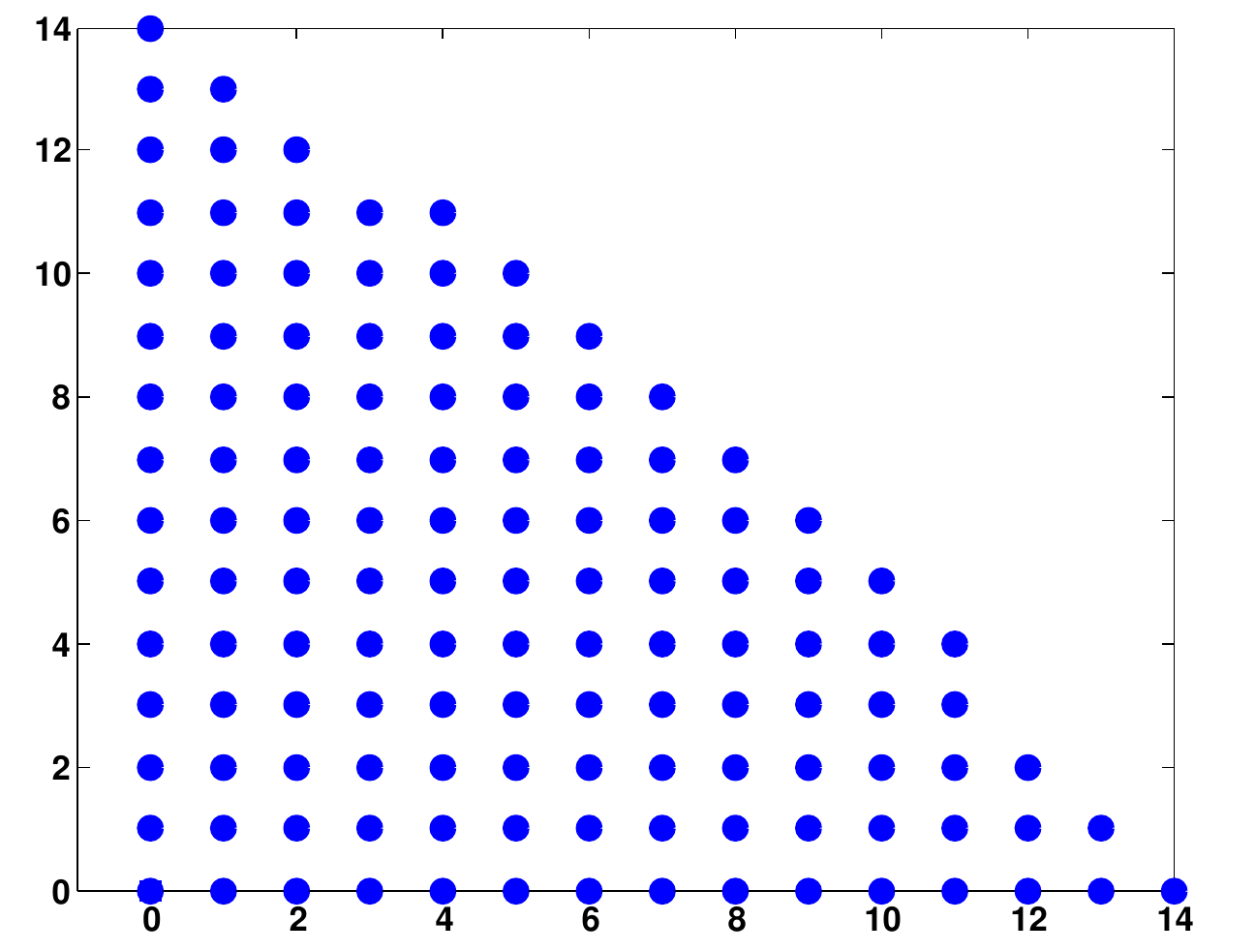}}
    \centering \caption{Two-dimensional constellations for $\mathcal{S}^{(2,7)}$.}
    \label{fig:constellation_figure2d}
\end{figure}

\begin{figure}[t]
    \centering
    \resizebox{9cm}{!}{\includegraphics{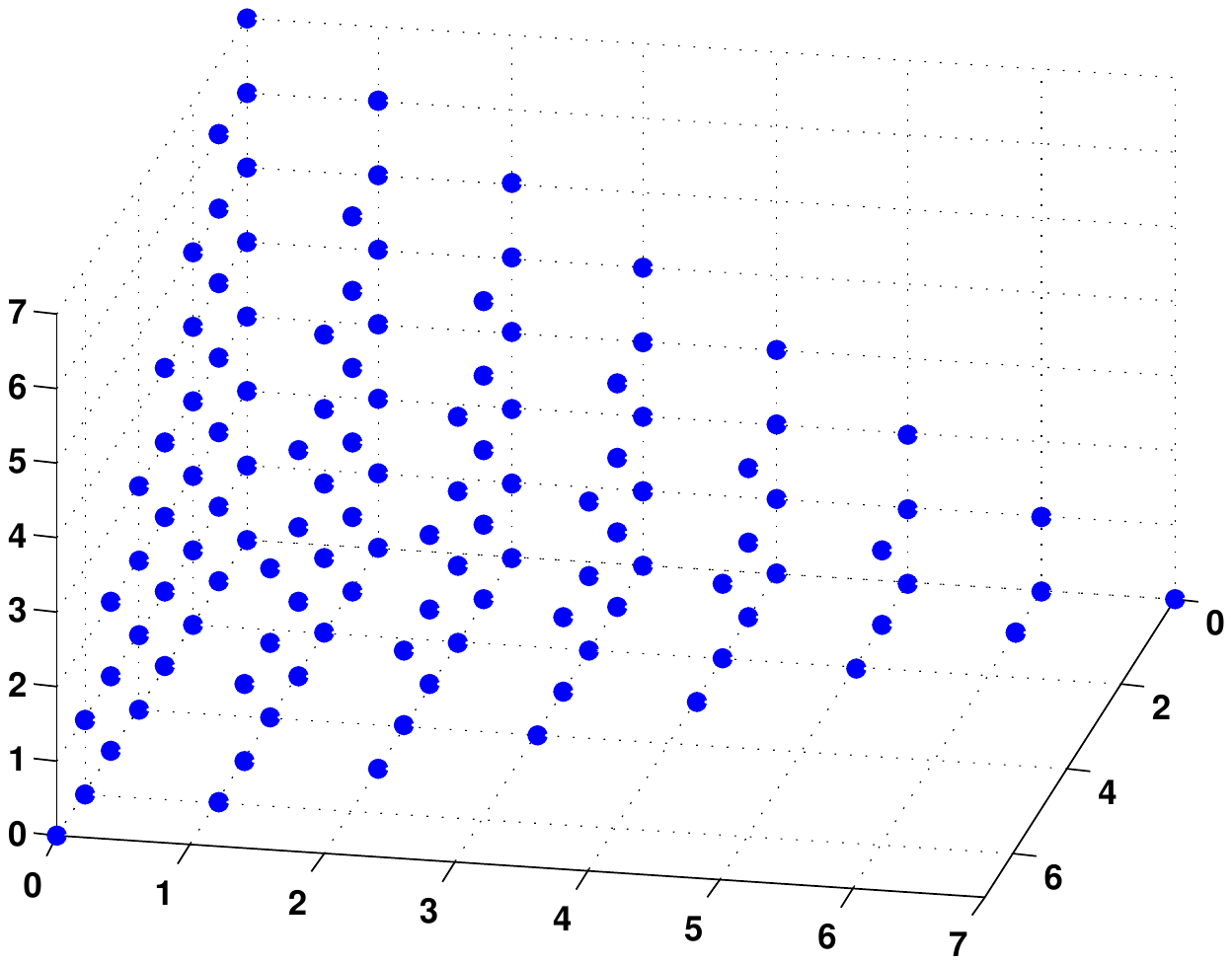}}
    \centering \caption{Three-dimensional constellations for $\mathcal{S}^{(3,7)}$.}
    \label{fig:constellation_figure3d}
\end{figure}

\begin{example}\label{example:integer} For any positive integers $L$ and $K$,  let $\mathcal{Z}^{(L,K)}$ be defined by
\begin{eqnarray*}
\mathcal{Z}^{(L,K)}=\cup_{q=0}^{Q-1}\mathcal{Z}^{(L)}_{q}\cup \bar{\mathcal Z}^{(L)}_Q\subseteq\cup_{q=0}^{\infty}\mathcal{S}^{(L)}_{q}
\end{eqnarray*}
 with
 \begin{eqnarray*}
\mathcal{Z}_{q}^{(L)}=\left\{\mathbf{x}:\mathbf{1}^T\mathbf{x}\mathbf{1}=q,\mathbf{x}\in\mathbb{N}^L\right\}
 \end{eqnarray*}
 where $|\mathcal{Z}^{(L)}_{q}|=\frac{(q+L-1)!}{q!(L-1)!}$, $Q$ is the smallest positive integer that satisfies $\frac{\left(Q+L\right)!}{L!Q!}\ge2^K$ and
\begin{eqnarray*}
\bar{\mathcal Z}^{(L)}_Q=\left\{\mathbf{s}_q\in\mathbb{N}^M:\mathbf{1}^T\mathbf{s}_q=Q,q,~\cdots,~2^{2K}-\frac{\left(Q+L\right)!}{L!Q!}\right\}
\end{eqnarray*}
In addition, for any $L$ and $K$, $\min_{\mathbf{s}\neq\tilde{\mathbf{s}},\mathbf{s},\tilde{\mathbf{s}}\in\mathcal{Z}^{(L,K)}}\|\mathbf{s}-\tilde{\mathbf{s}}\|_2=1$ since $\mathcal{Z}^{(L,K)}\subseteq\cup_{q=0}^{\infty}\mathcal{S}^{(L)}_{q}$. Furthermore, when $L=1,2,3$, $\mathcal{Z}^{(L,K)}=\mathcal{S}^{(L,K)}$.
\begin{enumerate}
  \item When $L=2$ and $K=4$, the elements of $\mathcal{S}^{(2,4)}$ are given by.
 \begin{eqnarray*}
   && \left(\begin{array}{llll}0\\0\end{array}\right),\left(\begin{array}{llll}1\\0\end{array}\right),\left(\begin{array}{llll}0\\1\end{array}\right),
   \left(\begin{array}{llll}1\\1\end{array}\right),
\nonumber\\&&
   \left(\begin{array}{llll}2\\0\end{array}\right),\left(\begin{array}{llll}0\\2\end{array}\right)
   \left(\begin{array}{llll}3\\0\end{array}\right),\left(\begin{array}{llll}0\\3\end{array}\right),
\nonumber\\ &&
  \left(\begin{array}{llll}1\\2\end{array}\right),\left(\begin{array}{llll}2\\1\end{array}\right),\left(\begin{array}{llll}4\\0\end{array}\right),\left(\begin{array}{llll}0\\4\end{array}\right),
\nonumber\\ && \left(\begin{array}{llll}1\\3\end{array}\right),\left(\begin{array}{llll}3\\1\end{array}\right),\left(\begin{array}{llll}2\\2\end{array}\right),\left(\begin{array}{llll}5\\0\end{array}\right).
  \end{eqnarray*}
  \item When $L=3$ and $K=5$, the elements of $\mathcal{S}^{(3,5)}$ are explicitly listed as follows.
 \begin{eqnarray*}
   && \left(\begin{array}{llll}0\\0\\0\end{array}\right),\left(\begin{array}{llll}1\\0\\0\end{array}\right),\left(\begin{array}{llll}0\\1\\0\end{array}\right),\left(\begin{array}{llll}0\\0\\1\end{array}\right),
   \nonumber\\ &&
   \left(\begin{array}{llll}2\\0\\0\end{array}\right),\left(\begin{array}{llll}0\\2\\0\end{array}\right)
   \left(\begin{array}{llll}0\\0\\2\end{array}\right),\left(\begin{array}{llll}1\\1\\0\end{array}\right),\nonumber\\
  &&
  \left(\begin{array}{llll}1\\0\\1\end{array}\right),\left(\begin{array}{llll}0\\1\\1\end{array}\right),\left(\begin{array}{llll}3\\0\\0\end{array}\right),\left(\begin{array}{llll}0\\3\\0\end{array}\right) \nonumber\\ &&\left(\begin{array}{llll}0\\0\\3\end{array}\right),\left(\begin{array}{llll}1\\2\\0\end{array}\right),\left(\begin{array}{llll}1\\0\\2\end{array}\right),\left(\begin{array}{llll}0\\1\\2\end{array}\right),
  \nonumber\\
  &&
  \left(\begin{array}{llll}0\\2\\1\end{array}\right),\left(\begin{array}{llll}2\\1\\0\end{array}\right),\left(\begin{array}{llll}2\\0\\1\end{array}\right),\left(\begin{array}{llll}1\\1\\1\end{array}\right),
  \nonumber\\ &&
  \left(\begin{array}{llll}4\\0\\0\end{array}\right),\left(\begin{array}{llll}0\\4\\0\end{array}\right),\left(\begin{array}{llll}0\\0\\4\end{array}\right),\left(\begin{array}{llll}3\\1\\0\end{array}\right)\nonumber\\
  &&\left(\begin{array}{llll}3\\0\\1\end{array}\right),\left(\begin{array}{llll}1\\3\\0\end{array}\right),\left(\begin{array}{llll}1\\0\\3\end{array}\right),\left(\begin{array}{llll}0\\1\\3\end{array}\right),
  \nonumber\\ &&
  \left(\begin{array}{llll}0\\3\\1\end{array}\right),\left(\begin{array}{llll}2\\2\\0\end{array}\right),
  \left(\begin{array}{llll}2\\0\\2\end{array}\right),\left(\begin{array}{llll}0\\2\\2\end{array}\right).
   \end{eqnarray*}
\end{enumerate}
In addition, the above constellation in Example~\ref{example:integer} is shaped like an isosceles right triangle in a 2-D case and an isosceles right vertebral in a 3-D case, respectively. To put the geometrical structure of the proposed constellation into perspective, we illustrate the constellations of $\mathcal{S}^{(2,7)}$ and $\mathcal{S}^{(3,7)}$ in Fig.~\ref{fig:constellation_figure2d} and Fig.~\ref{fig:constellation_figure3d}, respectively.
~\hfill\QED
\end{example}

\begin{example}
When $L=4$ and $K=6$, the elements of $\mathcal{S}^{(4,6)}$ are attained in the following.
 \begin{eqnarray*}
   && \left(\begin{array}{llll}0\\0\\0\\0\end{array}\right),\left(\begin{array}{llll}1\\0\\0\\0\end{array}\right),\left(\begin{array}{llll}0\\1\\0\\0\end{array}\right),\left(\begin{array}{llll}0\\0\\1\\0\end{array}\right),
   \nonumber\\ &&
   \left(\begin{array}{llll}0\\0\\0\\1\end{array}\right),\left(\begin{array}{llll}2\\0\\0\\0\end{array}\right)
   \left(\begin{array}{llll}0\\2\\0\\0\end{array}\right),\left(\begin{array}{llll}0\\0\\2\\0\end{array}\right),\nonumber\\
  &&
  \left(\begin{array}{llll}0\\0\\0\\2\end{array}\right),\left(\begin{array}{llll}1\\1\\0\\0\end{array}\right),\left(\begin{array}{llll}1\\0\\1\\0\end{array}\right),\left(\begin{array}{llll}1\\0\\0\\1\end{array}\right) \nonumber\\ &&\left(\begin{array}{llll}0\\1\\1\\0\end{array}\right),\left(\begin{array}{llll}0\\1\\0\\1\end{array}\right),\left(\begin{array}{llll}0\\0\\1\\1\end{array}\right),\frac{1}{2}\left(\begin{array}{llll}1\\1\\1\\1\end{array}\right),
  \nonumber\\
  &&
  \left(\begin{array}{llll}3\\0\\0\\0\end{array}\right),\left(\begin{array}{llll}0\\3\\0\\0\end{array}\right),\left(\begin{array}{llll}0\\0\\3\\0\end{array}\right),\left(\begin{array}{llll}0\\0\\0\\3\end{array}\right),
  \nonumber\\ &&
  \left(\begin{array}{llll}1\\2\\0\\0\end{array}\right),\left(\begin{array}{llll}1\\0\\2\\0\end{array}\right),\left(\begin{array}{llll}1\\0\\0\\2\end{array}\right),\left(\begin{array}{llll}2\\1\\0\\0\end{array}\right)\nonumber\\
  &&\left(\begin{array}{llll}2\\0\\1\\0\end{array}\right),\left(\begin{array}{llll}2\\0\\0\\1\end{array}\right),\left(\begin{array}{llll}0\\1\\2\\0\end{array}\right),\left(\begin{array}{llll}0\\1\\0\\2\end{array}\right),
  \nonumber\\ &&
  \left(\begin{array}{llll}0\\2\\1\\0\end{array}\right),\left(\begin{array}{llll}0\\2\\0\\1\end{array}\right),
  \left(\begin{array}{llll}0\\0\\2\\1\end{array}\right),\left(\begin{array}{llll}1\\1\\1\\0\end{array}\right),
\nonumber\\
  &&
  \left(\begin{array}{llll}1\\1\\0\\1\end{array}\right),\left(\begin{array}{llll}1\\0\\1\\1\end{array}\right),\left(\begin{array}{llll}0\\1\\1\\1\end{array}\right),\frac{1}{2}\left(\begin{array}{llll}3\\1\\1\\1\end{array}\right)
 \nonumber\\ &&\frac{1}{2}\left(\begin{array}{llll}1\\3\\1\\1\end{array}\right),\frac{1}{2}\left(\begin{array}{llll}1\\1\\3\\1\end{array}\right),\frac{1}{2}\left(\begin{array}{llll}1\\1\\1\\3\end{array}\right),\left(\begin{array}{llll}4\\0\\0\\0\end{array}\right),
\nonumber\\
  && \left(\begin{array}{llll}0\\4\\0\\0\end{array}\right),\left(\begin{array}{llll}0\\0\\4\\0\end{array}\right),\left(\begin{array}{llll}0\\0\\0\\4\end{array}\right),\left(\begin{array}{llll}1\\3\\0\\0\end{array}\right),
  \nonumber\\ &&
  \left(\begin{array}{llll}1\\0\\3\\0\end{array}\right),\left(\begin{array}{llll}1\\0\\0\\3\end{array}\right),\left(\begin{array}{llll}0\\1\\3\\0\end{array}\right),\left(\begin{array}{llll}0\\1\\0\\3\end{array}\right),\nonumber\\
    &&\left(\begin{array}{llll}0\\0\\1\\3\end{array}\right),\left(\begin{array}{llll}3\\1\\0\\0\end{array}\right),\left(\begin{array}{llll}3\\0\\1\\0\end{array}\right),\left(\begin{array}{llll}3\\0\\0\\1\end{array}\right),
    \nonumber\\ &&
    \left(\begin{array}{llll}0\\3\\1\\0\end{array}\right),\left(\begin{array}{llll}0\\3\\0\\1\end{array}\right),
\left(\begin{array}{llll}2\\2\\0\\0\end{array}\right),\left(\begin{array}{llll}2\\0\\2\\0\end{array}\right),
   \nonumber\\   &&
   \left(\begin{array}{llll}2\\0\\0\\2\end{array}\right),\frac{1}{2}\left(\begin{array}{llll}5\\1\\1\\1\end{array}\right),
\frac{1}{2}\left(\begin{array}{llll}1\\5\\1\\1\end{array}\right),\frac{1}{2}\left(\begin{array}{llll}1\\1\\5\\1\end{array}\right),
\nonumber\\ &&
\frac{1}{2}\left(\begin{array}{llll}1\\1\\1\\5\end{array}\right),\frac{1}{2}\left(\begin{array}{llll}3\\3\\1\\1\end{array}\right),\frac{1}{2}\left(\begin{array}{llll}3\\1\\3\\1\end{array}\right),\frac{1}{2}\left(\begin{array}{llll}3\\1\\1\\3\end{array}\right).
   \end{eqnarray*}
   ~\hfill\QED
\end{example}

\begin{figure}[t]
    \centering
    \resizebox{9cm}{!}{\includegraphics{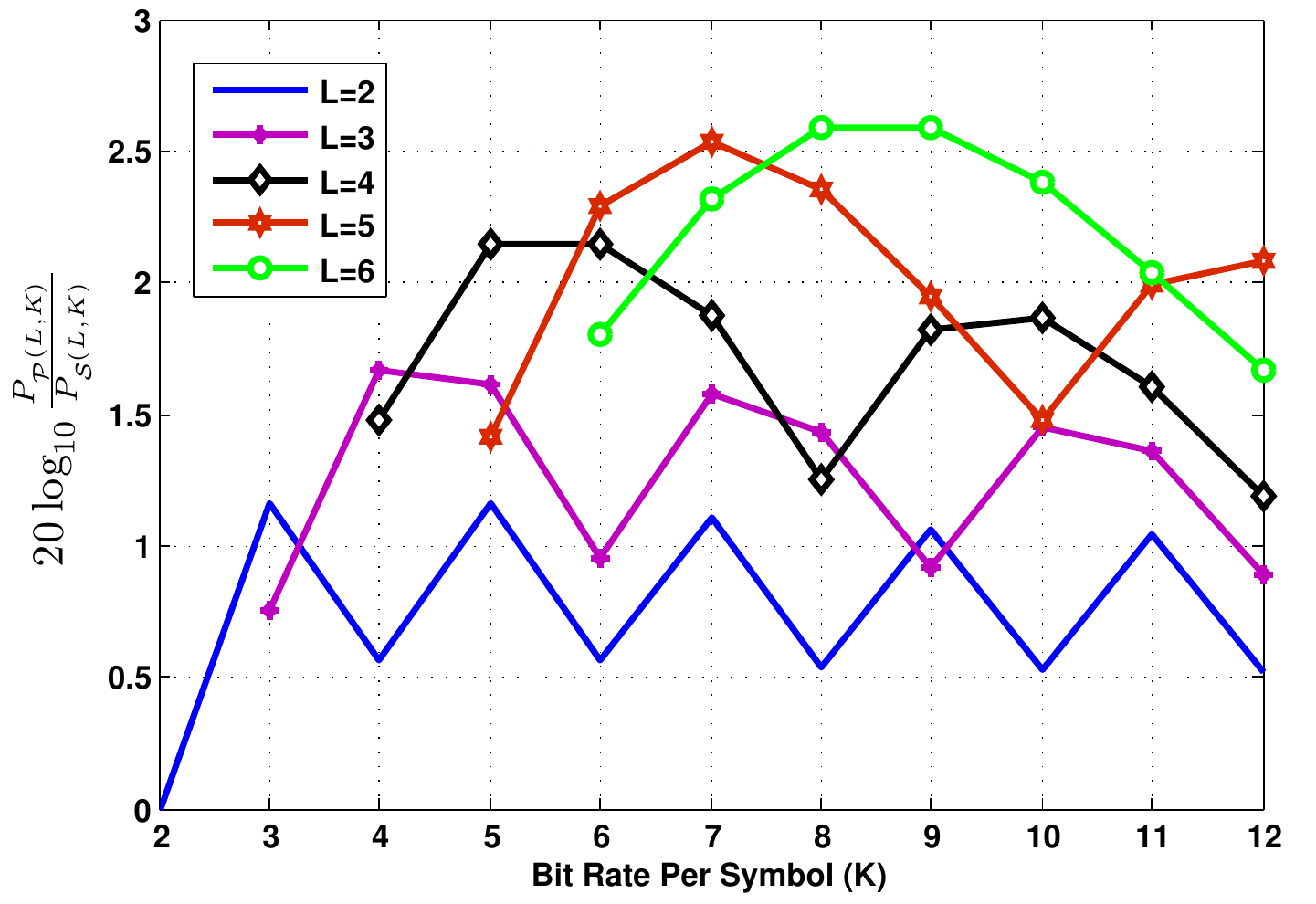}}
    \centering \caption{Energy efficiency comparisons between $\mathcal{S}^{(L,K)}$ and $\mathcal{P}^{(L,K)}$ with $P_{\mathcal{X}^{(L,K)}}=\frac{1}{2^K}\sum_{\mathbf{x}\in\mathcal{X}^{(L,K)}}\mathbf{x}^T\mathbf{1}$.}
    \label{fig:energy_efficiency}
\end{figure}

In Fig.~\ref{fig:energy_efficiency}, we show the energy efficiency superiority of $\mathcal{S}^{(L,K)}$ to $\mathcal{P}^{(L,K)}$ in the sense of producing a fixed minimum Euclidean distance. Moreover, we compare some numerical values of $P_{\mathcal{S}^{(L,K)}}=\frac{1}{2^K}\sum_{\mathbf{s}\in\mathcal{S}^{(L,K)}}\mathbf{s}^T\mathbf{1}$ and  $P_{\mathcal{P}^{(L,K)}}=\frac{1}{2^K}\sum_{\mathbf{p}\in\mathcal{P}^{(L,K)}}\mathbf{p}^T\mathbf{1}$.
From Fig.~\ref{fig:energy_efficiency},
 it can be seen that the energy advantage of our proposed Diophantine constellation over multi-dimensional PAM is dependent on the time dimension $L$  and the bit rate $K$ per symbol. Furthermore, when $L$ is sufficiently large, the advantage is substantial for any given $K$, as shown in Fig.~\ref{fig:energy_efficiency}.

\subsection{Golden Codes For Fast Fading Channels }\label{sec:golden_code}
In this section, we specifically design a linear STBC for $M\times N$ MIMO-OWC over two successive and independent channel uses. By solving the max-min design problem, we will prove that the optimal STBC is based on the Golden ratio $\frac{\sqrt{5}+1}{2}$ and thus, name this code the Golden Code.
\subsubsection{Design Problem}
 The symbols to be encoded are randomly and equally likely selected from the  unipolar PAM constellations. Specifically, we assume that $s_1\in\{0,1,~\cdots,~2^{K_1}-1\}$ and $s_2\in\{0,1,~\cdots,~2^{K_2}-1\}$, where  $K_1$ and $K_2$ are positive integers. For the first time slot and the second time slot, the respective transmitted signals are given as follows.
\begin{eqnarray*}
\mathbf{F}\mathbf{s}=\left(\begin{array}{ll}
f_{11}&f_{12}\\
\vdots&\vdots\\
f_{N1}&f_{N2}
\end{array}
\right)\left(\begin{array}{ll}
s_1\\
s_2
\end{array}
\right)
\end{eqnarray*}
and
\begin{eqnarray*}
\mathbf{G}\mathbf{s}=\left(\begin{array}{ll}
g_{11}&g_{12}\\
\vdots&\vdots\\
g_{N1}&g_{N2}
\end{array}
\right)\left(\begin{array}{ll}
s_1\\
s_2
\end{array}
\right)
\end{eqnarray*}
Accordingly, the received signals are given, respectively,  by
\begin{eqnarray*}
\mathbf{y}_1=\left(\mathbf{F}\mathbf{s}\right)^T\mathbf{H}_1+\mathbf{n}_1
\end{eqnarray*}
and
\begin{eqnarray*}
\mathbf{y}_2=\left(\mathbf{G}\mathbf{s}\right)^T\mathbf{H}_2+\mathbf{n}_2
\end{eqnarray*}
Hence, the equivalent channel model of these two channel uses can be written into
\begin{eqnarray*}
\left(
\begin{array}{ll}
\mathbf{y}_1\\
\mathbf{y}_2
\end{array}
\right)=
\left(\begin{array}{ll}
\left(\mathbf{F}\mathbf{s}\right)^T&\mathbf{0}\\
\mathbf{0}&\left(\mathbf{G}\mathbf{s}\right)^T
\end{array}
\right)
\left(\begin{array}{ll}
\mathbf{H}_1\\
\mathbf{H}_2
\end{array}
\right)+\left(\begin{array}{ll}
\mathbf{n}_1\\
\mathbf{n}_2
\end{array}
\right)
\end{eqnarray*}
Then, the resulting autocorrelation error coding matrix is determined by
\begin{eqnarray*}
\mathbf{X}^T\left(\mathbf{e}\right)\mathbf{X}\left(\mathbf{e}\right)=\left(\begin{array}{ll}
\mathbf{F}\mathbf{e}\mathbf{e}^T\mathbf{F}^T&\mathbf{0}\\
\mathbf{0}&\mathbf{G}\mathbf{e}\mathbf{e}^T\mathbf{G}^T
\end{array}
\right)
\end{eqnarray*}
By Property~\ref{property:block_diagonal}, we have the fact that $\mathbf{X}^T\left(\mathbf{e}\right)\mathbf{X}\left(\mathbf{e}\right)$ has full-cover if and only if both $\mathbf{F}\mathbf{e}\mathbf{e}^T\mathbf{F}^T$ and $\mathbf{G}\mathbf{e}\mathbf{e}^T\mathbf{G}^T$ have full-cover. In addition, each of nonzero $\mathbf{F}\mathbf{e}\mathbf{e}^T\mathbf{F}^T$ and $\mathbf{G}\mathbf{e}\mathbf{e}^T\mathbf{G}^T$ has rank-one. By Property~\ref{property:rank_one}, both $\mathbf{F}\mathbf{e}\mathbf{e}^T\mathbf{F}^T$ and $\mathbf{G}\mathbf{e}\mathbf{e}^T\mathbf{G}^T$ have full-cover if and only if all the entries of  $\mathbf{F}\mathbf{e}\mathbf{e}^T\mathbf{F}^T$ and $\mathbf{G}\mathbf{e}\mathbf{e}^T\mathbf{G}^T$ are positive.  Furthermore, by Theorem~\ref{theorem:rank_one_cover_length}, the cover lengths are given by $\frac{1}{|f_{i1}e_1+f_{i2}e_2|}$ and $\frac{1}{|g_{i1}e_1+g_{i2}e_2|}$. Therefore, to minimize the worst-case small-scale diversity loss, the optimal design problem can be equivalently formulated below:

\begin{problem}\label{prob:golden_code} (Golden STBC Design).
Given any positive integers $K_1$, $K_2$, $N$ and $M$, devise two $N\times 2$ matrices $\mathbf{F}$ and $\mathbf{G}$  with all their entries being nonnegative real-valued numbers such that
\begin{eqnarray}\label{eqn:design_criterion_variance}
&&\max_{\mathbf{F},\mathbf{G}}\min_{e_1,e_2} \prod_{i=1}^N\left(f_{i1}e_1+f_{i2}e_2\right)^{2\Omega_i}\left(g_{i1}e_1+g_{i2}e_2\right)^{2\Omega_i}
\nonumber\\
&&s.t.
\left\{\begin{array}{ll}
 f_{ij},g_{ij}>0,e_1^2+e_2^2\neq0,\\
 \sum_{i=1}^N\sum_{j=1}^2(f_{ij}+g_{ij})=1,\\
e_1\in\left\{0,\pm1,\ldots,\pm\left(2^{K_1}-1\right)\right\},\\
e_2\in\left\{0,\pm1,\ldots,\pm\left(2^{K_2}-1\right)\right\},\\
\forall i\neq j,\left(f_{i1}e_1+f_{i2}e_2\right)\left(f_{j1}e_1+f_{j2}e_2\right)>0,\\
\forall i\neq j,\left(g_{i1}e_1+g_{i2}e_2\right)\left(g_{j1}e_1+g_{j2}e_2\right)>0.
 \end{array}
\right.
\end{eqnarray}
where the constraint conditions
\begin{eqnarray*}
\left\{\begin{array}{ll}
\forall i\neq j,\left(f_{i1}e_1+f_{i2}e_2\right)\left(f_{j1}e_1+f_{j2}e_2\right)>0,\\
\forall i\neq j,\left(g_{i1}e_1+g_{i2}e_2\right)\left(g_{j1}e_1+g_{j2}e_2\right)>0.
 \end{array}
\right.
\end{eqnarray*}
are to assure a full large-scale diversity achievement.
~\hfill\QED
\end{problem}

\subsubsection{Optimal Design}
It turns out that the optimal solution to~\eqref{eqn:design_criterion_variance} is related to the Golden ratio $\frac{\sqrt{5}+1}{2}$.
\begin{theorem} \label{theorem:golden_code_optical}
 An optimal solution to Problem~\ref{prob:golden_code} is given by
\begin{eqnarray}\label{eqn:goledn_optical}
\left\{\begin{array}{llll}
\mathbf{F}=\frac{\sqrt{5}\mathbf{V}}{10}
\left(
\begin{array}{lll}
\Phi-1&\Phi\\
\vdots&\vdots\\
\Phi-1&\Phi
\end{array}
\right)
,\\
\mathbf{G}=\frac{\sqrt{5}\mathbf{V}}{10}
\left(
\begin{array}{lll}
\Phi&\Phi-1\\
\vdots&\vdots\\
\Phi&\Phi-1
\end{array}
\right)
\end{array}.
\right.
\end{eqnarray}
or equivalently,
\begin{eqnarray}\label{eqn:goledn_optical_equivalent}
\left\{\begin{array}{llll}
\mathbf{F}=\frac{\sqrt{5}\mathbf{V}}{10}
\left(
\begin{array}{lll}
\Phi&\Phi-1\\
\vdots&\vdots\\
\Phi&\Phi-1
\end{array}
\right)
,\\
\mathbf{G}=\frac{\sqrt{5}\mathbf{V}}{10}
\left(
\begin{array}{lll}
\Phi-1&\Phi\\
\vdots&\vdots\\
\Phi-1&\Phi
\end{array}\right)
\end{array}
\right.
\end{eqnarray}
where
\begin{eqnarray*}
\mathbf{V}=\frac{1}{\sum_{i=1}^N\Omega_i}\left(\begin{array}{lllll}
\Omega_1&0&\cdots&0\\
0&\Omega_2&\cdots&0\\
\mathbf{0}&\mathbf{0}&\ddots&\mathbf{0}\\
0&\mathbf{0}&\cdots&\Omega_N
\end{array}
\right)_{N\times N}
\end{eqnarray*}
and $\Phi=\frac{\sqrt{5}+1}{2}$ is known as the Golden ratio. ~\hfill\QED
 \end{theorem}
On Theorem~\ref{theorem:golden_code_optical}, whose proof is given in Appendix~\ref{app:proof_golden_optical}, the following remarks are made.
\begin{figure}[t]
    \centering
    \resizebox{8cm}{!}{\includegraphics{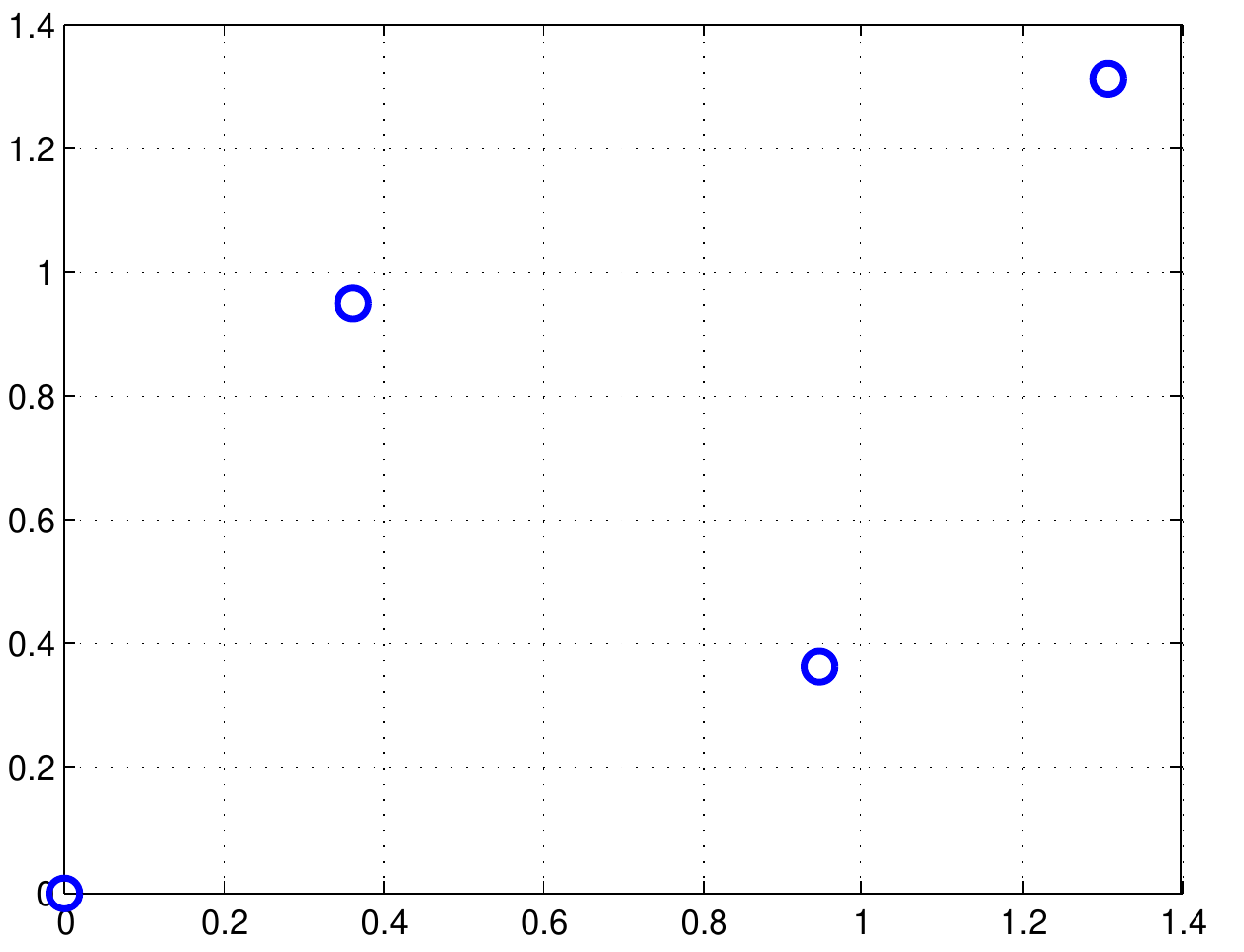}}
    \centering \caption{ Transmitted equivalent constellation with 1 bit pcu.}
    \label{fig:golden_constellation_1bit}
\end{figure}

\begin{figure}[t]
    \centering
    \resizebox{8cm}{!}{\includegraphics{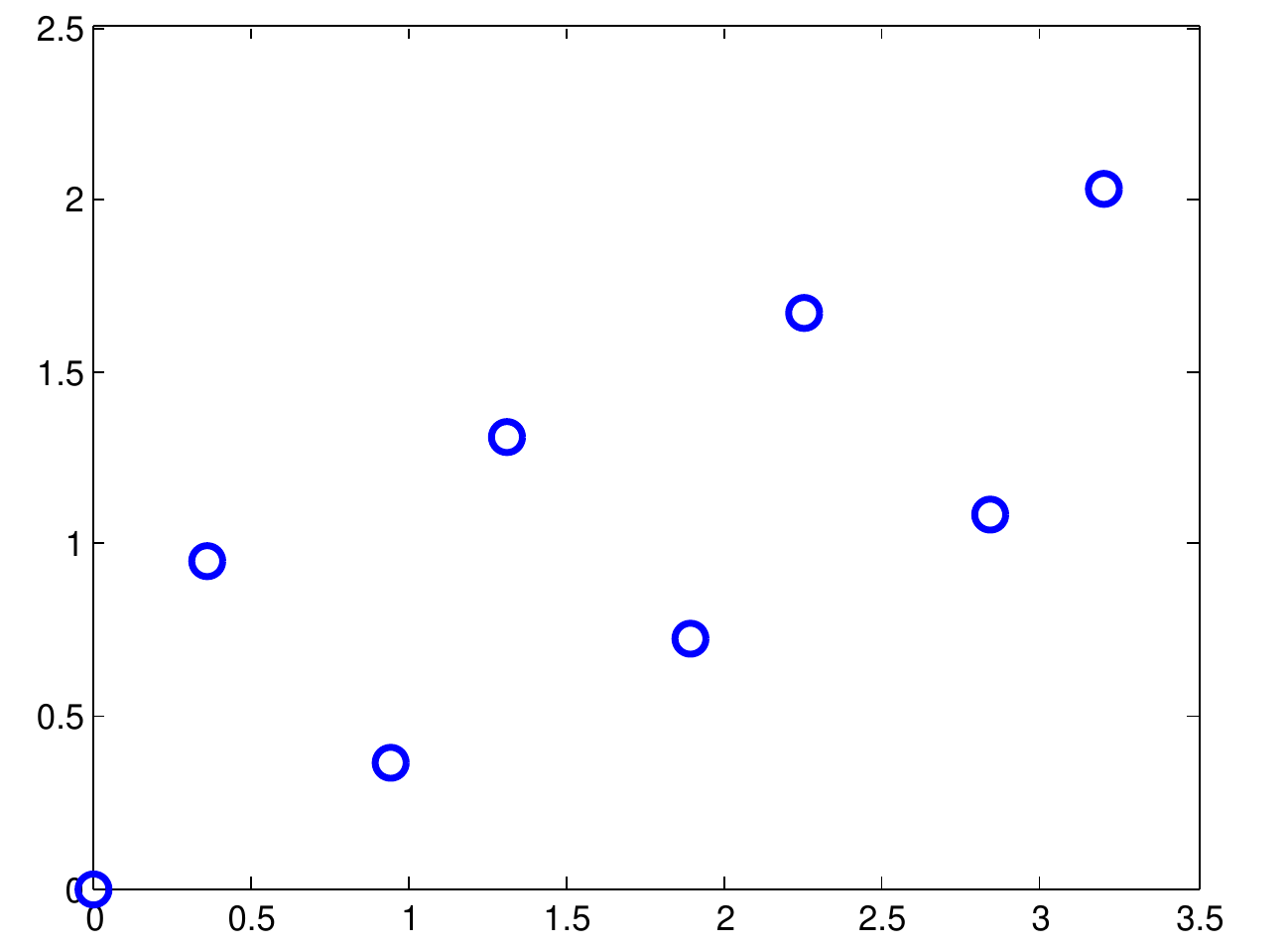}}
    \centering \caption{ Transmitted equivalent constellation with 1.5 bits pcu.}
    \label{fig:golden_constellation_15bit}
\end{figure}

\begin{figure}[t]
    \centering
    \resizebox{8cm}{!}{\includegraphics{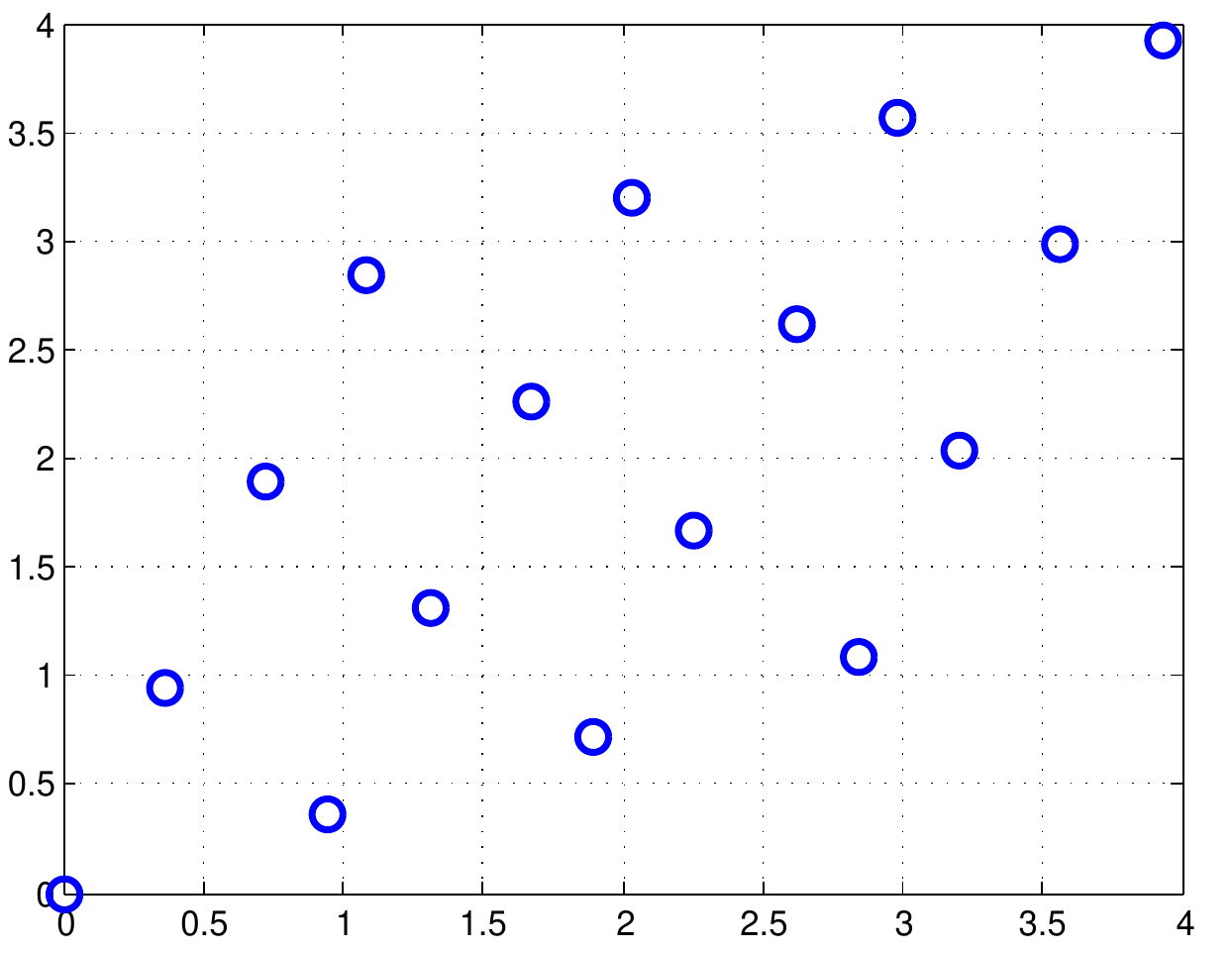}}
    \centering \caption{ Transmitted equivalent constellation with 2 bits pcu.}
    \label{fig:golden_constellation_2bit}
\end{figure}

\begin{figure}[t]
    \centering
    \resizebox{8cm}{!}{\includegraphics{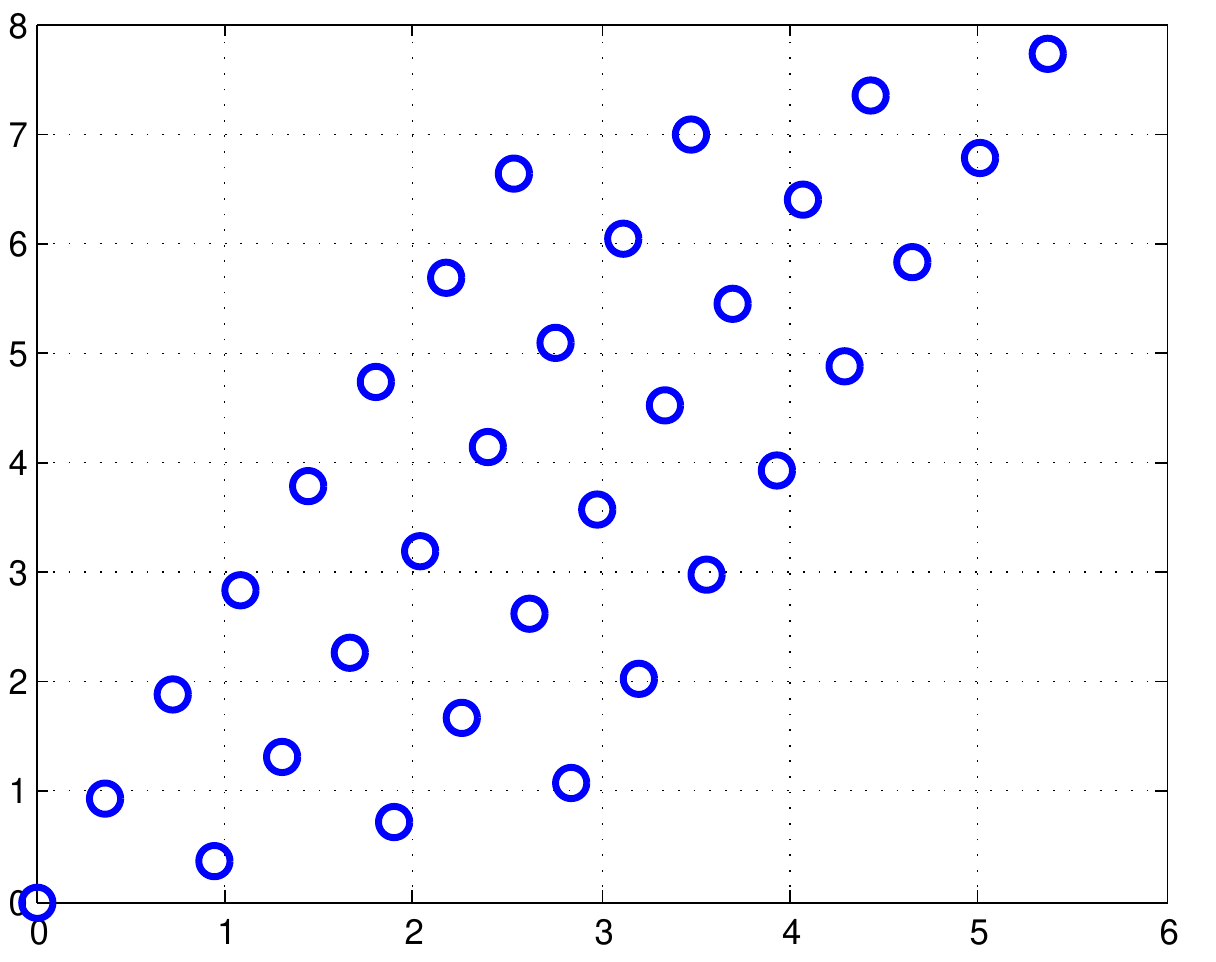}}
    \centering \caption{ Transmitted equivalent constellation with 2.5 bits pcu.}
    \label{fig:golden_constellation_25bit}
\end{figure}

\begin{figure}[t]
    \centering
    \resizebox{8cm}{!}{\includegraphics{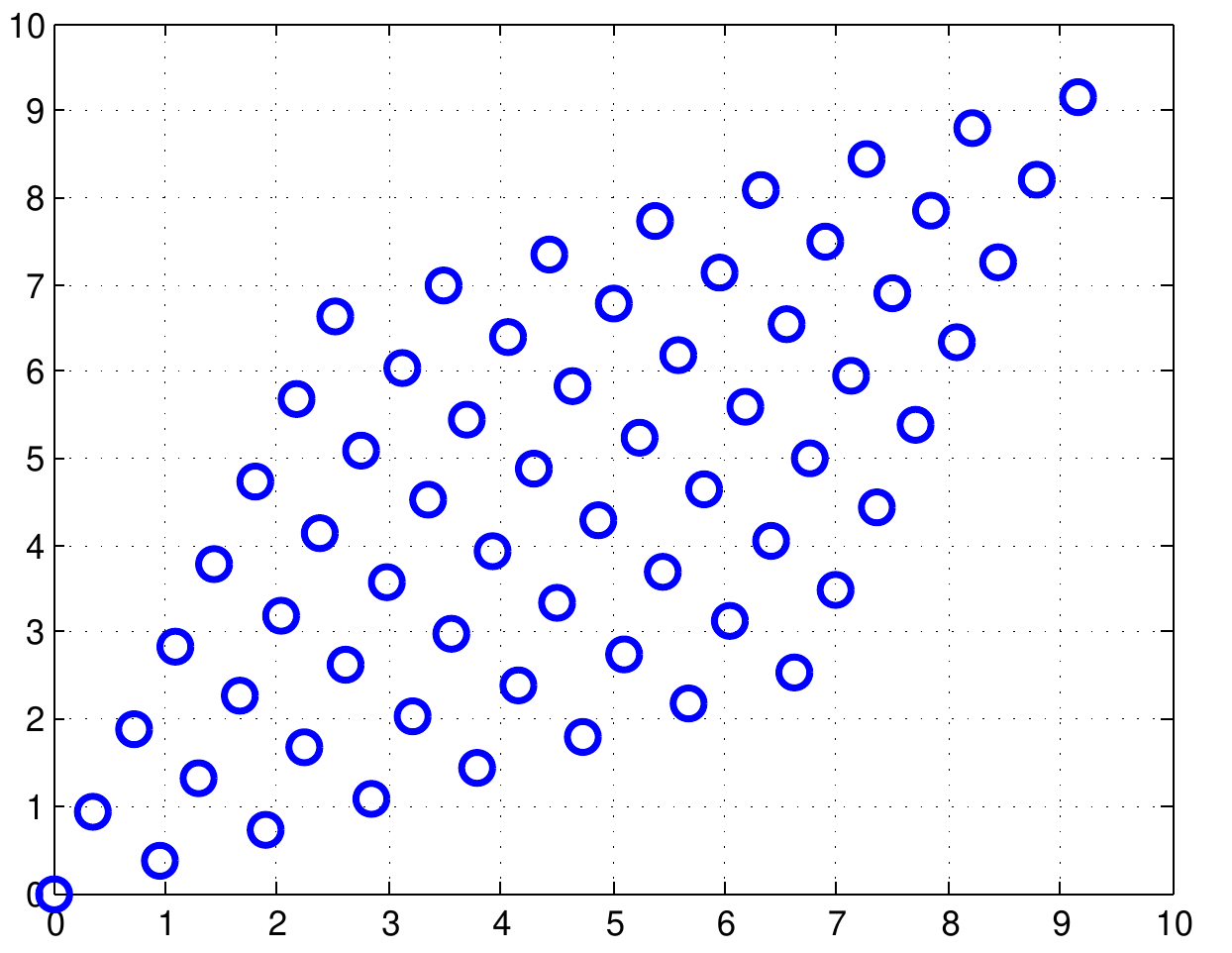}}
    \centering \caption{ Transmitted equivalent constellation with 3 bits pcu.}
    \label{fig:golden_constellation_3bit}
\end{figure}
\begin{figure}[t]
    \centering
    \resizebox{8cm}{!}{\includegraphics{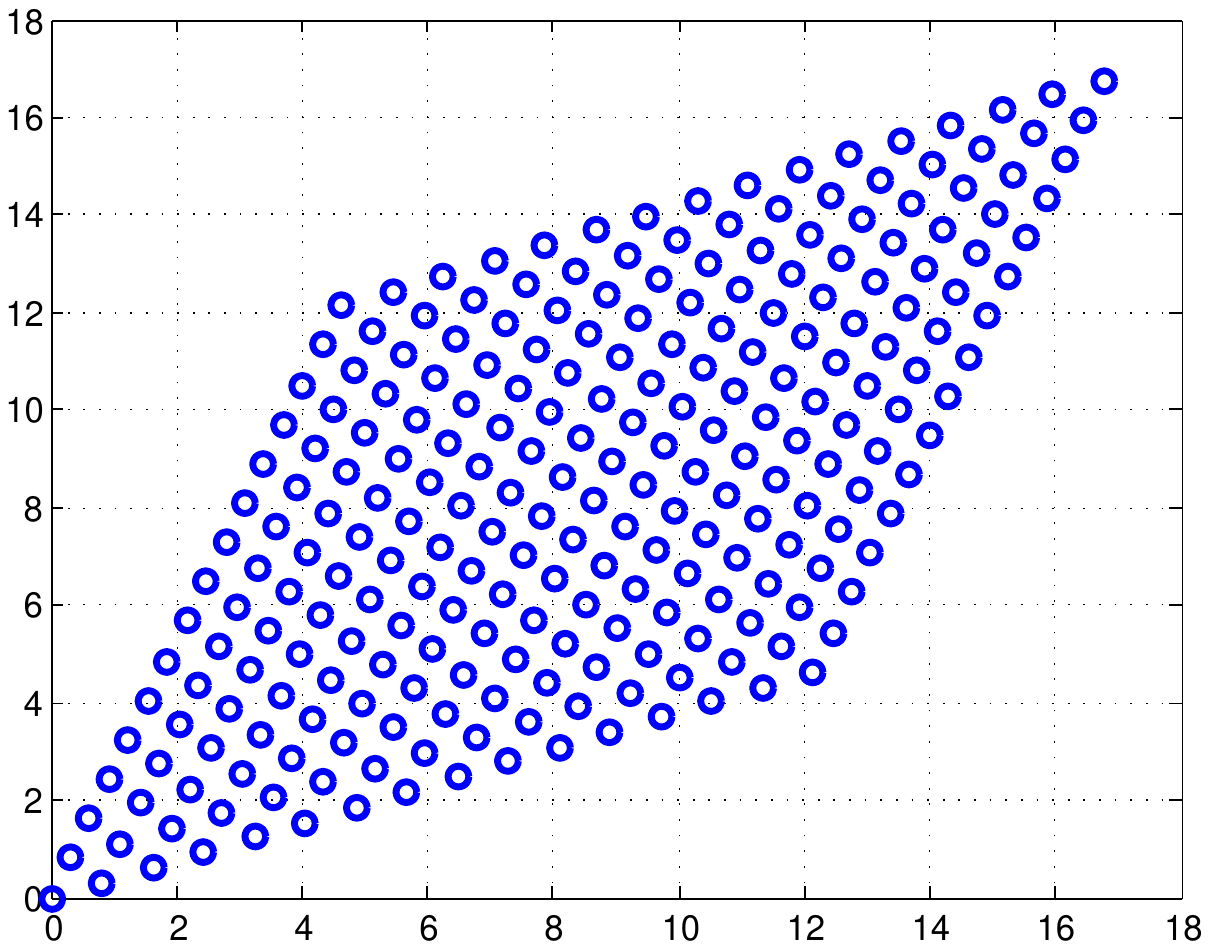}}
    \centering \caption{ Transmitted equivalent constellation with 4 bits pcu.}
    \label{fig:golden_constellation_4bit}
\end{figure}

\begin{figure}[t]
    \centering
    \resizebox{8cm}{!}{\includegraphics{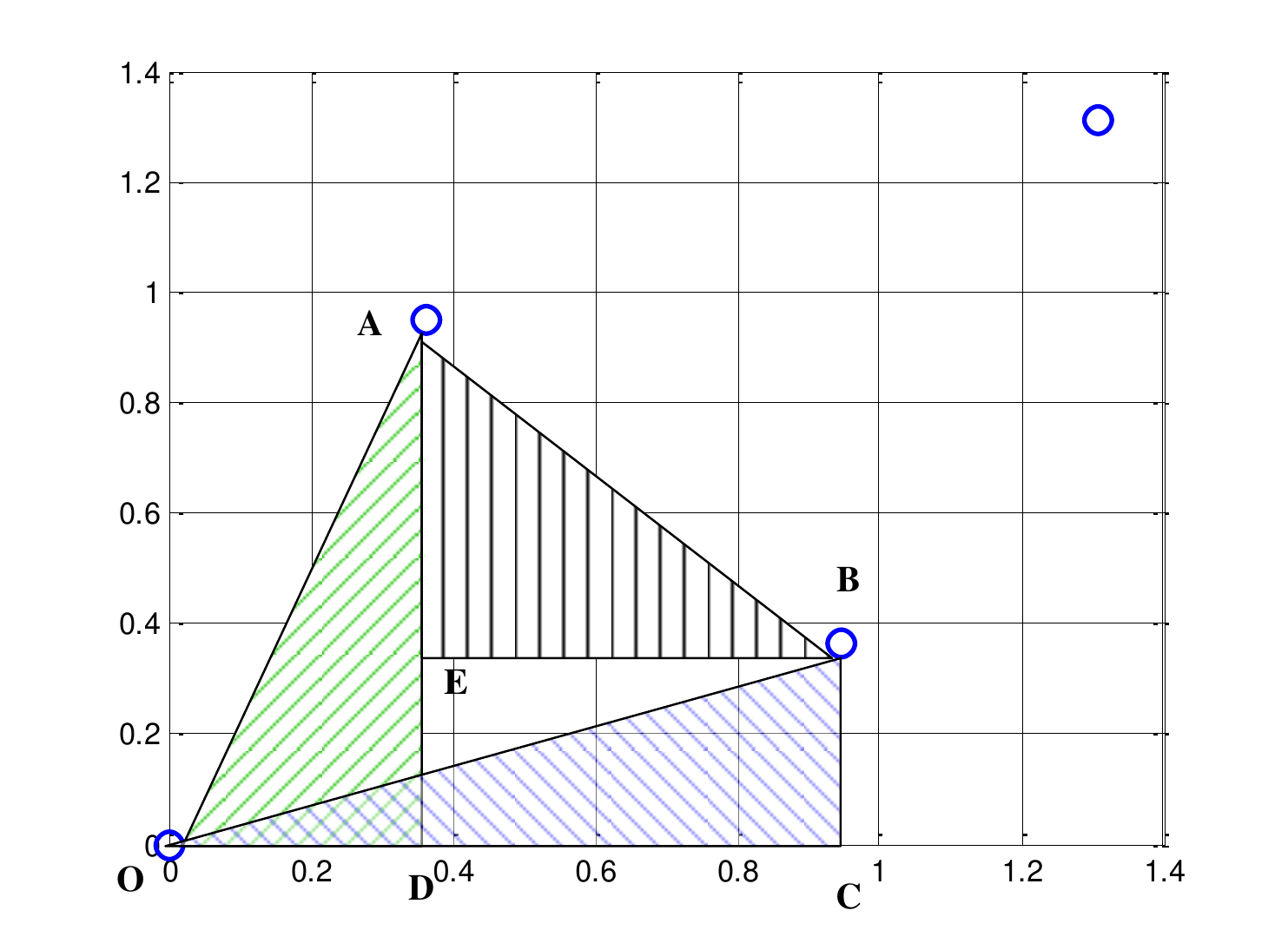}}
    \centering \caption{ Geometrical interpretation of Golden Code property.}
    \label{fig:geometry_interpretation}
\end{figure}

\begin{enumerate}
    \item \textit{Equivalent Constellation}. Our specific design is for the case with $L=2$. When $\Omega_1=~\cdots~=\Omega_N$, we denote the respective symbols for the first time slot and the second time slot by $x_1=(\Phi-1)s_1+\Phi s_2$ and $x_2=\Phi s_1+(\Phi-1)s_2$, where $s_1\in\{0,1,~\cdots,~2^{K_1}-1\}$ and $s_2\in\{0,1,~\cdots,~2^{K_2}-1\}$. Then, for the optical power constraint case, the equivalent constellations generated by $\left(x_1,x_2\right)$ are illustrated in Figs.~\ref{fig:golden_constellation_1bit}-\ref{fig:golden_constellation_4bit}, for 1 bit pcu ($K_1=K_2=1$), 1.5 bits pcu ($K_1=1,K_2=2$), 2 bits pcu ($K_1=K_2=2$), 2.5 bits pcu ($K_1=2,K_2=3$), 3 bits pcu ($K_1=K_2=3$) and 4 bits pcu ($K_1=K_2=4$) , respectively. We can see that the shape of the equivalent constellation is almost regularly spaced.
  \item \textit{Non-Increasing Small-Scale Diversity Loss}. From the proof of Theorem~\ref{theorem:golden_code_optical} in Appendix~\ref{app:proof_golden_optical}, the reciprocal of the  optimized small-scale diversity loss is equal to
      \begin{eqnarray*}
      &&\max_{\mathbf{F},\mathbf{G}}\min_{e_1,e_2} \prod_{i=1}^N\left(f_{i1}e_1+f_{i2}e_2\right)^{2\Omega_i}\left(g_{i1}e_1+g_{i2}e_2\right)^{2\Omega_i}\nonumber\\
&&=\left(\frac{1}{400}\right)^{\Omega}\prod_{i=1}^{N}\left(\frac{\Omega_i}{\Omega}\right)^{4\Omega_i}
      \end{eqnarray*}
Note that the maximum objective function in~\eqref{eqn:design_criterion} does not depend on $K_1$ and $K_2$ and thus, the attained small-scale diversity loss is non-increasing with increasing constellation size. This property of our Golden Code is similar to the Golden Code for MIMO-RF in~\cite{belfiore04,belfiore05}. Although both our proposed Golden Code and the Golden Code in~\cite{belfiore04} are based on the Golden ratio $\frac{\sqrt{5}+1}{2}$, our proposed Golden Code is attained by solving the max-min optimization problem in terms of the small-scale diversity loss  and satisfies the unipolarity requirement of intensity modulation.

\item \textit{Geometrical Interpretation}. In Fig.~\ref{fig:golden_constellation_1bit}, we show the equivalent transmitted constellation for four points. Based on Fig.~\ref{fig:golden_constellation_1bit}, we geometrically illustrate the property of non-increasing small-scale diversity loss by Fig.~\ref{fig:geometry_interpretation}.  As illustrated by Fig.~\ref{fig:geometry_interpretation}, there are the right triangles $\triangle OAD$, $\triangle AEB$ and $\triangle OCB$, formed by the signal points, where $\angle ADO=\angle AEB=\angle OCB=90^{\circ}$. For single input single output optical wireless communication systems, it is noticed that the areas of $\triangle OAD$, $\triangle AEB$ and $\triangle OCB$ are respectively given by $S_{\triangle OAD}=\frac{f_{11}g_{11}}{2}$, $S_{\triangle AEB}=\frac{(f_{11}-g_{11})(g_{12}-f_{12})}{2}$ and $S_{\triangle OCB}=\frac{g_{11}g_{12}}{2}$. If $f_{11}=\frac{\sqrt{5}\Phi}{10}$, $f_{12}=\frac{\sqrt{5}(\Phi-1)}{10}$,  $g_{11}=\frac{\sqrt{5}(\Phi-1)}{10}$, $g_{12}=\frac{\sqrt{5}\Phi}{10}$, then, we can have
    $S_{\triangle OAD}=S_{\triangle AEB}=S_{\triangle OCB}=\frac{1}{40}$. Therefore, it is this geometrical property that assures the resulting small-scale diversity loss does not increase against increasing modulation orders.
\end{enumerate}
\subsection{Remarks on the STBC Designs}
In the above sections, we have designed several classes of STBCs. To further appreciate our signal designs, we would like to make the following remarks:
\begin{enumerate}
  \item \textbf{ Design techniques}. It can be seen that for our systematical code constructions, it is the nonnegativity of the real-valued signals and the linear power constraint on the signal set that allow us to attain closed-form solution to the max-min optimization design problems. However, for MIMO-RF systems, where the signals are complex-valued, the max-min design problem is usually hard to be solved numerically. It goes without saying the closed-form solution in general. Therefore, for perspective of design techniques, IM/DD MIMO-OWC systems are remarkably different from the well-established MIMO-RF systems.
  \item \textbf{Universal Optimal STBC Structure}. For MIMO-RF systems, numerous STBC schemes have been designed based on diverse coding structures, which are usually dependent on the number of the transmitter and receiver antennas. Therefore, it is difficult to say what the optimal structure is. However, for IM/DD MIMO-OWC systems over block fading channels, we have characterized a universal optimal space-time coding structure based on our established super-rectangular cover criterion by solving the optimal design problem in the signal set within the positive orthants of a multidimensional space without additional assumption.
  \item \textbf{From MIMO To SISO}. It is also noticed that all the code constructions for IM/DD MIMO-OWC presented in this section indicates that the time design and space design can be done in an independent manner. This observation tells us that the time design for IM/DD MIMO-OWC can be generalized to the IM/DD SISO-OWC systems in a straightforward manner without losing any transmission rate, which is impossible for MIMO-RF systems.
\end{enumerate}
The above-mentioned observations obtained in this paper reveal useful insights into how the space-time coded IM/DD MIMO-OWC  systems are remarkably different from,  rather than ``mimic'', the   conventional  space-time coded MIMO-RF systems~\cite{tarokh98}.

\section{Simulation Results}\label{sec:code_costruction}

\subsection{Performance Comparisons of STBCs over Block Fading Channels}
In this subsection, we carry out extensive computer simulations to examine the error performance of our designed STBCs compared with RC, which is the best available code for IM/DD MIMO-OWC systems. All the schemes we would to compare are described as follows.
\begin{enumerate}
  \item \textit{Repetition Code (RC)}. The codeword matrix of RC is given by
  \begin{eqnarray*}
  \mathbf{X}\left(\mathbf{p}\right)&=&\frac{2L}{N\sum_{\ell=1}^L\left(2^{K_\ell}-1\right)}
  \left(\begin{array}{llll}
  p_1&\ldots&p_1\\
   \vdots&\ddots&\vdots\\
  p_L&\ldots&p_L
  \end{array}
  \right)
\end{eqnarray*}
where $\mathbf{p}\in\mathcal{P}^{(L,K)}=\mathcal{P}^{(1,K_1)}\times\cdots\times\mathcal{P}^{(1,K_L)}$ with $\mathcal{P}^{(1,K_\ell)}=\{0,1,~\cdots,~2^{K_\ell}-1\}$. In addition, for RC, $\mathcal{P}=\{m\}_{m=0}^{m=2^{K_1}-1}\times\cdots\times\{m\}_{m=0}^{m=2^{K_T}-1}$ is determined by properly selecting positive integer $K_i$ such that $\sum_{\mathbf{p}\in\mathcal{P}}\mathbf{1}^{\rm T}\mathbf{p}$ is minimized.
  \item \textit{Optimal Linear STBCs}. The optimal linear STBC design is proposed in Theorem~\ref{theorem:worst_case_pep} and the codeword matrix is of the following form
  \begin{eqnarray*}
  \mathbf{X}\left(\mathbf{p}\right)&=&\frac{2L}{\sum_{\ell=1}^L\left(2^{K_\ell}-1\right)}
  \left(\begin{array}{llll}
  \Omega_1p_1&\ldots&\Omega_Np_1\\
  \vdots&\ddots&\vdots\\
  \Omega_1p_L&\ldots&\Omega_Np_L
  \end{array}
  \right)
\end{eqnarray*}
where $\mathbf{p}\in\mathcal{P}^{(L,K)}=\mathcal{P}^{(1,K_1)}\times\cdots\times\mathcal{P}^{(1,K_L)}$. Note that $\sigma _{ij}^{2}$ depends on altitude-dependent, the light wavelength, the link distance and the root mean square wind speed~\cite{Karp1988,andrews2001laser}, giving us that $\Omega_i=\sum_{j=1}^M\sigma_{ij}^{2}$ may be different in practical scenarios.
  \item \textit{Collaborative STBC}. The collaborative STBC is constructed in Subsection~\ref{sec:collaborative_code} with codeword matrix being given as follows.
  \begin{eqnarray*}
  \mathbf{X}\left(\mathbf{s}\right)&=&\frac{L2^K}{\sum_{\mathbf{s}\in \mathcal{S}^{(L,K)}}\mathbf{1}^T\mathbf{s}}
  \left(\begin{array}{llll}
  \Omega_1s_1&\ldots&\Omega_Ns_1\\
  \vdots& \ddots&\vdots\\
  \Omega_1s_L&\ldots&\Omega_Ns_L
  \end{array}
  \right)
  \end{eqnarray*}
where $\mathbf{s}\in\mathcal{S}^{(L,K)}$ and $\mathcal{S}^{(L,K)}$ is proposed in Theorem~\ref{theorem:diophantine_conste}.
\end{enumerate}

It can be seen that all the above schemes have the normalized average optical power. Therefore, the SNR is defined by $\frac{1}{\sigma_{\mathbf{N}}^2}$. To make fair comparisons, the receivers of all these schemes are ML detectors and more details are provided in the following examples.

\begin{figure}[t]
    \centering
    \resizebox{8cm}{!}{\includegraphics{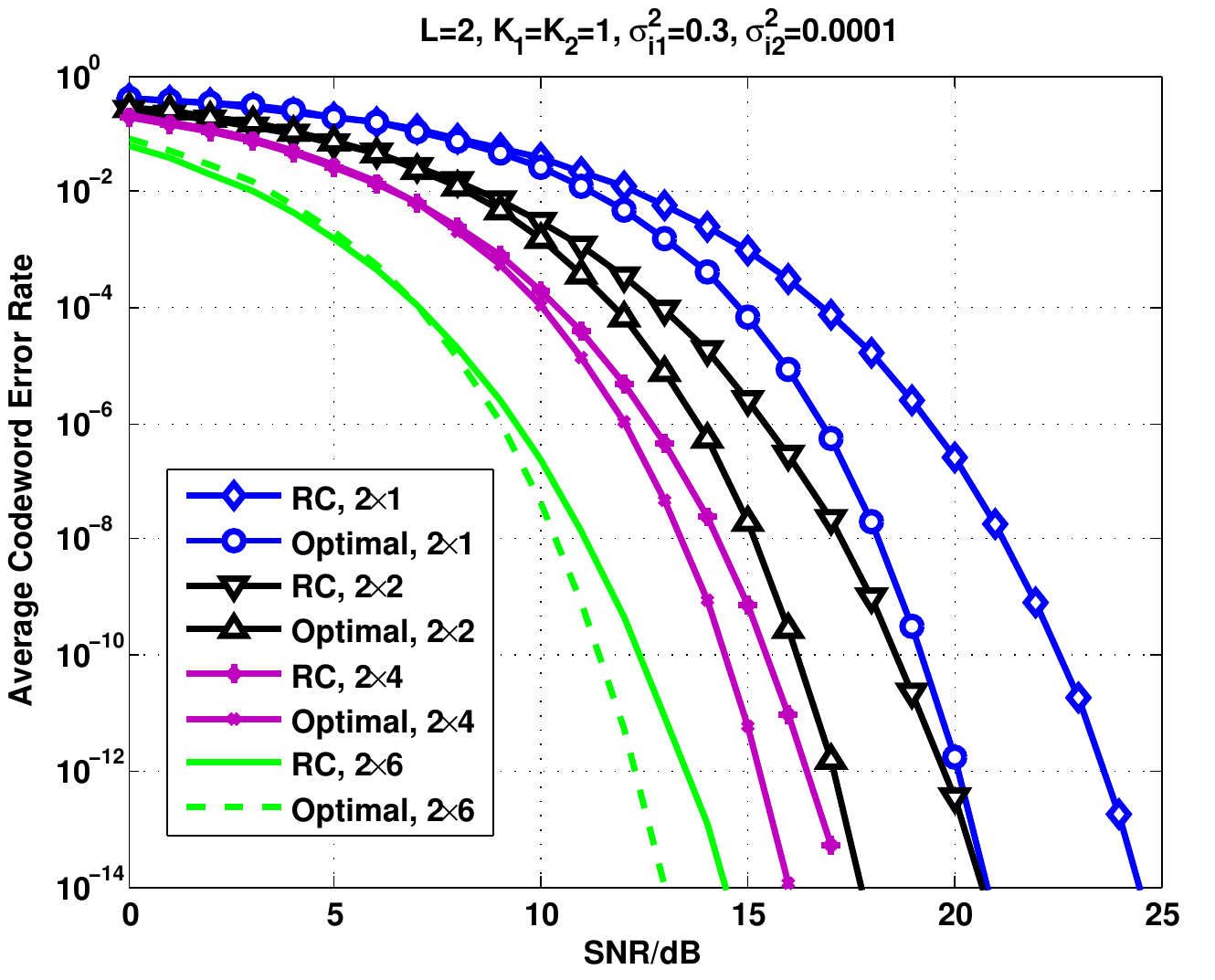}}
    \centering \caption{Average codeword ($\mathbf{X}(\mathbf{p})$) performances of optimal linear STBCs (optimal) and RC for $K_1=K_2=1$, $N=2$ and different $M$.}
    \label{fig:scsit_two_tx_bit1_bit1}
\end{figure}

\begin{figure}[t]
    \centering
    \resizebox{8cm}{!}{\includegraphics{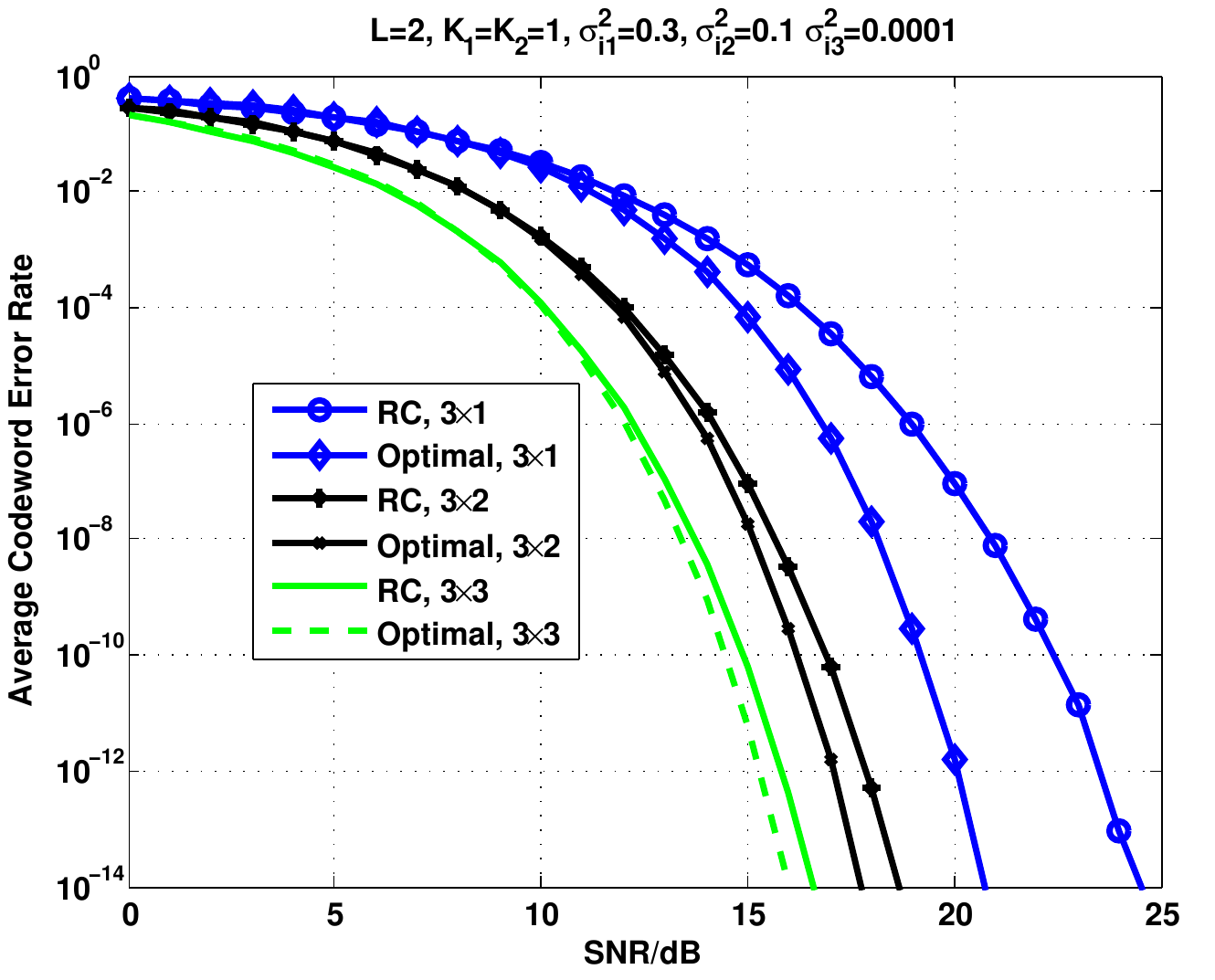}}
    \centering \caption{Average codeword ($\mathbf{X}(\mathbf{p})$) performances of optimal linear STBCs (optimal) and RC for $K_1=K_2=1$, $N=3$ and different $M$.}
    \label{fig:scsit_three_tx_bit1_bit1}
\end{figure}

\begin{example} We now present the error performance comparisons of our optimal designed linear STBCs and RC with nonequal $\Omega_i$ for $i=1,2,~\cdots,~N$. In addition, notice that we \textit{numerically calculate the average codeword error probability} based on~\eqref{eqn:codeword_error}, which allows us to scratch the error behaviour of our optimal linear STBC and RC even at an extremely high SNR.
From the numerical results shown by Figs.~\ref{fig:scsit_two_tx_bit1_bit1} and~\ref{fig:scsit_three_tx_bit1_bit1}, we find that for fixed transmitter and receiver numbers, the attained gain becomes larger against increasing SNR. This is because that the small-scale diversity loss of the optimal linear STBC is smaller than that of RC. Therefore, the polynomial decaying speed of our optimal design is faster than that of RC. For example, for $N=2$ and $M=1$, at the target average codeword error probability of $10^{-4}$, the attained gain by the optimal linear STBC is abound 2 dB. For this system, the attained gain becomes 3 dB at the target average codeword error probability of $10^{-8}$. However, as the number of receiver apertures increases, the attained gain decreases for the same target error probability. The reason is that the large-scale diversity gain governs the error curves with exponential decaying  and the decreased small-scale diversity loss resulted from our optimal designs only affects the polynomial decaying speed in a high SNR. When $M$ is sufficiently large, the effect of the increased large-scale diversity gain on the decaying speed resulted from the increasing $M$ is dominant, whereas  the decreased small-scale diversity loss produced by our optimal designs remains constant, since we fix $\frac{\Omega_1}{\Omega_2}$ to be $\frac{1}{300}$. Therefore, to make the performance gain attained by our optimal designs more noticeable, SNR is required to be much higher for increasing $M$.~\hfill\QED
\end{example}

\begin{figure}[t]
    \centering
    \resizebox{8cm}{!}{\includegraphics{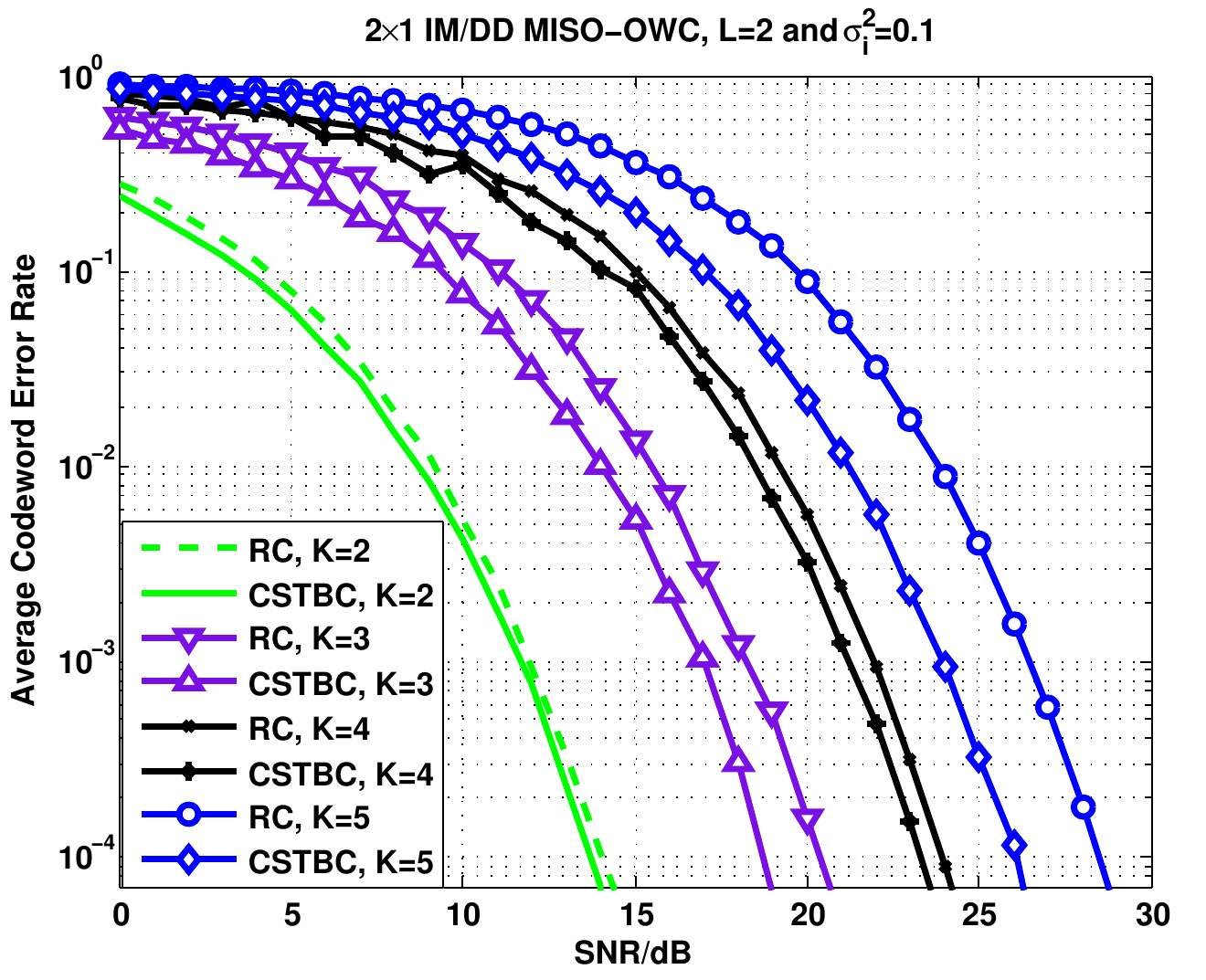}}
    \centering \caption{Average codeword ($\mathbf{X}(\mathbf{p})$) performances of CSTBCs and RC for $2\times1$ IM/DD MIMO-OWC.}
    \label{fig:time2_two_by_one}
\end{figure}

\begin{figure}[t]
    \centering
    \resizebox{8cm}{!}{\includegraphics{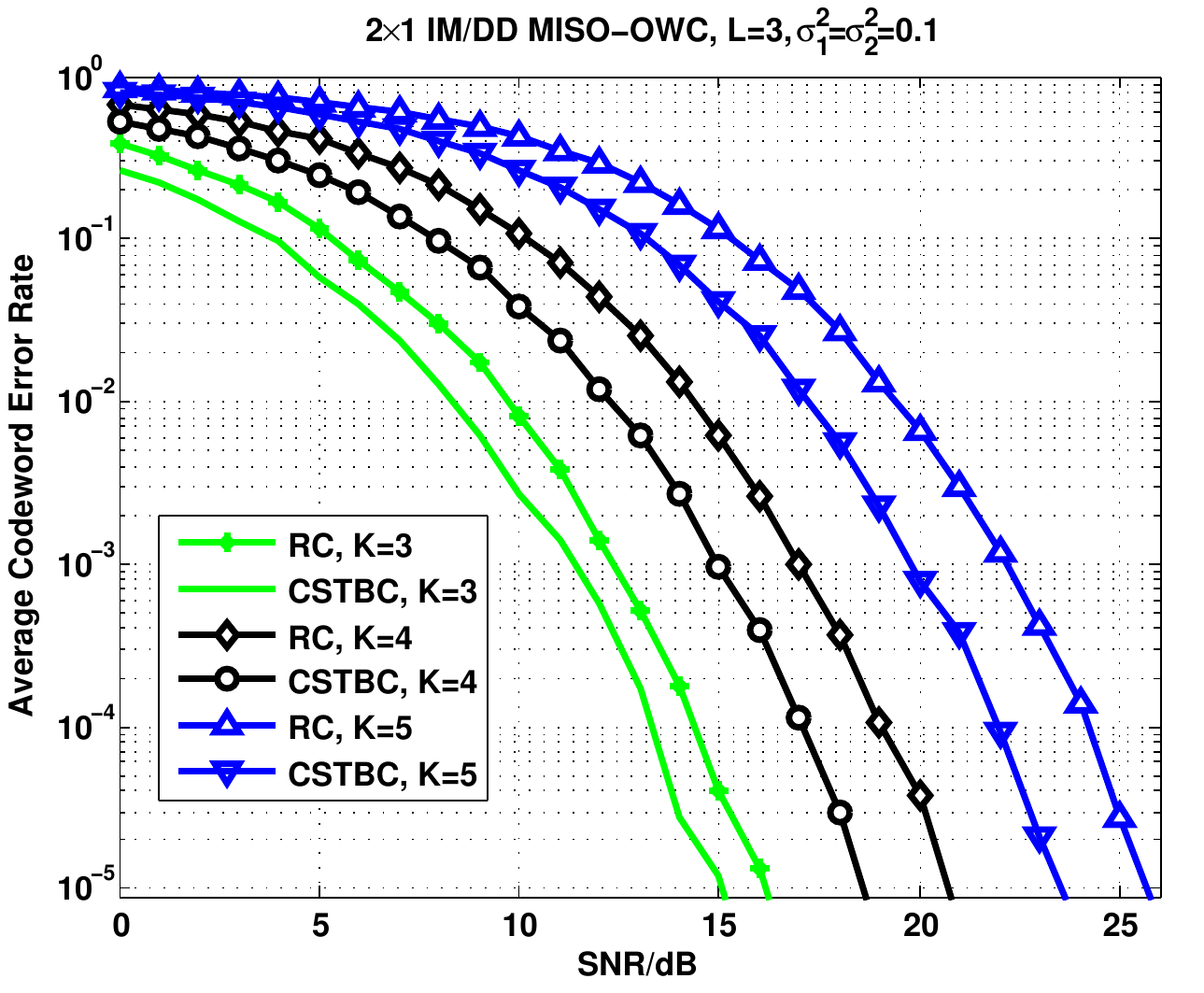}}
    \centering \caption{Average codeword ($\mathbf{X}(\mathbf{s})$) performances of CSTBCs and RC for $2\times1$ IM/DD MIMO-OWC with $L=3$.}
    \label{fig:three_time_12miso}
\end{figure}

\begin{figure}[t]
    \centering
    \resizebox{8cm}{!}{\includegraphics{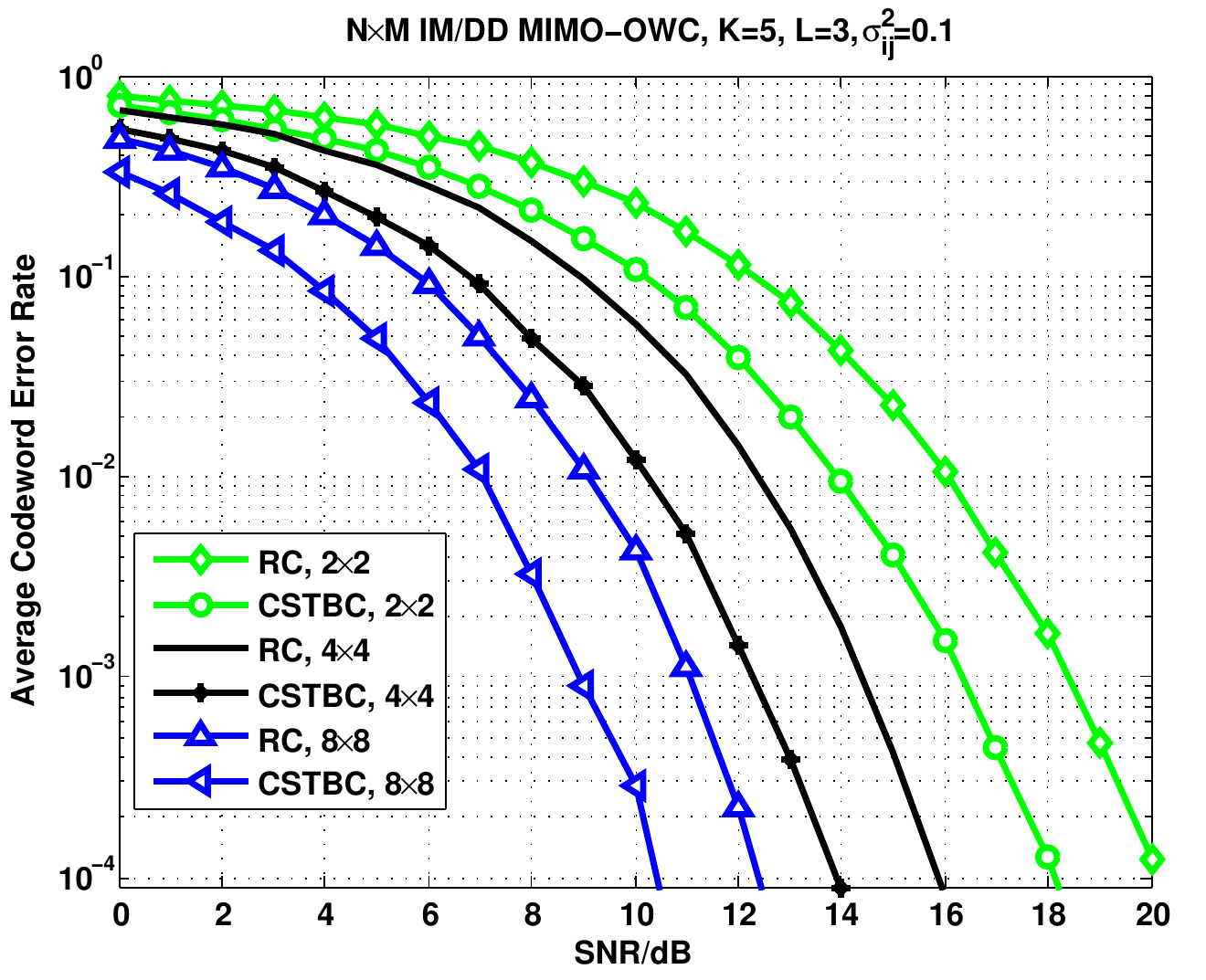}}
    \centering \caption{Average codeword ($\mathbf{X}(\mathbf{s})$) performances of CSTBCs and RC for $N\times M$ IM/DD MIMO-OWC with $L=3$ and $K=5$.}
    \label{fig:three_time_bit5_mimo}
\end{figure}

\begin{figure}[t]
    \centering
    \resizebox{8cm}{!}{\includegraphics{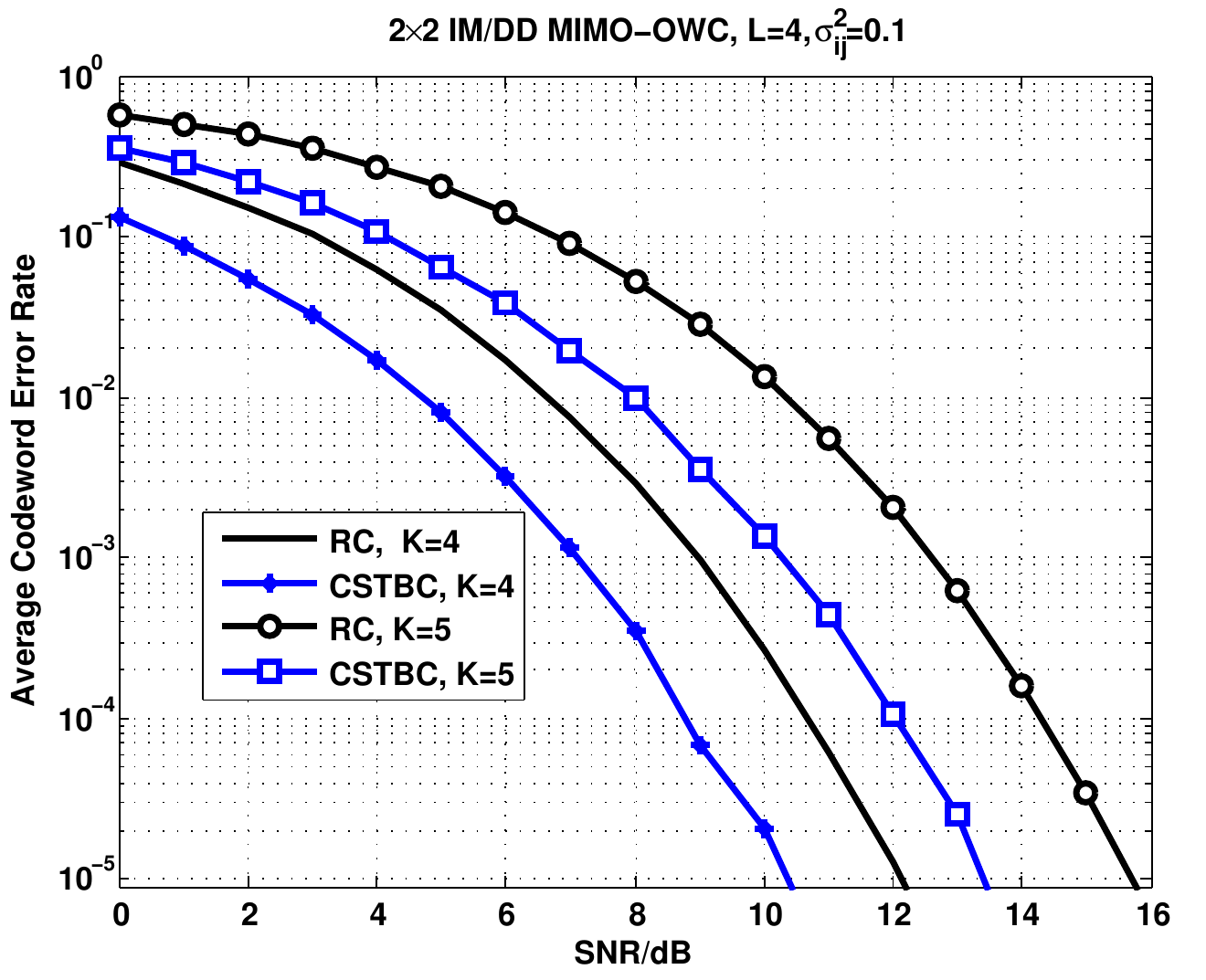}}
    \centering \caption{Average codeword ($\mathbf{X}(\mathbf{s})$) performances of CSTBCs and RC for $2\times2$ IM/DD MIMO-OWC with $L=4$.}
    \label{fig:four_time_2by2}
\end{figure}

\begin{figure}[t]
    \centering
    \resizebox{8cm}{!}{\includegraphics{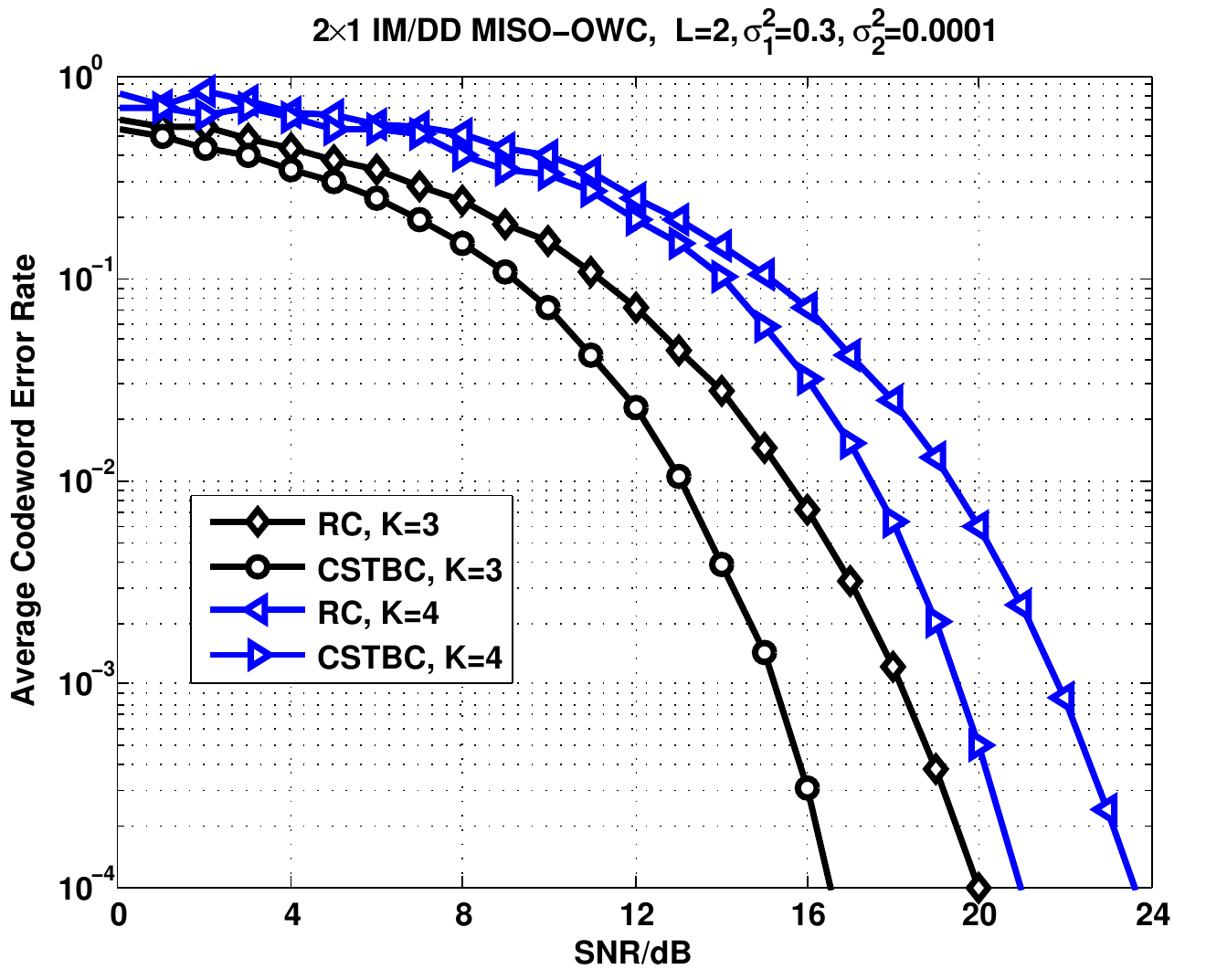}}
    \centering \caption{Average codeword ($\mathbf{X}(\mathbf{s})$) performances of CSTBCs and RC for $2\times1$ IM/DD MISO-OWC with different variances.}
    \label{fig:scist_cstbc_time2_two_by_one}
\end{figure}

\begin{example} This example compares  the error performance of our CSTBC with RC. On condition that  $\Omega_1=\Omega_2=~\cdots~=\Omega_N$, the simulation results are illustrated by Figs.~\ref{fig:time2_two_by_one}-\ref{fig:four_time_2by2}, from which we have the following observations. On one hand, our designed CSTBCs always have better error performance than RC. For the fixed time dimension number $L$, the  attained gain depends on the collaborative constellation size $2^K$. However, this gain will not decrease with increasing receiver aperture number $M$, as shown by Fig.~\ref{fig:three_time_bit5_mimo}. It can also be seen that when $L$ is increasing, the attain gains by our CSTBCs are substantial. For example, when $L=4$, for $K=4$ and $K=5$, the respective attained gains are above 2 dB at the target error rate $10^{-4}$.  However, the decaying speeds of the error curves in Figs.~\ref{fig:time2_two_by_one},~\ref{fig:three_time_12miso},~\ref{fig:three_time_bit5_mimo} and~\ref{fig:four_time_2by2} are almost the same. Recall that the large-scale diversity gain and small-scale diversity loss are defined for high SNRs and here, the upper end of the SNR range is not sufficiently high. To show the performance behaviour at higher SNR, the corresponding evaluation demands time-consuming computation in this example. To do this, we choose $L=2$, $\sigma_1^2=0.3$ and $\sigma_2^2=0.0001$ for $2\times1$ IM/DD MISO-OWC systems. Then, we compare the simulation results in Fig.~\ref{fig:scist_cstbc_time2_two_by_one}. It can be noticed that the decaying speed of the error curves of our CSTBCs are much faster than that of the optimal linear STBCs with RC as the same benchmark scheme. ~\hfill\QED
\end{example}

\subsection{Performance Comparisons of STBCs over Fast Fading Channels}
In this subsection, we carry out computer simulations and compare the error performance of the  Golden Codes proposed in this paper and other schemes for this application. All the schemes that we would like to compare here are described as follows:
\begin{enumerate}
  \item \textit{Space-time repetition coding (STRC)}.
   \begin{eqnarray*}
    \mathbf{X}\left(\mathbf{s}\right) &=& \frac{s_1+2^{K_1}s_2}{\left(2^{K_1+K_2}-1\right)}\left(
    \begin{array}{ll}
    1&1\\
    1&1
\end{array}
    \right)
  \end{eqnarray*}
  where $s_1\in\{0,1,\ldots,2^{K_1}-1\}$ and $s_2\in\{0,1,\ldots,2^{K_2}-1\}$ to form a $2^{K_1+K_2}$-ary PAM constellation, and the normalized constant $\left(2^{K_1+K_2}-1\right)$ assures that the average optical power $E\left[\mathbf{1}^T\mathbf{X}\left(\mathbf{s}\right)\mathbf{1}\right]=2$.

  It should be noted that RC can also be applicable to this scenario. However, the total large-scale diversity gain attained by our designed Golden Code is twice than that of RC. For sake of fairness consideration, we only compare the error performance of the Golden Code with the so-called STRC, which is a natural extension of RC.
  \item \textit{Golden Code}. Golden Code is proposed in Theorem~\ref{theorem:golden_code_optical} and the codeword matrix is of the following form.
  \begin{eqnarray*}
   &&\mathbf{X}\left(\mathbf{s}\right)=\frac{4}{\Omega\left(2\Phi-1\right)\left(2^{K_1}+2^{K_2}-2\right)}
   \nonumber\\
&&\times\left(
    \begin{array}{ll}
    \Omega_1\left(\Phi s_1+\left(\Phi-1\right)s_2\right)\times\mathbf{1}_{1\times2}\\
   \Omega_1\left(\left(\Phi-1\right) s_1+\Phi s_2\right)\times\mathbf{1}_{1\times2}
    \end{array}
    \right)
  \end{eqnarray*}
  where $\Phi=\frac{\sqrt{5}+1}{2}$, $s_1\in\{0,1,~\cdots,~2^{K_1}-1\}$, $s_2\in\{0,1,~\cdots,~2^{K_2}-1\}$ and the normalized constant $\frac{4}{\left(2\Phi-1\right)\left(2^{K_1}+2^{K_2}-2\right)}$ assures that the average optical power satisfies $E\left[\mathbf{1}^T\mathbf{X}\left(\mathbf{s}\right)\mathbf{1}\right]=2$.
\end{enumerate}
It can be seen that the above two transmission schemes have the same average bit rate, i.e., each transmission rate is $\frac{K_1+K_2}{2}$ bits per channel use. To make all the error performance comparisons as fair as possible, we decode all the codes using the ML detector.

\begin{figure}[t]
    \centering
    \resizebox{8cm}{!}{\includegraphics{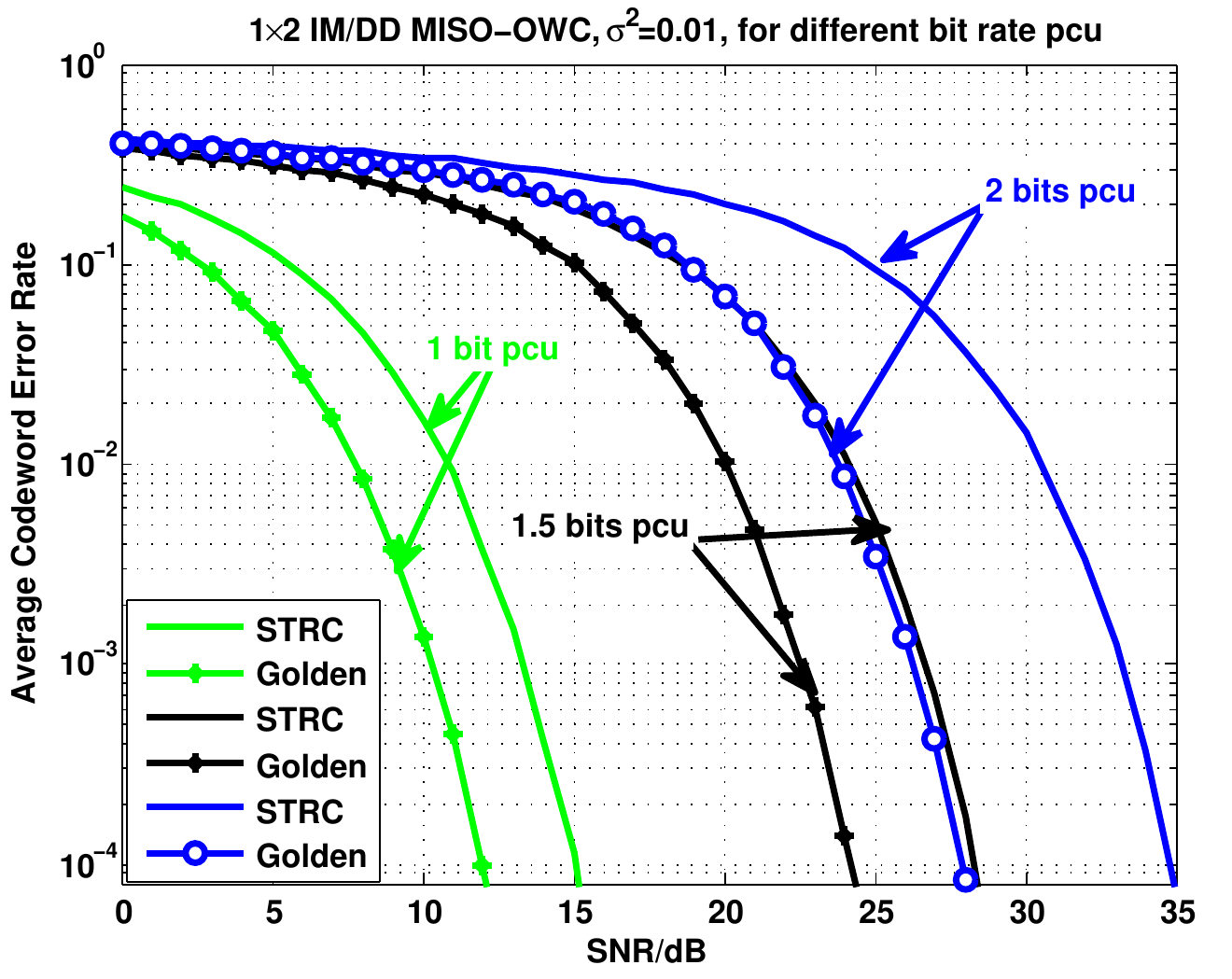}}
    \centering \caption{Average codeword error rate comparison of Golden Code and STRC for $2\times1 $ MISO-OWC with different bits pcu.}
    \label{fig:golden_rc_different_bit_miso21}
\end{figure}

\begin{figure}[t]
    \centering
    \resizebox{8cm}{!}{\includegraphics{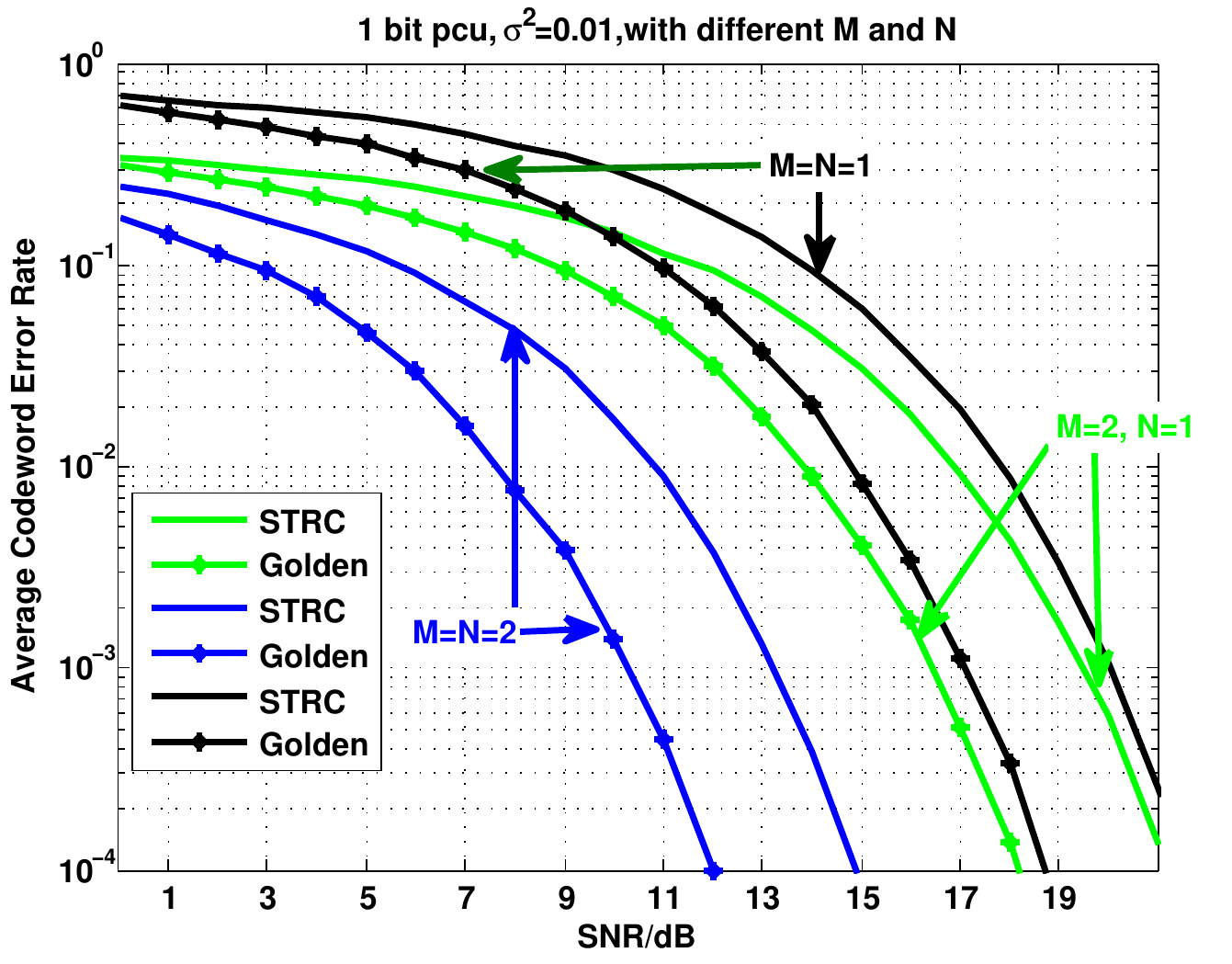}}
    \centering \caption{Average codeword error rate comparison of Golden Code and STRC for 1 bit pcu ($K_1=K_2=1$) with different aperture configurations.}
    \label{fig:golden1bitdifferentnumber}
\end{figure}

The simulation results are shown in Figs.~\ref{fig:golden_rc_different_bit_miso21} and~\ref{fig:golden1bitdifferentnumber}. It can be seen that substantial performance gains are attained by our proposed Golden Code over STRC for different bit rates and aperture numbers. In addition, the attained gain becomes larger with increasing bit rate compared with STRC. For example, in Fig.~\ref{fig:golden_rc_different_bit_miso21}, when the bit rates pcu are 1, 1.5 and 2, the respective attained gains are 3 dB, 4 dB and 6 dB  at the target error rate $10^{-4}$.  Furthermore, as illustrated in Fig.~\ref{fig:golden1bitdifferentnumber}, the attained gain is independent of the transmitter and receiver aperture numbers.

\section{Conclusion and Future Directions}
In this paper, we have considered the space-time coded IM/DD MIMO-OWC systems. From the viewpoint of detection theory, a novel super-rectangular cover theory has been developed to characterize the signal identification for IM/DD MIMO-OWC, where the channel coefficients are located at the nonnegative orthants of a real space. In particular, by fully utilizing this theory for the diversity analysis for IM/DD MIMO-OWC systems over log-normal fading channels, a general error performance design criterion of STBC with an ML detector has been established. The large-scale diversity gain and small-scale diversity loss have a geometrical interpretation as the maximal side number and minimal volume of the super-rectangle, respectively. Using this established design criterion, several classes of STBCs have been constructed.
\begin{enumerate}
  \item \textit{Optimal linear STBCs structure for block fading channels}. The optimal linear STBCs for block fading channels have been shown to be RC with an optimal power allocation by maximizing the large-scale diversity gain and minimizing the small-scale diversity loss.  Specifically, when $\Omega_1=~\cdots~=\Omega_N$, the optimal linear STBC is proved to be RC. Despite the fact that all the experimental evidences~\cite{majid2008twc,bayaki2010space}, thus far,  have strongly demonstrated that RC indeed provides encouraging error performance, the corresponding mathematical proof remains a long-standing open problem mainly because of the lack of an explicit signal design criterion like MIMO-RF communications. Hence, we actually solve this long-standing open problem on the RC optimality as a linear STBC in the sense of optimizing the large scale diversity gain and minimizing the small-scale diversity loss.
  \item \textit{Optimal non-linear STBCs structure for block fading channels}. It has been proved that constructing the optimal non-linear STBC is equivalent to designing the optimal multi-dimensional constellation, which is a classical and long-standing problem, remaining open thus far. For this reason, a specific energy-efficient multiple-dimensional constellation from Diophantine equations have been proposed.
  \item \textit{Golden Codes for fast fading channels}. For IM/DD MIMO-OWC over time-independent fast fading channels, a linear STBC is constructed by linearly collaborating the signals of two successive channel uses. Then, via analytical optimization, the optimal solution is shown to be related with the Golden number $\frac{\sqrt{5}+1}{2}$.
\end{enumerate}
However, the work in this paper just scratches the tip of the iceberg, and numerous important and interesting questions remain to be answered.
\begin{enumerate}
  \item \textit{Universal space-time coding structure ?} Despite the fact that the optimal coding structure given by Theorem~\ref{theorem:optimal_structure} is based on our super-rectangular cover criterion and specifically for IM/DD MIMO-OWC over log-normal fading channels, this structure also has the ability to assure the unique identification for any nonzero nonnegative channels. In addition, this unique identification is unrelated to the PDF of the channel coefficients. One natural question is that does our optimal coding structure guarantee high energy efficiency for arbitrary nonnegative channels? We conjecture that the PDF of the channel coefficient only affects the optimal power-loaded vector. This topic is indeed interesting and significant.
  \item \textit{Diophantine Coded Modulation}. The essence of CSTBC is to allow the transmitted signals in time dimensions to be cooperative for increasing channel reliability. At this point, when the bit strings are properly mapped into the multidimensional Diophantine constellation such that the bit string pairs with large Hamming distances correspond to the constellation point pairs with large Euclidean distances, our designed CSTBC can be viewed as a new kind of Diophantine coded modulation for the specific IM/DD MIMO-OWC systems, which is different from   the trellis coded modulation proposed by Ungerboeck in~\cite{Ungerboeck1982Channel,Ungerboeck1,Ungerboeck2} for the RF digital communication systems. So, is there a structured coded modulation scheme with low-complexity?

  \item \textit{Optimal multi-dimensional constellation}. It is well-known that this problem in modern RF wireless communication is extremely challenging. Despite the fact that the power constraint for IM/DD MIMO-OWC is different from that of MIMO-RF, the optimal design of multi-dimensional constellation for IM/DD MIMO-OWC is as hard as that for MIMO-RF. However, the Diophantine equation has played an important role in designing the energy-efficient multi-dimensional constellation. Then, can we construct the optimal multi-dimensional constellation from Diophantine equation?
  \item \textit{Nonnegative nonvanshing STBC.} In this paper, a Golden Code has been constructed via analytical optimization and shown to be related to the Golden number. It is noticed that the design criterion of minimizing small-scale diversity loss for fast fading channels is parallel to the product distance criterion in MIMO-RF and the proposed Golden Code has the  property of nonvanishing product distance (the reciprocal of the small-scale diversity loss). This property has been extensively investigated for space-time block coded  MIMO-RF in terms of  nonvanshing ``lattice code''. However, the signal matrix of IM/DD MIMO-OWC is required to be nonnegative. So, how can we construct the nonnegative STBC carved from the nonvanshing ``lattice code'' for IM/DD MIMO-OWC by coding over multiple blocks?
\item \textit{Other Fading Channels.} Our cover theory is general for the signal unique identification of IM/DD MIMO-OWC and robust to the PDF of the channel fading. Then, can the techniques in this paper be generalized  to IM/DD MIMO-OWC systems over other fading channels, such as Gamma-Gamma~\cite{vetelino2007fade,andrews1999theory,al2001mathematical}, K-distributed~\cite{andrews2001laser,sandalidis2008outage} ?
\item \textit{Correlated Log-normal MIMO Channels}. The main reason for using the super-rectangular cover instead of the other geometrical shape cover is that channel fading is independent. Hence, how can we generalize this theory to deal with a general correlated log-normal MIMO-OWC channel?
\end{enumerate}

\section{Appendix}

\subsection{Proof of Theorem~\ref{theorem:full_cover_uniquely}}\label{app:full_cover_identification}
Let us consider how to prove the sufficient condition first.
 Since $\frac{\mathbf{h}^T\mathbf{A}^T\mathbf{A}\mathbf{h}}{\|\mathbf{h}\|_2}$ is continuous in a closed and bounded feasible set $\{\frac{\mathbf{h}}{\|\mathbf{h}\|_2}: \mathbf{h}\in \mathbb{R}_+^N,\mathbf{h}\neq\mathbf{0}\}$, it has the minimum value, which is denoted by $C_{\min}$. In addition, utilizing the assumption that for any $\mathbf{h}\in\mathbb{R}_+^N$ leads to $\mathbf{h}^T\mathbf{A}^T\mathbf{A}\mathbf{h}\neq0$ for $\mathbf{h}\neq\mathbf{0}$ and thus, $C_{\min}>0$. It follows that  $C_{\min}\|\mathbf{h}\|_2\le\mathbf{h}^T\mathbf{A}^T\mathbf{A}\mathbf{h}$. Therefore, $\forall\tau>0$, if $\mathbf{h}^T\mathbf{A}^T\mathbf{A}\mathbf{h}\le \tau^2$, then, $C_{\min}\|\mathbf{h}\|_2\le\tau^2$ and as a consequence, $0\le h_n\le\frac{\tau}{\sqrt{C_{\min}}}$ for $n=1, 2, \cdots, N$, which indicates that $\mathbf{A}^T\mathbf{A}$ has full-cover.

 Now, let us prove the necessary condition. Suppose that there exists some $\mathbf{h}_{0}\in\mathbb{R}^{N}_{+}$ satisfying  $\mathbf{A}\mathbf{h}_{0}= \mathbf{0}$ for non-zero matrix $\mathbf{A}$. Then, $\forall \varepsilon>0$, we have $\varepsilon\mathbf{h}_{0}\in\mathbb{R}^{N}_{+}$ and  $(\varepsilon\mathbf{h}_{0})^T\mathbf{A}^T\mathbf{A}
  (\varepsilon\mathbf{h}_{0})= \varepsilon^2\mathbf{h}_{0}^T\mathbf{A}^T\mathbf{A}\mathbf{h}_{0} =0$.
  Let ${\mathcal H}_0=\{\mathbf{h}:\mathbf{h}=\varepsilon\mathbf{h}_{0},\forall \varepsilon\in \mathbb{R}_{+}\}$, which is an unbounded set.
  Therefore, there exists no outer super-rectangle covering ${\mathcal H}_0$. Let alone say $\{\mathbf{h}:\mathbf{h}^T\mathbf{A}^T\mathbf{A}\mathbf{h}\le\tau^2,\mathbf{h}\in\mathbb{R}^{N}_{+},\tau>0\}$ has full-cover.  This contradicts with the assumption that $\mathbf{A}^T\mathbf{A}$ has full-cover and thus, completes the proof of the necessary condition as well as Theorem~\ref{theorem:full_cover_uniquely}.~\hfill $\Box$

  \subsection{Proof of Theorem~\ref{theorem:full_cover_algebraic}}\label{app:sufficient_necessary}
Our proof of Theorem~\ref{theorem:full_cover_algebraic} is based on the following equality
\begin{eqnarray*}
\mathbf{h}^T\mathbf{P}\mathbf{h}=p_{ii}h_i^2+2h_i\bar{\mathbf{p}}_i^T\bar{\mathbf{h}}_i+
\bar{\mathbf{h}}_i^T\bar{\mathbf{P}}_{ii}\bar{\mathbf{h}}_i
\end{eqnarray*}

\textbf{\underline{Proof of Statement 1)}}: Let us first prove the sufficient condition. If there exists $i$ such that $\bar{\mathbf{p}}_i$ is nonnegative, then, for any $\mathbf{h}\in\mathbb{R}_+^N$, we have $p_{ii}h_i^2\ge0$, $2h_i\bar{\mathbf{p}}_i^T\bar{\mathbf{h}}_i\ge0$ and $\bar{\mathbf{h}}_i^T\bar{\mathbf{P}}_{ii}\bar{\mathbf{h}}_i\ge0$. Thus, $\mathbf{h}^T\mathbf{P}\mathbf{h}=0$ if and only if $p_{ii}h_i^2=0$, $2h_i\bar{\mathbf{p}}_i^T\bar{\mathbf{h}}_i=0$ and $\bar{\mathbf{h}}_i^T\bar{\mathbf{P}}_{ii}\bar{\mathbf{h}}_i=0$. Since $p_{ii}>0$, $p_{ii}h_i^2=0$ if and only if $h_i=0$. In addition, if $\bar{\mathbf{P}}_{ii}$ is full-cover, then, Theorem~\ref{theorem:full_cover_uniquely} tells us that $\bar{\mathbf{h}}_i^T\bar{\mathbf{P}}_{ii}\bar{\mathbf{h}}_i=0$ if and only if $\bar{\mathbf{h}}_i=\mathbf{0}$. In this case, $\mathbf{h}^T\mathbf{P}\mathbf{h}=0$ if and only if $\mathbf{h}=\mathbf{0}$. By Theorem~\ref{theorem:full_cover_uniquely}, we can conclude that $\mathbf{P}$ is full-cover and thus, the sufficiency proof of Statement 1) is complete.

To prove the necessity of Statement 1) by contradiction, we consider the following two possibilities.
 \begin{enumerate}
\item Let us suppose that there exists one $i$ such that $p_{ii}=0$. Then, there exists a vector $\mathbf{h}_0$ with the $i$-th entry being one and the other $(N-1)$ entries being zeros such that $\mathbf{h}_0^T\mathbf{P}\mathbf{h}_0=0$. This observation tells us  that $\mathbf{A}^T\mathbf{A}$ does not have full-cover by Theorem~\ref{theorem:full_cover_uniquely}. Thus, the positiveness of all the diagonal entries of $\mathbf{P}$ is necessary for $\mathbf{P}$ to have full-cover.
  \item 
  Suppose that there exists $i$ such that $\mathbf{\mathbf{p}}_i\in\mathbf{R}_+^{N-1}$ and $\bar{\mathbf{P}}_{ii}$ does not have full-cover. Then, by  Theorem~\ref{theorem:full_cover_uniquely}, we can always find an $(N-1)\times 1$ nonzero vector $\bar{\mathbf{h}}\in\mathbb{R}_+^{N-1}$ such that $\bar{\mathbf{h}}^T\bar{\mathbf{P}}_{ii}^T\bar{\mathbf{P}}_{ii}\bar{\mathbf{h}}=0$. Then, we form an $N\times1$ vector $\mathbf{h}$ by letting $h_i=0$ and $\bar{\mathbf{h}}_i=\bar{\mathbf{h}}$. As a result, this nonzero vector $\mathbf{h}\in\mathbb{R}_+^N$ assures that
      \begin{eqnarray*}
    \mathbf{h}^T\mathbf{P}\mathbf{h}&=&p_{ii}h_i^2+2h_i\bar{\mathbf{p}}_i^T\bar{\mathbf{h}}_i+
\bar{\mathbf{h}}_i^T\bar{\mathbf{P}}_{ii}\bar{\mathbf{h}}_i\nonumber\\
&=&\bar{\mathbf{h}}_i^T\bar{\mathbf{P}}_{ii}\bar{\mathbf{h}}_i=0
      \end{eqnarray*}
      which implies that $\mathbf{P}$ does not have full-cover  by Theorem~\ref{theorem:full_cover_uniquely}.
 \end{enumerate}
Thus, Statement 1) is necessary for $\mathbf{P}$ to have full-cover. Then, the proof of Statement 1) is complete.

\textbf{\underline{Proof of Statement 2)}}:  Since for any $i=1,2,~\cdots,~N$, $\Delta_i$ defined by~\eqref{eqn:discreminat} is either negative or equal to zero, we consider the following two cases.
      \begin{enumerate}
        \item $\Delta_i<0$. In this case, $\bar{\mathbf{h}}_i\neq\mathbf{0}_{N\times1}$. Otherwise,  $\Delta_i=0$. Then, $\forall\mathbf{h}\in\mathbb{R}_+^N$,  equation $p_{ii}h_i^2+2h_i\bar{\mathbf{p}}_i^T\bar{\mathbf{h}}_i+
\bar{\mathbf{h}}_i^T\bar{\mathbf{P}}_{ii}\bar{\mathbf{h}}_i=0$ with respect to $h_i$ has no solution and thus, $\mathbf{h}^T\mathbf{P}\mathbf{h}\neq0$.
        \item $\Delta_i=0$.   In this situation, $\bar{\mathbf{h}}_i^T(p_{ii}\bar{\mathbf{P}}_{ii}-\bar{\mathbf{p}}_i\bar{\mathbf{p}}_i^T)\bar{\mathbf{h}}_i=0$.  Since $(a_{ii}\bar{\mathbf{P}}_{ii}-\bar{\mathbf{p}}_i\bar{\mathbf{p}}_i^T)$ has full-cover, by Theorem~\ref{theorem:full_cover_uniquely}, $\bar{\mathbf{h}}_i^T(p_{ii}\bar{\mathbf{P}}_{ii}-\bar{\mathbf{p}}_i\bar{\mathbf{p}}_i^T)\bar{\mathbf{h}}_i=0$ if and only if $\bar{\mathbf{h}}_i=\mathbf{0}_{N-1}$. Then,
            \begin{eqnarray*}
         \mathbf{h}^T\mathbf{P}\mathbf{h}=p_{ii}h_i^2+2h_i\bar{\mathbf{p}}_i^T\bar{\mathbf{h}}_i+
\bar{\mathbf{h}}_i^T\bar{\mathbf{P}}_{ii}\bar{\mathbf{h}}_i=p_{ii}h_i^2
            \end{eqnarray*}
            Therefore, $\mathbf{h}^T\mathbf{P}\mathbf{h}=0$ if and only if $h_i=0$. Combining this with $\bar{\mathbf{h}}_i=\mathbf{0}_{N-1}$ leads to the fact that $\mathbf{h}^T\mathbf{P}\mathbf{h}=0$ if and only if $\mathbf{h}=\mathbf{0}_{N\times1}$. By Theorem~\ref{theorem:full_cover_uniquely}, we attain that $\mathbf{P}$ has full-cover.
\end{enumerate}
This completes the sufficiency proof of Statement 2).

\textbf{\underline{Proof of Statement 3)}}: To prove the necessary condition by contradiction, we suppose that $\mathbf{P}$ has full-cover and  there exists a principal sub-matrix $\breve{\mathbf{P}}$ of $\mathbf{P}$ such that $\breve{\mathbf{P}}$ does not have full-cover. Without loss of generality, we assume that $\breve{\mathbf{P}}$ is formed by the entries $p_{i_k^{(-)}i_{l}^{(-)}}$, where $k,\ell=1,~\cdots,~n$ with $1\le n\le N$. When $n=N$, we arrive at a contradiction with our assumption that $\mathbf{P}$ has full-cover. When $1\le n<N$, by Theorem~\ref{theorem:full_cover_uniquely}, there exists an $n\times n$ nonzero vector $\breve{\mathbf{h}}$ with nonnegative entries such that $\breve{\mathbf{h}}^T\breve{\mathbf{P}}\breve{\mathbf{h}}=0$. Then, we construct an $N\times1$ vector $\mathbf{h}$ by letting $h_{i_k^{(-)}}=\breve{h}_{i_k^{(-)}}$ for $k=1,~\cdots,~n$ and the other $(N-n)$ entries of $\mathbf{h}$ be zeros. Now, we attain $\mathbf{h}^T\mathbf{P}\mathbf{h}=\breve{\mathbf{h}}^T\breve{\mathbf{P}}\breve{\mathbf{h}}=0$, contradicting with our assumption that $\mathbf{P}$ has full-cover. In addition, the sufficiency holds since $\mathbf{P}$ itself is a principal matrix. Therefore, Statement 3) is indeed true.

\textbf{\underline{Proof of Statement 4)}}: To prove by contradiction, we suppose that $\mathbf{P}$ has full-cover and
    there exists $i$ such that $\mathbf{\mathbf{p}}_i\notin\mathbf{R}_+^{N-1}$ and  $\bar{\mathbf{C}}_{i_1 i_2\cdots i_n}^{(-)}$ does not have full-cover.
    Then, by Theorem~\ref{theorem:full_cover_uniquely}, there exists an $n\times1$ non-zero vector $\bar{\mathbf{v}}$ with positive entries such that  in $\bar{\mathbf{v}}^T\bar{\mathbf{C}}_{i_1 i_2\cdots i_n}^{(-)}\bar{\mathbf{v}}=0$. Now, we form an $\left(N-1\right)\times1$ vector $\bar{\mathbf{h}}_i$ by letting $\bar{h}_{i_1^{(-)}}=\bar{v}_{1},~\cdots~,\bar{h}_{i_n^{(-)}}=\bar{v}_n$ and the other $(N-n-1)$ entries be zeros. In this case,  we attain
        \begin{eqnarray*}
        \Delta_i&=&-4 \bar{\mathbf{h}}_i^T(p_{ii}\bar{\mathbf{P}}_{ii}-\bar{\mathbf{p}}_i\bar{\mathbf{p}}_i^T)\bar{\mathbf{h}}_i\nonumber\\
        &=&
    -4p_{ii}\bar{\mathbf{h}}_i^T\bar{\mathbf{P}}_{ii}\bar{\mathbf{h}}_i+4\left(\bar{\mathbf{p}}_i^T\bar{\mathbf{h}}_i\right)^2\nonumber\\
    &=& -4p_{ii}\bar{\mathbf{v}}^T\bar{\mathbf{C}}_{i_1 i_2\cdots i_n}^{(-)}\bar{\mathbf{v}}+4\left(\sum_{i=i_1}^{i_{(N-1)}}\bar{p}_{i}\bar{h}_i\right)^2\nonumber\\
    &=& 0+4\left(\sum_{i=i_1^{(-)}}^{i_n^{(-)}}\bar{p}_{i}\bar{v}_i\right)^2\ge0
           \end{eqnarray*}
  By our notations, we know that $\bar{v}_{i_1^{(-)}}<0,~\cdots,~\bar{v}_{i_n^{(-)}}<0$ and thus, $\bar{\mathbf{p}}_i^T\bar{\mathbf{h}}_i=\sum_{i=i_1^{(-)}}^{i_n^{(-)}}\bar{p}_{i}\bar{v}_i<0$. Since $a_{ii}>0$, $-\frac{\bar{\mathbf{p}}_i^T\bar{\mathbf{h}}_i}{a_{ii}}>0$ and $\Delta_i>0$, we can conclude that the quadratic equation $p_{ii}h_i^2+2h_i\bar{\mathbf{p}}_i^T\bar{\mathbf{h}}_i+
\bar{\mathbf{h}}_i^T\bar{\mathbf{P}}_{ii}\bar{\mathbf{h}}_i=0$ with respect to $h_i$ has one positive solution given by
\begin{eqnarray*}
h_i=\frac{-\sum_{i=i_1^{(-)}}^{i_n^{(-)}}\bar{p}_{i}\bar{v}_i+\sqrt{\Delta_i}}{2p_{ii}}
\end{eqnarray*}
Now, we see that  there exists an $N\times1$ non-zero vector $\mathbf{h}\in\mathbb{R}_+^N$ formed by $h_i$ and $\bar{\mathbf{h}}_i$ such that $\mathbf{h}^T\mathbf{P}\mathbf{h}=0$. By Theorem~\ref{theorem:full_cover_uniquely}, we know that $\mathbf{P}$ does not have full-cover, contradicting with our assumption that $\mathbf{P}$ has full-cover. Therefore, this  completes the proof of Statement 4) as well as of Theorem~\ref{theorem:full_cover_algebraic}.~\hfill $\Box$

\subsection{Proof of Property~\ref{property:block_diagonal}}\label{app:block_diagonal}
Let notation $\mathbf{h}_\ell$ denote an $N_\ell\times 1$ nonnegative vector. Let $\mathbf{h}=\left[\mathbf{h}_1^T,~\cdots,~\mathbf{h}_L^T\right]^T$. Then, $\mathbf{h}^T\mathbf{P}\mathbf{h}=\sum_{\ell=1}^{L}\mathbf{h}_\ell^T\mathbf{P}_l\mathbf{h}_l$. On one hand, if $\mathbf{P}_\ell$ has full-cover for any $\ell$, then, Theorem~\ref{theorem:full_cover_uniquely} tells us that  $\mathbf{h}_\ell^T\mathbf{P}_\ell\mathbf{h}_\ell=0$ if and only if $\mathbf{h}_\ell=\mathbf{0}_{N_\ell\times1}$. This implies that $\mathbf{h}^T\mathbf{P}\mathbf{h}=0$ if and only if $\mathbf{h}=\mathbf{0}$, giving us that $\mathbf{A}$ is full-cover by Theorem~\ref{theorem:full_cover_uniquely}. On the other hand, if $\mathbf{A}$ is full-cover, then, from Theorem~\ref{theorem:full_cover_uniquely}, $\mathbf{h}^T\mathbf{P}\mathbf{h}=\sum_{\ell=1}^{L}\mathbf{h}_\ell^T\mathbf{P}_\ell\mathbf{h}_\ell=0$  if and only if $\mathbf{h}_\ell=\mathbf{0}$ for any $\ell$. In addition,  $\mathbf{h}_\ell^T\mathbf{P}_\ell\mathbf{h}_\ell\ge0$ holds, implying that $\sum_{\ell=1}^{L}\mathbf{h}_\ell^T\mathbf{P}_\ell\mathbf{h}_\ell=0$ and $\mathbf{h}_\ell^T\mathbf{P}_\ell\mathbf{h}_\ell=0$ are equivalent. Putting things together produces that $\mathbf{P}_\ell$ has full-cover for any $\ell$. Therefore, the proof of Property~\ref{property:block_diagonal} is complete.~\hfill $\Box$

\subsection{Proof of Theorem~\ref{theorem:positve_vector}}\label{app:theorem_positive_vector}

 Our proof of Theorem~\ref{theorem:positve_vector}  is done by first verifying $R_c\ge \max_{\mathbb{S}_{\mathbf{A}}\cap\mathbb{\bar{R}}_{++}^{K}}$ and then proving $R_c\le \max_{\mathbb{S}_{\mathbf{A}}\cap\mathbb{\bar{R}}_{++}^{K}}$.

\subsubsection{\underline{\textbf{Proof of $R_c\ge \max_{\mathbb{S}_{\mathbf{A}}\cap\mathbb{\bar{R}}_{++}^{K}}$}}}
 Let us denote the $i$-th row of $\mathbf{A}$ by ${\mathbf a}^T_i$ for $i=1, 2,~\cdots,~M$ and first assume
 \begin{eqnarray}\label{assm:maxmum_dimension}
 K_0=\max_{\mathbb{S}_{\mathbf{A}}\cap\mathbb{\bar{R}}_{++}^{K}\neq\emptyset}K\ge1
 \end{eqnarray}
 Under this assumption, there exists an $M\times 1$ vector $\mathbf{v}$ such that
 \begin{eqnarray}\label{eqn:existence_vector}
\mathbf{v}^T {\mathbf A}={\mathbf p}^T\in\mathbb{S}_{\mathbf{A}}\cap\mathbb{\bar{R}}_{++}^{K_0}
 \end{eqnarray}
 For an arbitrarily given $\tau>0$, inequality $\|\mathbf{A}\mathbf{h}\|_2^2\le \tau^2$ gives $-\tau\le\mathbf{a}_i^T\mathbf{h}\le \tau$ for $i=1,\cdots, M$, where $\mathbf{h}\in\mathbb{R}_+^{N}$. Notice that $\forall v_i\in \mathbb{R}, i=1,2,~\cdots,~M$, we have
\begin{eqnarray*}
 -|v_i|\tau\le v_i\mathbf{a}_i^T\mathbf{h}\le |v_i|\tau
\end{eqnarray*}
Summing  the above $M$ inequalities produces
\begin{eqnarray}\label{eq:p}
-\tau\sum_{i=1}^M |v_{i}|\le \mathbf{p}^T\mathbf{h}\le \tau\sum_{i=1}^M |v_{i}|
\end{eqnarray}
 Since $\mathbf{p}^T\in\mathbb{S}_{\mathbf{A}}\cap\mathbb{\bar{R}}_{++}^{K_0}$, there exist $K_0$ integers $\ell_k^{(+)}\in\{1,\ldots,N\}$ for $k=1,~\cdots,~K_0$ such that $p_{\ell_k^{(+)}}>0$ by our definition of $\mathbb{\bar{R}}_{++}^{K_0}$ and thus, we have $\mathbf{p}^T\mathbf{h}\ge0$ for $\mathbf{h}\ge\mathbf{0}$. Combining this observation with~\eqref{eq:p} yields
\begin{eqnarray}\label{eqn:K_positive_vector}
0\le h_{\ell_k^{(+)}}\le\frac{ \tau}{p_{\ell_k^{(+)}}}\sum_{i=1}^M |v_{i}|, k=1,~\cdots,~K_0
\end{eqnarray}
Thus,
\begin{eqnarray*}
&&\{\mathbf{h}:\mathbf{h}\in\mathbb{R}_{+}^N, \mathbf{h}^T\mathbf{A}^T\mathbf{A}\mathbf{h}\le\tau^2 \}\nonumber\\
&&\subseteq\left\{\mathbf{h}:0\le h_{\ell_k^{(+)}}\le\frac{\sum_{i=1}^M |v_{i}|}{p_{\ell_k^{(+)}}}\tau,k=1,~\cdots,~K_0\right\}
\end{eqnarray*}
Then, by the definition of cover order in Definition~\ref{definition:cover_order}, we have
\begin{eqnarray}\label{eqn:cover_sufficient_small}
  R_{c}\ge K_0
\end{eqnarray}
Therefore, when $K_0=N$, we can have $R_c=N$ since definition of cover order in Definition~\ref{definition:cover_order} tells us that $0\le R_c\le N$.

\subsubsection{\underline{\textbf{Proof of $R_c\le \max_{\mathbb{S}_{\mathbf{A}}\cap\mathbb{\bar{R}}_{++}^{K}}$}}}\label{app_sub:theorem3}
In the following, we consider the case with $R_c<N$. Let $\mathbf{p}$ be defined in~\eqref{eqn:existence_vector}, which is an $N\times 1$ nonnegative vector with $K_0$ positive entries. Then, we denote the indexes of the zero-valued entries of $\mathbf{p}$ by $\ell_k^{(0)}$, where $k=1,~\cdots,~(N-K_0)$. This notation tells us that $\{\ell_k^{(0)},k=1,~\cdots,~(N-K_0)\}\cap\{\ell_k^{(+)},k=1,~\cdots~K_0\}=\emptyset$. Then, we can generate an $M\times (N-K_0)$ sub-matrix of $\mathbf{A}$ by using the $(N-K_0)$ columns of $\mathbf{A}$ indexed by $\ell_{k}^{(0)},k=1,~\cdots,~(N-K_0)$ and denote this sub-matrix by $\bar{\mathbf{\mathbf{A}}}^{(0)}$. We claim that $\mathbb{S}_{\bar{\mathbf{A}}^{(0)}}\cap\mathbb{R}_+^{N-K_0}=\emptyset$. To prove this claim by contradiction, let us first suppose that
 \begin{eqnarray}\label{assmp:null_space}
 \mathbb{S}_{\bar{\mathbf{A}}^{(0)}}\cap\mathbb{R}_+^{N-K_0}\neq\emptyset
 \end{eqnarray}
 As a consequence, there exists an $(N-K_0)\times 1$ vector $\bar{\mathbf{q}}^{(0)}$ such that $\bar{\mathbf{q}}^{(0)}\in\mathbb{S}_{\bar{\mathbf{A}}^{(0)}}\cap\mathbb{R}_+^{N-K_0}$. Then, by the definition of $\bar{\mathbf{A}}^{(0)}$, there exists  a vector $\tilde{\mathbf{p}}$ such that the $\ell_k^{(0)}$-th entry of $\tilde{\mathbf{p}}$ is given by the $k$-th entry of $\bar{\mathbf{q}}^{(0)}$ for all $k=1,~\cdots,~(N-K_0)$  and $\tilde{\mathbf{p}}\in\mathbb{S}_{\mathbf{A}}$. Then, we denote the minimum of the $K_0$ positive entries of $\mathbf{p}$ (defined in~\eqref{eqn:existence_vector}) indexed by $\ell_k^{(+)},k=1,~\cdots,~K_0$ by $p_{\min}^{(+)}$. Moreover, let us denote the minimum and the maximum of the entries of $\tilde{\mathbf{p}}$ indexed by $\ell_k^{(+)},k=1,~\cdots,~K_0$ by $\tilde{p}_{\min}^{(+)}$ and $\tilde{p}_{\max}^{(+)}$, respectively. We notice that for $\ell_k^{(+)}$ with $k=1,~\cdots,~K_0$,
  \begin{eqnarray*}
&&\frac{|\tilde{p}_{\min}^{(+)}|+|\tilde{p}_{\max}^{(+)}|+1}{p_{\min}^{(+)}}p_{\ell_k^{(+)}}+\tilde{p}_{\ell_k^{(+)}}\nonumber\\
&&\ge|\tilde{p}_{\min}^{(+)}| +|\tilde{p}_{\max}^{(+)}|+1+\tilde{p}_{\ell_k^{(+)}}\ge1>0
  \end{eqnarray*}
  by the definitions of $p_{\min}^{(+)}$, $\tilde{p}_{\min}^{(+)}$ and $\tilde{p}_{\max}^{(+)}$. From $\bar{\mathbf{q}}^{(0)}\in\mathbb{S}_{\bar{\mathbf{A}}^{(0)}}\cap\mathbb{R}_+^{N-K_0}$, where $\mathbb{R}_+^{N-K_0}$ is defined by the $(N-K_0)\times1$ nonnegative vector with at least one positive entry, we can have that among all the $(N-K_0)$ nonnegative entries of $\tilde{\mathbf{p}}$ indexed by $\ell_k^{(0)}$ for $k=1,~\cdots,~(N-K_0)$, one of them at least is positive. Combining this observation with the above established facts that
   \begin{eqnarray*}
  \{\ell_k^{(0)},1\le k\le (N-K_0)\}\cap\{\ell_k^{(+)},1\le k\le K_0\}=\emptyset
   \end{eqnarray*}
   and the entries of $\frac{|\tilde{p}_{\min}^{(+)}|+|\tilde{p}_{\max}^{(+)}|+1}{p_{\min}^{(+)}}\mathbf{p}+\tilde{\mathbf{p}}$ indexed by  $\ell_k^{(+)},k=1,~\cdots,~K_0$ are positive, we can conclude that at least $(K_0+1)$  entries of  $\frac{|\tilde{p}_{\min}^{(+)}|+|\tilde{p}_{\max}^{(+)}|+1}{p_{\min}^{(+)}}\mathbf{p}+\tilde{\mathbf{p}}$ are  positive.
Putting  this result and  $\mathbf{p},\tilde{\mathbf{p}}\in\mathbb{S}_{\mathbf{A}}$ together leads us to
 \begin{eqnarray*}
\frac{|\tilde{p}_{\min}^{(+)}|+|\tilde{p}_{\max}^{(+)}|+1}{p_{\min}^{(+)}}\mathbf{p}+\tilde{\mathbf{p}}\in\mathbb{S}_{\mathbf{A}}\cap\bar{\mathbb{R}}_{++}^{K}
 \end{eqnarray*}
 with $K\ge K_0+1$, which is a contradiction with our assumption that $K_0=\max_{\mathbb{S}_{\mathbf{A}}\cap\mathbb{\bar{R}}_{++}^{K}\neq\emptyset}K$ in~\eqref{assm:maxmum_dimension}. Thus, our assumption that $\mathbb{S}_{\bar{\mathbf{A}}^{(0)}}\cap\mathbb{R}_+^{N-K_0}\neq\emptyset$ in~\eqref{assmp:null_space} is not true. As a consequence, we arrive at $\mathbb{S}_{\bar{\mathbf{A}}^{(0)}}\cap\mathbb{R}_+^{N-K_0}=\emptyset$. In addition, by Proposition~\ref{proposition:non_full_cover}, if $\mathbb{S}_{\bar{\mathbf{A}}^{(0)}}\cap\mathbb{R}_+^{N-K_0}=\emptyset$ and $\mathbb{S}_{\bar{\mathbf{A}}^{(0)}}\cap\mathbb{R}_+^{N-K_0}=\emptyset$, then, there exists an $(N-K_0)\times 1$ vector $\tilde{\mathbf{q}}^{(0)}$ such that $\tilde{\mathbf{q}}^{(0)}\in\mathbb{S}_{\bar{\mathbf{A}}^{(0)}}^{\bot}\cap\mathbb{R}_{++}^{N-K_0}$. Then, we generate an $N\times1$ vector $\mathbf{h}_0$ by letting $h_{\ell_k^{(+)}}=0$ for $k=1,~\cdots,~K_0$ and $h_{\ell_k^{(0)}}=\tilde{q}_k^{(0)}$ for $k=1,~\cdots,~(N-K_0)$, respectively. Since $\tilde{\mathbf{q}}^{(0)}\in\mathbb{S}_{\bar{\mathbf{A}}^{(0)}}^{\bot}\cap\mathbb{R}_{++}^{N-K_0}$ gives us $\bar{\mathbf{A}}^{(0)}\tilde{\mathbf{q}}^{(0)}=\mathbf{0}$, we attain that
\begin{eqnarray*}
\mathbf{h}_0^T\mathbf{A}^T\mathbf{A}\mathbf{h}_0=0+(\tilde{\mathbf{q}}^{(0)})^T(\bar{\mathbf{A}}^{(0)})^T\bar{\mathbf{A}}^{(0)}\tilde{\mathbf{q}}^{(0)}=0
\end{eqnarray*}
Now, we have that for any positive $\tau$, it holds that $\{\xi\mathbf{h}_0: \xi>0\}\subset\{\mathbf{h}:\mathbf{h}^T\mathbf{A}^T\mathbf{A}\mathbf{h}\le\tau^2,\mathbf{h}\in\mathbb{R}^{N}_{+}\}$. Then, the arbitrariness of positive $\xi$ gives us that for any $\tau>0$, there exists no positive constant $c_{\ell_k^{(0)}}$ such that $\xi h_{\ell_k^{(0)}}\le c_{\ell_k^{(0)}}\tau$. In other words, at least $(N-K_0)$ entries of $\mathbf{h}_0$ can not be covered. By the definition of cover order in Definition~\ref{definition:cover_order}, which says that the cover order is the maximum number of the covered entries of any given nonzero $\mathbf{h}$, we arrive at $R_{c}\le K_0$.

 Combining $R_{c}\le K_0$ with~\eqref{eqn:cover_sufficient_small} gives us that $R_c=K_0$. For the case $K_0=0$, following the same argument as $R_{c}\le K_0$, we can have $R_c\le0$. By Definition~\ref{definition:cover_order}, $R_c$ is a nonnegative integer and thus, we can have $R_c=K_0=0$. Therefore, the proof of Theorem~\ref{theorem:positve_vector} is complete. ~\hfill$\Box$

\subsection{Proof of Theorem~\ref{theorem:zero_cover}}\label{app:zero_cover}
\underline{\textbf{Proof of Statement 1)}}: Since $\mathbf{h}^T\mathbf{A}^T\mathbf{A}\mathbf{h}\le \lambda_{\max}\|\mathbf{h}\|_2^2$, we can have that if $\lambda_{\max}\|\mathbf{h}\|_2^2\le\tau^2$ for any positive $\tau$, then, $\mathbf{h}^T\mathbf{A}^T\mathbf{A}\mathbf{h}\le\tau^2$. This result gives us that
\begin{eqnarray*}
&&\left\{\mathbf{h}:0\le h_i\le\frac{\tau}{\sqrt{N\lambda_{\max}}},1\le i\le N,\xi>0\right\}
\nonumber\\
&&\subseteq\left\{\mathbf{h}:\lambda_{\max}\|\mathbf{h}\|_2^2\le\tau^2,\mathbf{h}\in\mathbb{R}_+^{N}\right\}
\nonumber\\
&&\subseteq\left\{\mathbf{h}:0\le
\mathbf{h}^T\mathbf{A}^T\mathbf{A}\mathbf{h}\le\tau^2,\mathbf{h}\in\mathbb{R}_+^{N}\right\}
\end{eqnarray*}
In the following, we construct the desired nonnegative vector $\mathbf{v}$.
By Theorem~\ref{theorem:positve_vector}, we know that if the cover order of $\mathbf{A}^T\mathbf{A}$ is $R_c$, then, there exists a nonnegative vector $\mathbf{p}$ with $R_c$ positive entries and $(N-R_c)$ zero-valued entries. Denote the indexes of these positive entries of $\mathbf{p}$ by $\ell^{(+)}_i$ and the indexes of the zero-valued entries by $\ell^{(0)}_k$, where $i=1,~\cdots,~R_c$ and $k=1,~\cdots,~(N-R_c)$. Then, similar to the proof of Theorem~\ref{theorem:positve_vector} in Appendix~\ref{app:theorem_positive_vector}, we form an $(N-R_c)\times (N-R_c)$ sub-matrix of $\mathbf{A}^T\mathbf{A}$ by using the columns of $\mathbf{A}$ indexed by $\ell_{i}^{(0)},i=1,~\cdots,~(N-K_0)$ and denote this sub-matrix by $\bar{\mathbf{\mathbf{A}}}^{(0)}$. From the second part proof of Theorem~\ref{theorem:positve_vector} (see Appendix~\ref{app_sub:theorem3}), it holds that
 $\mathbb{S}_{\bar{\mathbf{A}}^{(0)}}\cap\mathbb{R}_+^{N-R_c}=\emptyset$. In addition, from Proposition~\ref{proposition:non_full_cover}, there exists an $(N-R_c)\times1$ positive vector $\bar{\mathbf{v}}$ such that $\bar{\mathbf{v}}\in\mathbb{S}_{\bar{\mathbf{A}}^{(0)}}^{\bot}\cap\mathbb{R}_{++}^{N-R_c}$ satisfying $\bar{\mathbf{v}}^T\left(\bar{\mathbf{\mathbf{A}}}^{(0)}\right)^T\bar{\mathbf{\mathbf{A}}}^{(0)}\bar{\mathbf{v}}=0$. Then, we construct an $N\times1$ nonnegative vector $\mathbf{v}$ by letting the $\ell_{i}^{(0)}$-th entry be given by the $i$-th entry of $\bar{\mathbf{v}}$ and the other $R_c$ entries be zero. Such vector $\mathbf{v}$ satisfies $\mathbf{v}^T\mathbf{A}^T\mathbf{A}\mathbf{v}=0$. Therefore, if
 \begin{eqnarray*}
&&\mathbf{h}_0\in\left\{\mathbf{h}:0\le h_i\le\frac{\tau}{\sqrt{N\lambda_{\max}}},1\le i\le N,\xi>0\right\}\nonumber\\
&&\subseteq\left\{\mathbf{h}:0\le \mathbf{h}^T\mathbf{A}^T\mathbf{A}\mathbf{h}\le\tau^2,\mathbf{h}\in\mathbb{R}_+^{N}\right\}
 \end{eqnarray*}
 then,
 \begin{eqnarray*}
&&\left(\mathbf{h}_0+\xi\mathbf{v}\right)^T\mathbf{h}^T\mathbf{A}^T\mathbf{A}\left(\mathbf{h}_0+\xi\mathbf{v}\right)\nonumber\\
&&=\mathbf{h}_0^T\mathbf{A}^T\mathbf{A}\mathbf{h}_0\le\tau^2
 \end{eqnarray*}
 holds for any positive numbers $\xi$ and $\tau$. Therefore, $\left(\mathbf{h}_0+\xi\mathbf{v}\right)\in\left\{\mathbf{h}:0\le \mathbf{h}^T\mathbf{A}^T\mathbf{A}\mathbf{h}\le\tau^2,\mathbf{h}\in\mathbb{R}_+^{N}\right\}$.
 This completes the proof of Statement 1).

 \underline{\textbf{Proof of Statement 2)}}:
 In addition, from the  above arguments, we have $\mathbb{S}_{\bar{\mathbf{A}}^{(0)}}\cap\mathbb{R}_+^{N-R_c}=\emptyset$. Then, Theorem~\ref{theorem:positve_vector} indicates that $\left(\bar{\mathbf{\mathbf{A}}}^{(0)}\right)^T\bar{\mathbf{\mathbf{A}}}^{(0)}$ has zero-cover. This completes the proof of Statement 2) as well as of Theorem~\ref{theorem:zero_cover}. ~\hfill $\Box$

\subsection{Proof of Theorem~\ref{theorem:cover_link}}\label{app:cover_link}
Let $\mathbf{P}$  be an $N\times N$ PSD matrix with cover order being given by $1\le R_c\le N$. For $R_c=N$, Statement 2) indeed holds.  To prove by contradiction for $1\le R_c<N$, we assume that there exist two distinct cover links $i_1~\cdots~i_{R_c}$ and $j_1~\cdots~j_{R_c}$ such that $\{i_1,~\cdots,~i_{R_c}\}\neq\{j_1,~\cdots,~j_{R_c}\}$. Then, for the definition of cover link given in Definition~\ref{definition:cover_order}, we have that for any given positive constant $\tau$,
\begin{eqnarray*}
&&\{\mathbf{h}\in\mathbb{R}_+^N:\mathbf{h}^T\mathbf{P}\mathbf{h}\le \tau^2\}\nonumber\\
&&\subseteq\{\mathbf{h}\in\mathbb{R}_+^N: 0\le h_{i_k}\le c_{i_k}\tau,1\le k\le R_c\}
\end{eqnarray*}
and
\begin{eqnarray*}
&&\{\mathbf{h}\in\mathbb{R}_+^N:\mathbf{h}^T\mathbf{P}\mathbf{h}\le \tau^2\}\nonumber\\
&&\subseteq\{\mathbf{h}\in\mathbb{R}_+^N: 0\le h_{j_k}\le c_{j_k}\tau,1\le k\le R_c\}
\end{eqnarray*}
Therefore, we arrive at the following
\begin{eqnarray*}
&&\{\mathbf{h}\in\mathbb{R}_+^N:\mathbf{h}^T\mathbf{P}\mathbf{h}\le \tau^2\}
\nonumber\\
&&
\subseteq\{\mathbf{h}\in\mathbb{R}_+^N: 0\le h_{i_k}\le c_{i_k}\tau,1\le k\le R_c\}
\nonumber\\
&&
\cup\{\mathbf{h}\in\mathbb{R}_+^N: 0\le h_{j_k}\le c_{j_k}\tau,1\le k\le R_c\}
\end{eqnarray*}
Now, we consider the following cases:
\begin{enumerate}
  \item $\{i_1,~\cdots,~i_{R_c}\}\cap\{j_1,~\cdots,~j_{R_c}\}=\emptyset$. In this case,
  \begin{eqnarray*}
&&  \{\mathbf{h}\in\mathbb{R}_+^N: 0\le h_{i_k}\le c_{i_k}\tau,1\le k\le R_c\}
\nonumber\\
&&\cup\{\mathbf{h}\in\mathbb{R}_+^N: 0\le h_{j_k}\le c_{j_k}\tau,1\le k\le R_c\}\nonumber\\
&&=
\{\mathbf{h}\in\mathbb{R}_+^N: 0\le h_{i_k}\le c_{i_k}\tau, ,1\le k\le R_c
\nonumber\\
&&0\le h_{j_k}\le c_{j_k}\tau,1\le k\le R_c\}
  \end{eqnarray*}
  Then,
\begin{eqnarray*}
 && \{\mathbf{h}\in\mathbb{R}_+^N:\mathbf{h}^T\mathbf{P}\mathbf{h}\le \tau^2\}
 \nonumber\\
&&\{\mathbf{h}\in\mathbb{R}_+^N: 0\le h_{i_k}\le c_{i_k}\tau,
1\le k\le R_c
\nonumber\\
&& 0\le h_{j_k}\le c_{j_k}\tau,1\le k\le R_c\}
  \end{eqnarray*}
  implying that the cover order of $\mathbf{P}$ is not smaller than $2R_c$ and leading to a contradiction with our assumption that the cover order of $\mathbf{P}$ is $R_c$.  Therefore, $\{i_1,~\cdots,~i_{R_c}\}\cap\{j_1,~\cdots,~j_{R_c}\}=\emptyset$ can not happen.
  \item The cardinality of $\{i_1,~\cdots,~i_{R_c}\}\cap\{j_1,~\cdots,~j_{R_c}\}$ is equal to $\bar{R}$ where $1\le \bar{R}<R_c$. For presentation simplicity, we assume that $i_k=j_k$ for $k=1,\cdots,\bar{R}$. Furthermore, we redenote the entries by $\{i_1,~\cdots,~i_{R_c}\}\cap\{j_1,~\cdots,~j_{R_c}\}$ by $n_1~\cdots~n_{\bar{R}}$ and the the remaining $2R_c-\bar{R}$ entries of $\{i_1,~\cdots,~i_{R_c}\}\cup\{j_1,~\cdots,~j_{R_c}\}$ by $n_{\bar{R}+1}~\cdots~n_{2R_c-\bar{R}}$. Then,
        \begin{eqnarray*}
&&  \{\mathbf{h}\in\mathbb{R}_+^N: 0\le h_{i_k}\le c_{i_k}\tau,1\le k\le R_c\}
\nonumber\\
&&\cup\{\mathbf{h}\in\mathbb{R}_+^N: 0\le h_{j_k}\le c_{j_k}\tau,1\le k\le R_c\}\nonumber\\
&&=
\{\mathbf{h}\in\mathbb{R}_+^N: 0\le h_{n_\ell}\le \max(c_{i_\ell},c_{j_\ell})\tau, 1\le \ell\le \bar{R},
\nonumber\\
&&0\le h_{i_k}\le c_{i_k}\tau,R+1\le k\le R_c
\nonumber\\
&&0\le h_{j_k}\le c_{j_k}\tau,R+1\le k\le R_c\}
  \end{eqnarray*}
  telling us the cover order of $\mathbf{P}$ must be equal to or larger than $2R_c-\bar{R}\ge R_c+1$ and thus, contradicting our assumption that $R_c$ is the cover order of $\mathbf{P}$.
\end{enumerate}
The above discussions allow us to conclude that if the cover order of $\mathbf{P}$ is equal to $R_c$, then, there exists no two distinct cover links  $\{i_1,~\cdots,~i_{R_c}\}$ and $\{j_1,~\cdots,~j_{R_c}\}$. Therefore, Theorem~\ref{theorem:cover_link} is indeed true. ~\hfill$\Box$

\subsection{Proof of Theorem~\ref{theorem:2_by_2_full_diversity}}\label{app:two_by_two}
Let $\mathbf{P}=\left(
{\begin{array}{cc}
p_{11}&p_{12}\\
p_{12}&p_{22}\\
\end{array}
}\right)$.
Notice that
\begin{eqnarray*}
\mathbf{h}^T\mathbf{P}\mathbf{h}&=&p_{11}h_1^2+2p_{12}h_1h_2+p_{22}h_2^2\nonumber\\
&=&p_{11}\left(h_1+\frac{p_{12}}{p_{11}}h_2\right)^2+\frac{\left(p_{11}p_{22}-p_{12}^2\right)h_2^2}{p_{11}}\nonumber\\
&=&p_{11}\left(h_1+\frac{p_{12}}{p_{11}}h_2\right)^2+\frac{h_2^2\det\mathbf{P}}{p_{11}}
\end{eqnarray*}
We consider the following possibilities.
\begin{enumerate}
\item \underline{$p_{11},p_{12},p_{22}>0$}. In this case, $\mathbf{h}^T\mathbf{P}\mathbf{h}=p_{11}h_1^2+2p_{12}h_1h_2+p_{22}h_2^2\le \tau^2$ gives us $h_1\le \frac{\tau}{\sqrt{p_{11}}}$ and $h_2\le \frac{\tau}{\sqrt{p_{22}}}$. Therefore, $\mathbf{P}$ has full-cover and $c_i=\frac{\tau}{\sqrt{p_{ii}}}$ for $i=1,2$.

  \item \underline{$\det\mathbf{P}=0$ and $p_{12}\le0$}. By Property~\ref{property:rank_one}, we know that $\mathbf{P}$ is zero-cover.

  \item \underline{$\det\mathbf{P}>0$ and $p_{12}\le0$}. In this case, $\mathbf{h}^T\mathbf{P}\mathbf{h}\le \tau^2$ implies $h_2\le\frac{\sqrt{p_{11}}\tau}{\sqrt{\det\mathbf{P}}}$ and in the same token, $h_1\le \frac{\sqrt{p_{22}}\tau}{\sqrt{\det\mathbf{P}}}$. Therefore, $c_i=\sqrt{\left[\mathbf{P}^{-1}\right]_{ii}},i=1,2$.
 \end{enumerate}
 Therefore, the $2\times2$ matrix $\mathbf{P}$ is full-cover if and only if $\mathbf{P}$ is full-rank or all its entries are positive. If $\mathbf{P}$ is full-rank and $p_{12}\le 0$, then, $c_i=\sqrt{\left[\mathbf{P}^{-1}\right]_{ii}},i=1,2$. When $p_{11},p_{12},p_{22}>0$, the cover length of full-cover $\mathbf{P}$ is given by $c_i=\frac{1}{\sqrt{p_{ii}}},i=1,2$. This completes the proof of Theorem~\ref{theorem:2_by_2_full_diversity}.
~\hfill $\Box$

\subsection{Proof of Theorem~\ref{theorem:rank_one_cover_length}}\label{app:cover_length_bound}
 Let us define $\mathbf{e}_i$ by an $N\times1$ vector such that the $i$-th entry is nonzero and the other $(N-1)$ entries are zeros. By following the notations defined in Subsection~\ref{subsec:full_cover}, we have $\mathbf{h}^T\mathbf{P}\mathbf{h}=p_{ii}h_i^2+2h_i\bar{\mathbf{p}}_i^T\bar{\mathbf{h}}_i+
\bar{\mathbf{h}}_i^T\bar{\mathbf{P}}_{ii}\bar{\mathbf{h}}_i$. Notice that  $\frac{\tau}{\sqrt{p_{ii}}}\mathbf{e}_i$ is one solution to the equation $p_{ii}h_i^2+2h_i\bar{\mathbf{p}}_i^T\bar{\mathbf{h}}_i+
\bar{\mathbf{h}}_i^T\bar{\mathbf{P}}_{ii}\bar{\mathbf{h}}_i=\tau^2$ with respect to $\mathbf{h}$. Then, by Definition~\ref{definition:cover_order}, $\frac{\tau}{\sqrt{p_{ii}}}\mathbf{e}_i\in\{\mathbf{h}:\mathbf{h}\in\mathbb{R}_{+}^N, \mathbf{h}^T\mathbf{P}\mathbf{h}\le\tau^2 \}$ and further,
\begin{eqnarray*}
\frac{\tau}{\sqrt{p_{ii}}}\mathbf{e}_i\in \{\mathbf{h}:0\le h_{k}\le{c}_{k}\tau, k=1, 2,~\cdots,~N\}
\end{eqnarray*}
giving us $\frac{\tau}{\sqrt{p_{ii}}}\le c_{i}$ for any $1\le i\le N$.

 In addition, if all the entries of $\mathbf{P}$ are nonnegative, then, for any $\mathbf{h}\in\mathbb{R}_+^N$, we can always have that  $p_{ii}h_i^2\ge0, 2h_i\bar{\mathbf{p}}_i^T\bar{\mathbf{h}}_i\ge0$ and $\bar{\mathbf{h}}_i^T\bar{\mathbf{P}}_{ii}\bar{\mathbf{h}}_i\ge0$. Letting $\mathbf{h}^T\mathbf{P}\mathbf{h}\le\tau^2$ allows us to arrive at $p_{ii}h_i^2\le\tau^2$. This observation tells us that for any $\mathbf{h}\in\mathbb{R}_+^N$ and a given matrix $\mathbf{P}$, of which all the entries are nonnegative,  satisfying $\mathbf{h}^T\mathbf{P}\mathbf{h}\le\tau^2$, the maximum  achievable value of $h_i$ is $\frac{1}{\sqrt{p_{ii}}}$. By the definition of cover length in Definition~\ref{definition:cover_order}, the $i$-th cover length of the matrix $\mathbf{P}$ is given by $c_i=\frac{1}{\sqrt{p_{ii}}}$ if all the entries of full-cover $\mathbf{P}$ are nonnegative. Then, we prove that  $c_i=\frac{1}{\sqrt{p_{ii}}}$ holds for any $i$ only if all the entries of full-cover $\mathbf{P}$ are nonnegative. To prove by contradiction, we assume that there exists a PSD matrix $\mathbf{P}$ satisfying $c_i=\frac{1}{\sqrt{p_{ii}}}$ for $i$ and having a negative nondiagonal entry denoted by $p_{ij}$. Now, we consider the $2\times2$ subprincipal matrix $\mathbf{\bar{P}}=\left(\begin{array}{ll}p_{ii},p_{ij}\\p_{ij},p_{jj}\end{array}\right)$ of $\mathbf{P}$. From Theorem~~\ref{theorem:2_by_2_full_diversity}, we know that over the domain $\{\mathbf{h}\in\mathbb{R}_+^N:p_{ii}h_{i}^2+2p_{ij}h_ih_j+p_{jj}^2h_j^2\le\tau^2\}$, if $p_{ij}<0$ and $\left(\begin{array}{ll}p_{ii},p_{ij}\\p_{ij},p_{jj}\end{array}\right)$ has full-cover, then, $\max h_i=\tau\sqrt{\left[\mathbf{P}^{-1}\right]_{ii}}$, which  is larger than $\frac{\tau}{\sqrt{p_{ii}}}$ because  $p_{ii}h_{i}^2+2p_{ij}h_ih_j+p_{jj}^2h_j^2=\tau^2$ holds for $h_i=\frac{\tau}{\sqrt{p_{ii}}}$ and $h_i=0$. From the fact that  $\{\mathbf{h}\in\mathbb{R}_+^N:p_{ii}h_{i}^2+2p_{ij}h_ih_j+p_{jj}^2h_j^2\le\tau^2\}\subseteq
 \{\mathbf{h}\in\mathbb{R}_+^N:\mathbf{h}^T\mathbf{P}\mathbf{h}\le\tau^2\}$, we can conclude that over the domain that $ \{\mathbf{h}\in\mathbb{R}_+^N:\mathbf{h}^T\mathbf{P}\mathbf{h}\le\tau^2\}$, $c_i\tau=\max h_i>\frac{\tau}{\sqrt{p_{ii}}}$, contradicting with our assumption that  $c_n=\frac{1}{\sqrt{p_{nn}}}$ for any $n=1,\cdots,N$. Therefore, $c_i=\frac{1}{\sqrt{p_{ii}}}$ for any $i$ holds if and only if the PSD matrix $\mathbf{P}$ has full-cover and all its entries are nonnegative.
  Therefore, the proof of Theorem~\ref{theorem:rank_one_cover_length} is complete.
~\hfill $\Box$

\subsection{Proof of Theorem~\ref{theorem:full_diversity_condition}}\label{app:cover_diversity}
Let us suppose $0\le R_c\le N$ and $\tau$ is positive and the function of $\rho$. Then, $P(\mathbf{X}\rightarrow\hat{\mathbf{X}})$ can be fragmented as
\begin{eqnarray} \label{eqn:ml_detection_pep_fragment}
&&P(\mathbf{X}\rightarrow\hat{\mathbf{X}})
={\frac{1}{\pi}\int_{0}^{\frac{\pi}{2}}\int_{\sum_{j=1}^M \|\Delta\mathbf{X}\mathbf{h}_j\|_2^2\le\tau^2}
\frac{f_{\mathbf{H}}\left(\mathbf{H}\right)d\mathbf{H}d\theta}{e^{\frac{\rho\sum_{j=1}^M\|\Delta\mathbf{X}\mathbf{h}_j\|_2^2}{8N\sin^2\theta}}} }\nonumber\\
&&+\underbrace{\frac{1}{\pi}\int_{0}^{\frac{\pi}{2}}\int_{\sum_{j=1}^M \|\Delta\mathbf{X}\mathbf{h}_j\|_2^2>\tau^2}\frac{f_{\mathbf{H}}\left(\mathbf{H}\right)d\mathbf{H}d\theta}{e^{\frac{\rho\sum_{j=1}^M\|\Delta\mathbf{X}\mathbf{h}_j\|_2^2}{8N\sin^2\theta}}} }_{\bar{P}_{\tau}(\mathbf{X}\rightarrow\hat{\mathbf{X}})}\nonumber\\
&\le&\underbrace{\frac{1}{\pi}\int_{0}^{\frac{\pi}{2}}\int_{\|\Delta\mathbf{X}\mathbf{h}_j\|_2^2\le\tau^2}
\frac{f_{\mathbf{H}}\left(\mathbf{H}\right)d\mathbf{H}d\theta}{e^{\frac{\rho\sum_{j=1}^M\|\Delta\mathbf{X}\mathbf{h}_j\|_2^2}{8N\sin^2\theta}}} }_{P_{\tau}(\mathbf{X}\rightarrow\hat{\mathbf{X}})}+{\bar{P}_{\tau}(\mathbf{X}\rightarrow\hat{\mathbf{X}})}
\end{eqnarray}
since if $\sum_{j=1}^M \|\Delta\mathbf{X}\mathbf{h}_j\|_2^2\le\tau^2$, then, we have $\|\Delta\mathbf{X}\mathbf{h}_j\|_2^2\le\tau^2$ for $j=1, 2, \cdots, M$. If the cover order of $\Delta\mathbf{X}^T\Delta\mathbf{X}$ is $R_c$ ($0<R_c\le N$), then, without loss of generality, we fix the indexes of the covered columns of $\Delta\mathbf{X}$ to be $j=1, 2,~\cdots,~R_c$, if a proper permutation matrix is used. Then, by the definition of cover order, we attain
\begin{eqnarray}\label{eqn:cover_domain}
&&\{\mathbf{h}:\mathbf{h}\in\mathbb{R}_{+}^N,\mathbf{h}^T\Delta\mathbf{X}^T\Delta\mathbf{X}\mathbf{h}\le\tau^2 \}\nonumber\\
&&\subseteq\{\mathbf{h}:0\le h_{i}\le{c}_{i}\tau,i=1, 2, \cdots, R_c\}
\end{eqnarray}
Upper-bound and Lower-bound of $P(\mathbf{X}\rightarrow\hat{\mathbf{X}})$ are given as follows.
\subsubsection{Upper-bound of $P_{\tau}(\mathbf{X}\rightarrow\hat{\mathbf{X}})$}
$0\le \|\Delta\mathbf{X}\mathbf{h}_j\|_2^2\le\tau^2$ allows us to upper-bound $P_{\tau}(\mathbf{X}\rightarrow\hat{\mathbf{X}})$ by
\begin{eqnarray*}
P_{\tau}(\mathbf{X}\rightarrow\hat{\mathbf{X}})
&=&\frac{1}{\pi}\int_{0}^{\frac{\pi}{2}}\int_{\|\Delta\mathbf{X}\mathbf{h}_j\|_2^2\le\tau^2}
\frac{f_{\mathbf{H}}\left(\mathbf{H}\right)d\mathbf{H}d\theta}{e^{\frac{\rho\sum_{j=1}^M\|\Delta\mathbf{X}\mathbf{h}_j\|_2^2}{8N\sin^2\theta}}} \nonumber\\
&\le&
\frac{1}{\pi}\int_{0}^{\frac{\pi}{2}}\int_{\|\Delta\mathbf{X}\mathbf{h}_j\|_2^2\le\tau^2}f_{\mathbf{H}}\left(\mathbf{H}\right)d\mathbf{H}d\theta \nonumber\\
&=&\frac{1}{2}\int_{\|\Delta\mathbf{X}\mathbf{h}_j\|_2^2\le\tau^2}f_{\mathbf{H}}\left(\mathbf{H}\right)d\mathbf{H} \end{eqnarray*}
where the last inequality is resulted from the fact that $\forall x\ge0,\exp(-x)\le1$. By~\eqref{eqn:cover_domain}, we can further upper-bound $P_{\tau}(\mathbf{X}\rightarrow\hat{\mathbf{X}})$ by
\begin{eqnarray}\label{eqn:0_tau}
P_{\tau}(\mathbf{X}\rightarrow\hat{\mathbf{X}})&\le&
\frac{1}{2}\prod_{j=1}^M\prod_{i=1}^{R_c}\int_{0}^{c_i\tau}f_{H_{ij}}\left(h_{ij}\right)dh_{ij}
\nonumber\\
&=&\frac{1}{2}\prod_{j=1}^{M}\prod_{i=1}^{R_c}Q\left(-\frac{\ln\left(c_i\tau\right)-\mu_{ij}}{\sigma_{ij}}\right)
\end{eqnarray}

\subsubsection{Upper-bound of $\bar{P}_\tau(\mathbf{X}\rightarrow\hat{\mathbf{X}})$}
Since $e^{-x}$ is monotonically decreasing with respect to nonnegative $x$,
  $\bar{P}_\tau(\mathbf{X}\rightarrow\hat{\mathbf{X}})$ can be upper-bounded  by
    \begin{eqnarray}\label{eqn:tau_infty}
&&\bar{P}_\tau(\mathbf{X}\rightarrow\hat{\mathbf{X}})
=\frac{1}{\pi}\int_{0}^{\frac{\pi}{2}}\int_{\sum_{j=1}^M \|\Delta\mathbf{X}\mathbf{h}_j\|_2^2>\tau^2}
\frac{f_{\mathbf{H}}\left(\mathbf{H}\right)d\mathbf{H}d\theta}{e^{\frac{\rho\sum_{j=1}^M\|\Delta\mathbf{X}\mathbf{h}_j\|_2^2}{8N\sin^2\theta}}} \nonumber\\
&&\le\frac{1}{2}\exp\left(-\frac{\rho\tau^2}{8N}\right)\int_{\sum_{j=1}^M \|\Delta\mathbf{X}\mathbf{h}_j\|_2^2\ge0}f_{\mathbf{H}}\left(\mathbf{H}\right)d\mathbf{H}
\nonumber\\
&&=\frac{1}{2}\exp\left(-\frac{\rho\tau^2}{8N}\right)
\end{eqnarray}
Combining~\eqref{eqn:tau_infty} and~\eqref{eqn:0_tau} leads us to the upper-bound of $P(\mathbf{X}\rightarrow\hat{\mathbf{X}})$ as follows.
\begin{eqnarray}\label{eqn:upper_bound11}
P(\mathbf{X}\rightarrow\hat{\mathbf{X}})
&\le&\frac{1}{2}\prod_{j=1}^{M}\prod_{i=1}^{R_c}Q\left(-\frac{\ln\left(c_i\tau\right)-\mu_{ij}}{\sigma_{ij}}\right)\nonumber\\
&+&\frac{1}{2}\exp\left(-\frac{\rho\tau^2}{8N}\right)
\end{eqnarray}
\subsubsection{Lower-bound of $P(\mathbf{X}\rightarrow\hat{\mathbf{X}})$}\label{sec:lower_bound}
We have
\begin{eqnarray}\label{eqn:lower_bound_non}
&&P(\mathbf{X}\rightarrow\hat{\mathbf{X}})\ge
\frac{1}{\pi}\int_{0}^{\frac{\pi}{2}}\int_{\|\Delta\mathbf{X}\mathbf{h}_j\|_2^2\le\frac{1}{\rho}}\frac{f_{\mathbf{H}}\left(\mathbf{H}\right)d\mathbf{H}d\theta }{e^{\frac{\rho\sum_{j=1}^M\|\Delta\mathbf{X}\mathbf{h}_j\|_2^2}{8N\sin^2\theta}}}
\nonumber \\
&&\ge
\frac{1}{\pi}\int_{0}^{\frac{\pi}{2}}\int_{\|\Delta\mathbf{X}\mathbf{h}_j\|_2^2\le\frac{1}{\rho}}\frac{f_{\mathbf{H}}\left(\mathbf{H}\right)d\mathbf{H}d\theta }{e^{\frac{M}{8N\sin^2\theta}}}
\nonumber \\
&&=
Q\left(\sqrt{\frac{M}{4N}}\right)\int_{\|\Delta\mathbf{X}\mathbf{h}_j\|_2^2\le\frac{1}{\rho}}f_{\mathbf{H}}\left(\mathbf{H}\right)d\mathbf{H}
\end{eqnarray}
By  Theorem~\ref{theorem:zero_cover}, we know that if the cover order $\Delta\mathbf{X}^T\Delta\mathbf{X}$ is $R_c$, then, there exists a nonnegative vector $\mathbf{v}$ such that
 \begin{eqnarray*}
&&\left\{\mathbf{h}+\xi\mathbf{v}:0\le h_i\le\frac{\tau}{\sqrt{N\lambda_{\max}}},1\le i\le N,\xi>0\right\}\nonumber\\
&&\subseteq\left\{\mathbf{h}:0\le \mathbf{h}^T\mathbf{A}^T\mathbf{A}\mathbf{h}\le\tau^2,\mathbf{h}\in\mathbb{R}_+^{N}\right\}
 \end{eqnarray*}
 holds for any given positive constant $\tau$, where $\lambda_{\max}$ denotes the maximum eigenvalue of $\mathbf{A}^T\mathbf{A}$. Denote the indexes of the positive entries of $\mathbf{v}$ by $\ell_k^{(+)}$ where $k=1,~\cdots,~(N-R_c)$ and the indexes of the zero-valued entries of $\mathbf{v}$ by $\ell_k^{(0)}$ with $k=1,~\cdots,~R_c$. Then, for any positive number $\xi$, it holds that
\begin{eqnarray*}
&&\bigg{\{}\mathbf{h}:0\le h_{\ell_k^{(0)}}\le\frac{1}{\sqrt{N\rho\lambda_{\max}}},1\le k \le R_c,
\xi v_{\ell_k^{(+)}}\le\nonumber\\
&& h_{\ell_k^{(+)}}\le\xi v_{\ell_k^{(+)}}+\frac{1}{\sqrt{\rho\lambda_{\max}}},1\le k\le(N-R_c),
\xi>0\bigg{\}}
\nonumber\\
&&
\subseteq\left\{\mathbf{h}:0\le \mathbf{h}^T\mathbf{A}^T\mathbf{A}\mathbf{h}\le\frac{1}{\rho},\mathbf{h}\in\mathbb{R}_+^{N}\right\}
\end{eqnarray*}
Then, combining this result with~\eqref{eqn:lower_bound_non} allows us to further lower-bound $P(\mathbf{X}\rightarrow\hat{\mathbf{X}})$ below.
\begin{eqnarray*}
&&P(\mathbf{X}\rightarrow\hat{\mathbf{X}})\ge
Q\left(\sqrt{\frac{M}{4N}}\right)\nonumber \\
&&\times
\prod_{k=1}^{R_c}\prod_{j=1}^M
\int_{0}^{\frac{1}{\sqrt{N\rho\lambda_{\max}}}}f_{H_{\ell_k^{(0)}j}}\left(h_{\ell_k^{(0)}j}\right)dH_{\ell_k^{(0)}j}
\nonumber \\
&&\times \prod_{k=1}^{N-R_c}\prod_{j=1}^M
\int_{\xi v_{\ell_k^{(+)}}}^{\xi v_{\ell_k^{(+)}}+\frac{1}{\sqrt{N\rho\lambda_{\max}}}}f_{H_{\ell_k^{(+)}j}}\left(h_{\ell_k^{(+)}j}\right)dH_{\ell_k^{(+)}j}
\nonumber \\
&&=
Q\left(\sqrt{\frac{M}{4N}}\right)
\prod_{k=1}^{R_c}\prod_{j=1}^MQ\left(\frac{\ln\left(\sqrt{N\lambda_{\max}\rho}\right)+\mu_{\ell_k^{(0)}j}}{\sigma_{\ell_k^{(0)}j}}\right)
\nonumber \\
&&\times \prod_{k=1}^{N-R_c}\prod_{j=1}^M
\int_{\xi v_{\ell_k^{(+)}}}^{\xi v_{\ell_k^{(+)}}+\frac{1}{\sqrt{N\rho\lambda_{\max}}}}f_{H_{\ell_k^{(+)}j}}\left(h_{\ell_k^{(+)}j}\right)dH_{\ell_k^{(+)}j}
\end{eqnarray*}
Now, let the maximum entry of $\mathbf{v}$ be denoted by $V_{\max}$, the peak point of the PDF function  $f_{H_{ij}}\left(h_{ij}\right)$ be denoted by $\tilde{H}_{ij}$ and thus, $H_{\min}=\min_{i,j} \tilde{H}_{ij}$. Then, we choose $\xi=\frac{H_{\min}}{2V_{\max}}$ so that $\xi v_{\ell_k^{(+)}}\le \frac{H_{\min}}{2}$. Hence, in the high SNR regimes, it holds that $\xi v_{\ell_k^{(+)}}+\frac{1}{\sqrt{N\rho\lambda_{\max}}}\in\left(0,H_{\min}\right)$. In addition, the PDF of log-normal distribution is monotonically increasing in $\left(0,H_{\min}\right)$. Therefore, when SNR is sufficiently high, $\int_{\xi v_{\ell_k^{(+)}}}^{\xi v_{\ell_k^{(+)}}+\frac{1}{\sqrt{N\rho\lambda_{\max}}}}f_{H_{\ell_k^{(+)}j}}\left(h_{\ell_k^{(+)}j}\right)dH_{\ell_k^{(+)}j}$ can be lower-bounded by $\frac{1}{\sqrt{N\rho\lambda_{\max}}}f_{H_{\ell_k^{(+)}j}}\left(\xi v_{\ell_k^{(+)}}\right)$.  This observation allows us to further lower-bound $P(\mathbf{X}\rightarrow\hat{\mathbf{X}})$ by
\begin{eqnarray}\label{eqn:lower_bound_over}
&&P(\mathbf{X}\rightarrow\hat{\mathbf{X}})\ge
Q\left(\sqrt{\frac{M}{4N}}\right)
\nonumber\\
&&\times \prod_{k=1}^{R_c}\prod_{j=1}^MQ\left(\frac{\ln\left(\sqrt{N\lambda_{\max}\rho}\right)+\mu_{\ell_k^{(0)}j}}{\sigma_{\ell_k^{(0)}j}}\right)
\nonumber \\
&&\times \prod_{k=1}^{N-R_c}\prod_{j=1}^M
\frac{1}{\sqrt{N\rho\lambda_{\max}}}f_{H_{\ell_k^{(+)}j}}\left(\xi v_{\ell_k^{(+)}}\right)
\nonumber \\
&&=C_{L}\rho^{-\frac{M\left(N-R_c\right)}{2}}\prod_{k=1}^{R_c}\prod_{j=1}^M
Q\left(\frac{\ln\rho+\ln\left(N\lambda_{\max}\right)+2\mu_{\ell_k^{(0)}j}}{2\sigma_{\ell_k^{(0)}j}}\right)\nonumber \\
\end{eqnarray}
where
\begin{eqnarray*}
C_L=Q\left(\sqrt{\frac{M}{4N}}\right)\left(N\lambda_{\max}\right)^{-\frac{M\left(N-R_c\right)}{2}}
\prod_{k=1}^{N-R_c}\prod_{j=1}^M
f_{H_{\ell_k^{(+)}j}}\left(\xi v_{\ell_k^{(+)}}\right)
\end{eqnarray*}
is independent of SNR.

\subsubsection{Determination of  $\tau$}
Combining~\eqref{eqn:upper_bound11} and~\eqref{eqn:lower_bound_over} leads us to what follows
\begin{eqnarray}\label{eqn:pep_temp_non}
&&C_{L}\rho^{-\frac{M\left(N-R_c\right)}{2}}\prod_{k=1}^{R_c}\prod_{j=1}^M
Q\left(\frac{\ln\rho+\ln\left(N\lambda_{\max}\right)+2\mu_{\ell_k^{(0)}j}}{2\sigma_{\ell_k^{(0)}j}}\right)
\nonumber\\
&&\le  P(\mathbf{X}\rightarrow\hat{\mathbf{X}})
\nonumber\\
&&\le \frac{1}{2}\prod_{k=1}^{R_c}\prod_{j=1}^{M}Q\left(\frac{-\ln\left(c_{\ell_k^{(0)}}\tau\right)+\mu_{\ell_k^{(0)}j}}{\sigma_{\ell_k^{(0)}j}}\right)
\nonumber \\
&&+\frac{1}{2}\exp\left(-\frac{\rho\tau^2}{8N}\right)
\end{eqnarray}
We choose $\tau$ as follows.
\begin{eqnarray*}
\tau=\sqrt{\frac{N\ln^2\rho}{M\rho}\sum_{j=1}^M\sum_{k=1}^{R_c}\sigma_{{\ell_k^{(0)}}j}^{-2}}
\end{eqnarray*}
 Then,~\eqref{eqn:pep_temp_non} can be rewritten into
 \begin{eqnarray}\label{eqn:final_bounds}
&&C_{L}\rho^{-\frac{M\left(N-R_c\right)}{2}}\prod_{k=1}^{R_c}\prod_{j=1}^M
Q\left(\frac{\ln\rho+\ln\left(N\lambda_{\max}\right)+2\mu_{\ell_k^{(0)}j}}{2\sigma_{\ell_k^{(0)}j}}\right)\nonumber\\
&&\le  P(\mathbf{X}\rightarrow\hat{\mathbf{X}})
\nonumber\\
&&\le \frac{1}{2}\prod_{k=1}^{R_c}\prod_{j=1}^{M}Q\left(\frac{\ln\rho-2\ln\ln\rho-2\ln c_{\ell_k^{(0)}}+2\mu_{\ell_k^{(0)}j}}{2\sigma_{\ell_k^{(0)}j}}\right)
\nonumber\\&&+\frac{1}{2}\exp\left(-\sum_{j=1}^M\sum_{k=1}^{R_c}\frac{\ln^2\rho}{8\sigma_{{\ell_k^{(0)}}j}^{2}}\right)
\end{eqnarray}
From the lower-bound of Q-function $Q\left(x\right)\ge\frac{1}{\sqrt{2\pi}}\left(\frac{1}{x}-\frac{1}{x^3}\right)\exp\left(-\frac{x^2}{2}\right)$ for $x\ge\frac{\sqrt{2}}{2}$ and the definition of  the large-scale diversity gain $\mathcal{D}_{l}\left(\Delta\mathbf{X}\right)$ given by~\eqref{def:larg}, we can, without much difficulty, arrive at that $\sum_{j=1}^M \sum_{k=1}^{R_c}\sigma_{i_k, j}^{-2}\le\mathcal{D}_{l}\left(\Delta\mathbf{X}\right)\le \sum_{j=1}^M \sum_{k=1}^{R_c}\sigma_{i_k, j}^{-2}$. Therefore, this completes the proof of Theorem~\ref{theorem:full_diversity_condition}.~\hfill$\Box$

\subsection{Proof of Theorem~\ref{theorem:mimo_ovlc_pep}}\label{app:mimo_ovlc_pep}

Let us assume that when $\rho\rightarrow\infty$, $\tau\rightarrow0$. Then, $P(\mathbf{X}\rightarrow\hat{\mathbf{X}})$ can be fragmented into
\begin{eqnarray} \label{eqn:ml_detection_pep_fragment}
&&P(\mathbf{X}\rightarrow\hat{\mathbf{X}})
=\int_{\sum_{j=1}^M\|\Delta\mathbf{X}\mathbf{h}_j\|_2^2\le\tau^2}P\left(\mathbf{X}\rightarrow\hat{\mathbf{X}}|\mathbf{H}\right)f_{\mathbf{H}}\left(\mathbf{H}\right)d\mathbf{H}
\nonumber\\
&&+\int_{\sum_{j=1}^M\|\Delta\mathbf{X}\mathbf{h}_j\|_2^2>\tau^2}P\left(\mathbf{X}\rightarrow\hat{\mathbf{X}}|\mathbf{H}\right)f_{\mathbf{H}}\left(\mathbf{H}\right)d\mathbf{H}.
\nonumber\\
&&\le\int_{\|\Delta\mathbf{X}\mathbf{h}_j\|_2^2\le\tau^2}P\left(\mathbf{X}\rightarrow\hat{\mathbf{X}}|\mathbf{H}\right)f_{\mathbf{H}}\left(\mathbf{H}\right)d\mathbf{H}
\nonumber\\
&&+\int_{\sum_{j=1}^M\|\Delta\mathbf{X}\mathbf{h}_j\|_2^2>\tau^2}P\left(\mathbf{X}\rightarrow\hat{\mathbf{X}}|\mathbf{H}\right)f_{\mathbf{H}}\left(\mathbf{H}\right)d\mathbf{H}.
\end{eqnarray}
where the last inequality holds for the reason that if $\sum_{j=1}^M \|\Delta\mathbf{X}\mathbf{h}_j\|_2^2\le\tau^2$, then, we have $\|\Delta\mathbf{X}\mathbf{h}_j\|_2^2\le\tau^2$ for $j=1, 2, \cdots, M$.
 Three asymptotical bounds on $P(\mathbf{X}\rightarrow\hat{\mathbf{X}})$ in \eqref{eqn:ml_detection_pep_fragment} are given as follows.

\subsubsection{Upper-bound of PEP over $\left(0,\tau\right)$}
To begin with, we denote the first part of $P(\mathbf{X}\rightarrow\hat{\mathbf{X}})$ in \eqref{eqn:ml_detection_pep_fragment} by $P_{\tau}(\mathbf{X}\rightarrow\hat{\mathbf{X}})$.
  Since the PDF of log-normal is monotonically increasing over $\left(0,c_i\tau\right)$ when $c_i\tau\rightarrow0$ with increasing SNRs, it is true that $f_{H_{ij}}\left(h_{ij}\right)\le f_{H_{ij}}\left(c_i\tau\right)$ over $\left(0,c_i\tau\right)$ with high SNRs. This observation allows us to  upper-bound $P_{\tau}(\mathbf{X}\rightarrow\hat{\mathbf{X}})$ by
\begin{eqnarray}\label{eqn:pep_upper_bound}
&&P_{\tau}(\mathbf{X}\rightarrow\hat{\mathbf{X}})\le \frac{1}{2} \underbrace{\prod_{j=1}^M\prod_{i=1}^Nf_{H_{ij}}\left(c_i\tau\right)}_{P_{\tau}^{\left(1\right)}(\mathbf{X}\rightarrow\hat{\mathbf{X}})}
\nonumber\\
&&\times\prod_{j=1}^M\underbrace{
\int_{\|\Delta\mathbf{X}\mathbf{h}_j\|_2^2\le\tau^2}
\exp\left(-\frac{\rho \mathbf{h}_j^T\Delta\mathbf{X}^T\Delta\mathbf{X}\mathbf{h}_j}{8N}\right)d\mathbf{h}_j
}_{P_{\tau}^{\left(2\right)}(\mathbf{X}\rightarrow\hat{\mathbf{X}})}
 \end{eqnarray}
where
\begin{eqnarray*}
&&P_{\tau}^{\left(1\right)}(\mathbf{X}\rightarrow\hat{\mathbf{X}})\nonumber\\
&&=\prod_{j=1}^{M}\prod_{i=1}^N\frac{1}{\sqrt{2\pi\sigma_{ij}^2}c_i\tau}\exp\left(-\frac{\left(\ln\left(c_i\tau\right)-\mu_{ij}\right)^2}{2\sigma_{ij}^2}\right)
\end{eqnarray*}
By using hyper-sphere coordinate transformation, we transform $\mathbf{h}_j\in\mathbb{R}_+^N$ into $\left[r,\phi_1,~\cdots,~\phi_{N-1}\right]^T$, where $r\ge0$ and $\phi_{i}\in\left[0,\frac{\pi}{2}\right]$. The corresponding Jacobian matrix determinant is given by $r^{N-1}\prod_{i=1}^{N-2}\sin^{N-1-i}\phi_{i}$. For notational simplicity, we denote $d\phi_1~\cdots~d\phi_{N-1}$, $\prod_{i=1}^{N-2}\sin^{N-1-i}\phi_{i}$ and the transformed $\mathbf{h}^T_j\Delta\mathbf{X}^T\Delta\mathbf{X}\mathbf{h}_j$, by $d\Upsilon$, $\Phi$ and $\Theta$, respectively.  we have
\begin{eqnarray*}
&&P_\tau^{\left(2\right)}(\mathbf{X}\rightarrow\hat{\mathbf{X}})=\int_{\Upsilon}\int_0^{\frac{\tau}{\Phi}}
\exp\left(-\frac{\rho r^2\Theta}{8N}\right)r^{N-1}\Phi dr d\Upsilon\nonumber\\
&&=\frac{1}{2}\left(\frac{\rho}{8N}\right)^{-\frac{N}{2}}\int_0^{\frac{\rho\tau^2}{8N}}\exp\left(-x\right)x^{\frac{N}{2}-1}dx\int_{\Upsilon}\Theta^{-\frac{N}{2}}\Phi d\Upsilon
\end{eqnarray*}
Then, the following two possibilities are considered
\begin{enumerate}
  \item [(i)] $N$ is even.
  When $N=2$,
  \begin{eqnarray*}
\int_0^{\frac{\rho\tau^2}{8N}}\exp\left(-x\right)x^{\frac{N}{2}-1}dx=1-e^{-\frac{\rho\tau^2}{8N}}
\end{eqnarray*}
For $N=2m,m\ge2$,
  \begin{eqnarray*}
&&\int_0^{\frac{\rho\tau^2}{8N}}\exp\left(-x\right)x^{\frac{N}{2}-1}dx\nonumber\\
&&=-\left(\frac{\rho\tau^2}{8N}\right)^{\frac{N}{2}-1}
\exp\left(-\frac{\rho\tau^2}{8N}\right)\nonumber\\
&&+\left(\frac{N}{2}-1\right)\int_0^{\frac{\rho\tau^2}{8N}}\exp\left(-x\right)x^{\frac{N}{2}-2}dx \nonumber\\
&&=
\left(\frac{N}{2}-1\right)!\left(1-\exp\left(-\frac{\rho\tau^2}{8N}\right)\sum_{i=0}^{\frac{N}{2}-1}\frac{1}{i!}\left(\frac{\rho\tau^2}{8N}\right)^i\right)
\nonumber\\
&&\le\left(\frac{N}{2}-1\right)!
\end{eqnarray*}
which leads us to the following
\begin{eqnarray*}
P_\tau^{\left(2\right)}(\mathbf{X}\rightarrow\hat{\mathbf{X}})
&\le&\frac{1}{2}\left(\frac{N}{2}-1\right)!\left(1-\sum_{i=0}^{\frac{N}{2}-1}\frac{\left(\frac{\rho\tau^2}{8N}\right)^i}{e^{-\frac{\rho\tau^2}{8N}}i!}\right)\nonumber\\
&\times&\left(\frac{\rho}{8N}\right)^{-\frac{N}{2}}\int_{\Upsilon}\Theta^{-\frac{N}{2}}\Phi d\Upsilon
\end{eqnarray*}
  \item [(ii)] $N$ is even.
  When $N=1$,
  \begin{eqnarray*}
 \int_0^{\frac{\rho\tau^2}{8N}}e^{-x}x^{-\frac{1}{2}}dx=2\sqrt{\pi} \left(1-Q\left(\sqrt[4]{\frac{\rho\tau^2}{N}}\right)\right)
  \end{eqnarray*}
  When $N=2m+1,m\ge1$,
  \begin{eqnarray*}
&&\int_0^{\frac{\rho\tau^2}{8N}}e^{-x}x^{\frac{N}{2}-1}dx
=-\left(\frac{\rho\tau^2}{8N}\right)^{\frac{N}{2}-1}e^{-\frac{\rho\tau^2}{8N}}\nonumber\\
&&+\left(\frac{N}{2}-1\right)\int_0^{\frac{\rho\tau^2}{8N}}e^{-x}x^{\frac{N}{2}-2}dx \nonumber\\
&&=
\prod_{i=0}^{\frac{N-3}{2}}\left(i+\frac{1}{2}\right)\left(2\sqrt{\pi} \left(1-Q\left(\sqrt[4]{\frac{\rho\tau^2}{N}}\right)\right)\right)\nonumber\\
&&-\prod_{i=0}^{\frac{N-3}{2}}\left(i+\frac{1}{2}\right)\left(\sum_{i=0}^{\frac{N-3}{2}}\frac{ e^{-\frac{\rho\tau^2}{8N}}}{\prod_{k=0}^{i}\left(k+\frac{1}{2}\right)}\left(\frac{\rho\tau^2}{8N}\right)^{i+\frac{1}{2}}\right)
\nonumber\\
&&\le
\sqrt{\pi}\prod_{i=0}^{\frac{N-3}{2}}\left(i+\frac{1}{2}\right)
\end{eqnarray*}
\end{enumerate}
 Then, the common bound of $\int_0^{\frac{\rho\tau^2}{8N}}e^{-x}x^{\frac{N}{2}-1}dx$ can be given by $\Gamma\left(\frac{N}{2}\right)$, where $\Gamma\left(x\right)$ is the Gamma Function.
 \begin{eqnarray}\label{eqn:sphere_intergral}
&&P_\tau^{\left(2\right)}(\mathbf{X}\rightarrow\hat{\mathbf{X}})\le\frac{\Gamma\left(\frac{N}{2}\right)}{2}\left(\frac{\rho}{8N}\right)^{-\frac{N}{2}}\int_{\Upsilon}\Theta^{-\frac{N}{2}}\Phi d\Upsilon\nonumber\\
&&=\frac{\Gamma\left(\frac{N}{2}\right)}{2}\left(\frac{\rho}{8N}\right)^{-\frac{N}{2}}\int_{\mathbf{z}\in\mathbb{R}_+^N,\|\mathbf{z}\|_2^2=1}\left(\mathbf{z}^T\Delta\mathbf{X}^T\Delta\mathbf{X}\mathbf{z}\right)^{-\frac{N}{2}} d\mathbf{z}\nonumber \\
\end{eqnarray}
 It is true that $\mathbf{z}^T\Delta\mathbf{X}^T\Delta\mathbf{X}\mathbf{z}$ for PSD $\Delta\mathbf{X}^T\Delta\mathbf{X}$ is a continuous function with respect to $\mathbf{z}$ over the closed bounded feasible domain $\mathbb{R}_+^N\cap\left\{\mathbf{z}:\|\mathbf{z}\|_2^2=1\right\}$. In addition, by Theorem~\ref{theorem:full_cover_uniquely}, the full cover property of  $\Delta\mathbf{X}^T\Delta\mathbf{X}$ assures $\forall \mathbf{z}\in\mathbb{R}_+^N,\mathbf{z}^T\Delta\mathbf{X}^T\Delta\mathbf{X}\mathbf{z}>0$. Then, over $\mathbb{R}_+^N\cap\left\{\mathbf{z}:\|\mathbf{z}\|_2^2=1\right\}$, $\mathbf{z}^T\Delta\mathbf{X}^T\Delta\mathbf{X}\mathbf{z}$ has the minimum positive value denoted by $C_{\min}$. Then,
 \begin{eqnarray*}
\int_{\mathbf{z}\in\mathbb{R}_+^N,\|\mathbf{z}\|_2^2=1}\left(\mathbf{z}^T\Delta\mathbf{X}^T\Delta\mathbf{X}\mathbf{z}\right)^{-\frac{N}{2}} d\mathbf{z}\le C_{\min}^{-\frac{N}{2}}\int_{\mathbf{z}\in\mathbb{R}_+^N,\|\mathbf{z}\|_2^2=1} d\mathbf{z}
 \end{eqnarray*}
 We know that the surface area of a unit hyper-sphere is  given by $\frac{2\pi^{\frac{N}{2}}}{\Gamma\left(\frac{n}{2}\right)}$. Then, over $\mathbb{R}_+^N\cap\left\{\mathbf{z}:\|\mathbf{z}\|_2^2=1\right\}$, we can have $\int_{\mathbf{z}\in\mathbb{R}_+^N,\|\mathbf{z}\|_2^2=1} d\mathbf{z}=\frac{2\pi^{\frac{N}{2}}}{2^N\Gamma\left(\frac{n}{2}\right)}$. Substituting this result into~\eqref{eqn:sphere_intergral} gives us what follows.
 \begin{eqnarray}\label{eqn:sphere_intergral_final}
P_\tau^{\left(2\right)}(\mathbf{X}\rightarrow\hat{\mathbf{X}})\le\left(\frac{C_{\min}\rho}{2\pi N}\right)^{-\frac{N}{2}}
\end{eqnarray}
Thus, we arrive at the following upper-bound of $P_{\tau}(\mathbf{X}\rightarrow\hat{\mathbf{X}})$
\begin{eqnarray}\label{eqn:upper_bound_0_tau}
&&P_{\tau}(\mathbf{X}\rightarrow\hat{\mathbf{X}})\le\frac{1}{2}\left(\frac{C_{\min}\rho}{2\pi N}\right)^{-\frac{MN}{2}}\nonumber\\
&&\times
\prod_{j=1}^{M}\prod_{i=1}^N\frac{1}{\sqrt{2\pi\sigma_{ij}^2}c_i\tau}
\exp\left(-\frac{\left(\ln\left(c_i\tau\right)-\mu_{ij}\right)^2}{2\sigma_{ij}^2}\right)
\end{eqnarray}

\subsubsection{Upper-bound of PEP over $\left(\tau,\infty\right)$}
Now, we are in a position to analyze $P(\mathbf{X}\rightarrow\hat{\mathbf{X}})-P_{\tau}(\mathbf{X}\rightarrow\hat{\mathbf{X}})$, i.e., the second term of \eqref{eqn:ml_detection_pep_fragment}, which is denoted by $\bar{P}_\tau(\mathbf{X}\rightarrow\hat{\mathbf{X}})$. For $h_{ij}$, $f'_{H_{ij}}\left(h_{ij}\right)=0$ gives the extreme point $h_{{ij},0}=\exp\left(-\sigma_{ij}^2+\mu_{ij}\right)$  of $f_{H_{ij}}\left(h_{ij}\right)$, i.e.,
\begin{eqnarray*}
f_{H_{ij}}\left(h_{ij}\right)&\le& f_{H_{ij}}\left(h_{ij,0}\right)\nonumber\\
&=&\frac{1}{\sqrt{2\pi \sigma_{ij}^2}}\exp\left(-\frac{\sigma_{ij}^2}{2}\right)
\end{eqnarray*}
where $i=1,\ldots, N, j=1,\ldots,M$. If we let $\mathbf{H}_0=\left[h_{ij,0}\right]_{N\times M}$, then, we have $f_{\mathbf{H}}\left(\mathbf{H}\right)\le f_{\mathbf{H}}\left(\mathbf{H}_0\right)$ and therefore,
  $\bar{P}_\tau(\mathbf{X}\rightarrow\hat{\mathbf{X}})$ can be upper-bounded  by
    \begin{eqnarray*}
&&\bar{P}_\tau(\mathbf{X}\rightarrow\hat{\mathbf{X}})=
\int_{\sum_{j=1}^M\|\Delta\mathbf{X}\mathbf{h}_j\|_2^2>\tau^2}P\left(\mathbf{X}\rightarrow\hat{\mathbf{X}}|\mathbf{H}\right)f_{\mathbf{H}}\left(\mathbf{H}\right)d\mathbf{H} \nonumber \\
&&\le\frac{f_{\mathbf{H}}\left(\mathbf{H}_0\right)}{2}
\nonumber \\
&&\times \underbrace{\int_{\sum_{j=1}^M\|\Delta\mathbf{X}\mathbf{h}_j\|_2^2>\tau^2}\exp\left(-\frac{\rho\mathbf{h}_j^T\Delta\mathbf{X}^T\Delta\mathbf{X}\mathbf{h}_j}
{8N}\right)d\mathbf{H}}_{\bar{P}_\tau^{\left(Q\right)}(\mathbf{X}\rightarrow\hat{\mathbf{X}})}
\end{eqnarray*}
$\sum_{j=1}^M\|\Delta\mathbf{X}\mathbf{h}_j\|_2^2$ can be rewritten into $\bar{h}^T\mathbf{P}_x\bar{h}$, where $\mathbf{P}_x=\textrm{diag}\left(\Delta\mathbf{X}^T\Delta\mathbf{X},\cdots,\Delta\mathbf{X}^T\Delta\mathbf{X}\right)_{MN\times MN}$ and $\bar{h}=\left[h_{11},\cdots,h_{N1},\cdots,h_{M1},\cdots,h_{MN}\right]_{MN\times1}^T$.
By using hyper-sphere coordinate transformation, we arrive at what follows.
\begin{eqnarray*}
&&\bar{P}_\tau^{\left(Q\right)}(\mathbf{X}\rightarrow\hat{\mathbf{X}})= \int_{\Upsilon}\int_{r>\frac{\tau}{\Theta}}^{\infty}e^{-\frac{\rho r^2\Theta}{8N}}r^{MN-1}\Phi dr d\Upsilon\nonumber\\
&&=\left(\frac{\rho}{8N}\right)^{-\frac{MN}{2}}\int_{\frac{\rho\tau^2}{8N}}^{\infty}e^{-x}x^{\frac{MN}{2}-1}dx\int_{\Upsilon}\Theta^{-\frac{MN}{2}}\Phi d\Upsilon
\end{eqnarray*}
We know that
\begin{eqnarray*}
\int_0^{\frac{\rho\tau^2}{8N}}e^{-x}x^{\frac{MN}{2}-1}dx+\int_{\frac{\rho\tau^2}{8N}}^{\infty}e^{-x}x^{\frac{MN}{2}-1}dx=1
\end{eqnarray*}
From the results for $\int_0^{\frac{\rho\tau^2}{8N}}e^{-x}x^{\frac{MN}{2}-1}dx$, we can have the following equality.
\begin{eqnarray*}
&&\int_{\frac{\rho\tau^2}{8N}}^{\infty}\frac{x^{\frac{MN}{2}-1}dx}{e^{x}}=\nonumber\\
&&
\left\{\begin{array}{lll}
1-2\sqrt{\pi} \left(1-Q\left(\sqrt[4]{\frac{\rho\tau^2}{N}}\right)\right),~\hfill MN=1&&\\
\left(\frac{MN}{2}-1\right)!e^{-\frac{\rho\tau^2}{8N}}\sum_{i=0}^{\frac{MN}{2}-1}\frac{1}{i!}\left(\frac{\rho\tau^2}{8N}\right)^i+1-\left(\frac{N}{2}-1\right)!,&&\\
~\hfill MN=2m,m\in\mathbb{Z}^+&&\\
1+\frac{\left(\frac{\sum_{i=0}^{\frac{MN-3}{2}}\frac{\left(\frac{\rho\tau^2}{8N}\right)^{i+\frac{1}{2}}}{\prod_{k=0}^{i}\left(k+\frac{1}{2}\right)}}{e^{\frac{\rho\tau^2}{8N}}}-2\sqrt{\pi}Q\left(\sqrt[4]{\frac{\rho\tau^2}{N}}\right)\right)}{\prod_{i=0}^{\frac{MN-3}{2}}\left(i+\frac{1}{2}\right)^{-1}},&&\\ ~\hfill MN=2m+1,m\in\mathbb{Z}^+&&
\end{array}
\right.
\end{eqnarray*}
From our assumption that $\rho\tau^2\rightarrow\infty$ when $\rho\rightarrow\infty$, we know that among $\frac{1}{i!}\left(\frac{\rho\tau^2}{8N}\right)^i$, the dominant term is $\left(\frac{\rho\tau^2}{8N}\right)^{\frac{MN}{2}-1}$ for $MN>1$. Note that for $MN=1$, the dominant exponential term is $\left(\frac{\rho\tau^2}{8N}\right)^{\frac{1}{2}}$. Then, we can have a common bound of $\int_{\frac{\rho\tau^2}{8N}}^{\infty}\frac{x^{\frac{MN}{2}-1}dx}{e^{x}}$ as follows
\begin{eqnarray*}
\int_{\frac{\rho\tau^2}{8N}}^{\infty}\frac{x^{\frac{MN}{2}-1}dx}{e^{x}}\le
\Gamma\left(\frac{MN}{2}\right)\left(\frac{\rho\tau^2}{8N}\right)^{\frac{MN}{2}}e^{-\frac{\rho\tau^2}{8N}}
\end{eqnarray*}
Using the techniques for~\eqref{eqn:sphere_intergral}, we arrive at the following inequality
\begin{eqnarray*}
\bar{P}_\tau^{\left(Q\right)}(\mathbf{X}\rightarrow\hat{\mathbf{X}})&\le&\left(\frac{C_{\min}\rho}{2\pi N}\right)^{-\frac{N}{2}}\left(\frac{\rho\tau^2}{8N}\right)^{\frac{N}{2}}e^{-\frac{\rho\tau^2}{8N}}
\end{eqnarray*}
Then, we can upper-bound $\bar{P}_\tau(\mathbf{X}\rightarrow\hat{\mathbf{X}})$ by
\begin{eqnarray}\label{eqn:upper_bound_tau_infty}
\bar{P}_\tau(\mathbf{X}\rightarrow\hat{\mathbf{X}})&\le&\frac{1}{2}\prod_{j=1}^{M}\prod_{i=1}^N\frac{e^{-\frac{\sigma_{ij}^2}{2}}}{\sqrt{2\pi\sigma_{ij}^2}}
\left(\frac{C_{\min}\rho}{2\pi N}\right)^{-\frac{N}{2}}
\nonumber\\
&\times&\left(\frac{\rho\tau^2}{8N}\right)^{\frac{MN}{2}}\exp\left(-\frac{\rho\tau^2}{8N}\right)
\end{eqnarray}

\subsubsection{Determination of $\tau$ and $\nu$}
Combining~\eqref{eqn:upper_bound_0_tau},~\eqref{eqn:upper_bound_tau_infty} and~\eqref{eqn:lower_bound_over} gives us the following inequalities
\begin{eqnarray*}
&&Q\left(\sqrt{\frac{M}{4N}}\right)\prod_{i=1}^{N}\prod_{j=1}^M
Q\left(\frac{\ln\rho+\ln\lambda_{\max}+2\mu_{ij}}{2\sigma_{ij}}\right)
\nonumber\\
&&\le P(\mathbf{X}\rightarrow\hat{\mathbf{X}})\le \frac{1}{2}\prod_{j=1}^{M}\prod_{i=1}^N\frac{1}{\sigma_{ij}c_i}\left(\frac{C_{\min}}{ N}\right)^{-\frac{MN}{2}}
\nonumber\\
&&\times
\left(\rho\tau^2\right)^{-\frac{MN}{2}}
\exp\left(-\sum_{j=1}^{M}\sum_{i=1}^N\frac{\left(\ln\left(c_i\tau\right)-\mu_{ij}\right)^2}{2\sigma_{ij}^2}\right) \nonumber\\ &&+\frac{1}{2}\prod_{j=1}^{M}\prod_{i=1}^N\frac{\exp\left(-\frac{\sigma_{ij}^2}{2}\right)}{\sigma_{ij}}
\left({C_{\min}}/{ N}\right)^{-{MN}/{2}}
\nonumber\\
&&\times\left(\frac{\tau^2}{8N}\right)^{{MN}/{2}}\exp\left(-\frac{\rho\tau^2}{8N}\right)
\end{eqnarray*}
Since $Q\left(x\right)\ge\frac{1}{\sqrt{2\pi}}\left(\frac{1}{x}-\frac{1}{x^3}\right)\exp\left(-\frac{x^2}{2}\right)$ for $x\ge\frac{\sqrt{2}}{2}$, we can lower-bound $P_{L}(\mathbf{X}\rightarrow\hat{\mathbf{X}})$ by
\begin{small}
\begin{eqnarray*}
&&P_{L}(\mathbf{X}\rightarrow\hat{\mathbf{X}})\!\ge\!\! \frac{Q\left(\sqrt{\frac{M}{4N}}\right)}{\sqrt{2\pi}}\!\!\prod_{j=1}^{M}\prod_{i=1}^{N}\!\!\frac{1-\left(\frac{\ln\rho +2\ln \lambda_{\max}+2\mu_{ij}}{2\sigma_{ij}}\right)^{-2}}{\frac{\ln\rho +2\ln \lambda_{\max}+2\mu_{ij}}{2\sigma_{ij}}}
\nonumber\\
&&\times\exp\left(-\sum_{j=1}^{M}\sum_{i=1}^{N}\frac{\left(\ln\rho +2\ln \lambda_{\max}+2\mu_{ij}\right)^2}{8\sigma_{ij}^2}\right)
\end{eqnarray*}
\end{small}
We let
\begin{eqnarray*}
\tau=\sqrt{\frac{N\ln^2\rho\sum_{j=1}^M\sum_{i=1}^N\sigma_{ij}^{-2}}{\rho}}
\end{eqnarray*}
and denote $\Omega=\sum_{j=1}^M\sum_{i=1}^N\sigma_{ij}^{-2}$, $\Omega_i=\sum_{j=1}^M\sigma_{ij}^{-2}$, $\tilde{\Omega}=\sum_{j=1}^M\sum_{i=1}^N\mu_{ij}\sigma_{ij}^{-2}$, and $\tilde{\Omega}_i=\sum_{j=1}^M\mu_{ij}\sigma_{ij}^{-2}$. Then, we arrive at~\eqref{eqn:general_stbc_design_MIMO}.  This completes the proof of Theorem~\ref{theorem:mimo_ovlc_pep}. ~\hfill $\Box$

\subsection{Proof of Property~\ref{theorem:degenerate_code}}\label{app:mimo_degenerate}
Denote
\begin{eqnarray*}
\tilde{h}_i=\min\left(\arg\max_{h_{i1}}f_{H_{i1}}\left(h_{i1}\right),\arg\max_{h_{i2}}f_{H_{i2}}\left(h_{i2}\right)\right)
\end{eqnarray*}
for $i=1,2$. Then, over $\left(0,\tilde{h}_i\right)$, $f_{H_{i1}}\left(h_{i1}\right)$ and $f_{H_{i2}}\left(h_{i2}\right)$ are monotonically increasing with respect to $h_{i1}$ and $h_{i2}$, respectively. We suppose that $\tau$ is a positive and monotonically decreasing function with respect to $\rho$. When $\rho$ goes to infinity, $\tau$ approaches zero.
Then, in high SNR regimes, $0\le \frac{1}{2}\tilde{h}_i-\tau\le h_2\le h_{i1}\le h_{i2}+\tau\le\tilde{h}_i$ holds. Thus, in this situation,  we have
\begin{eqnarray*}
&&P\left(\mathbf{s}\rightarrow\mathbf{\hat{s}}\right)
=\frac{1}{\pi}\int_{0}^{\frac{\pi}{2}}\int \frac{\prod_{i,j=1}^2f_{H_{ij}}\left(h_{ij}\right)dh_{ij}d\theta}{e^{\frac{\rho\sum_{i=1}^2\left(h_{i1}-h_{i2}\right)^2} {8\sin^2\theta}}} \nonumber\\
&&\ge
\frac{1}{\pi}\int_{0}^{\frac{\pi}{2}}\int_{\frac{1}{2}\tilde{h}_i-\tau\le h_2\le h_{i1}\le h_{i2}+\tau\le\tilde{h}_i}\frac{\prod_{i,j=1}^2f_{H_{ij}}\left(h_{ij}\right)dh_{ij}d\theta}{e^{\frac{\rho\sum_{i=1}^2\left(h_{i1}-h_{i2}\right)^2} {8\sin^2\theta}}}
\end{eqnarray*}
where the inequality holds for the fact that
From $\frac{1}{2}\tilde{h}_i-\tau\le h_2\le h_{i1}\le h_{i2}+\tau\le\tilde{h}_i$, we have $\left(h_{i1}-h_{i2}\right)^2\le \tau^2$. The monotonically decreasing property of $e^{-x}$ with respect positive $x$ allows us to attain
\begin{eqnarray*}
&&P\left(\mathbf{s}\rightarrow\mathbf{\hat{s}}\right)\ge
\frac{1}{\pi}\int_{0}^{\frac{\pi}{2}}e^{-\frac{2\rho\tau^2} {8\sin^2\theta}}d\theta \nonumber\\
&&\times\prod_{i=1}^2\int_{\frac{1}{2}\tilde{h}_i}^{\tilde{h}_i-\tau}f_{H_{i2}}\left(h_{i2}\right)\int_{h_{i2}} ^{h_{i2}+\tau}f_{H_{i1}}\left(h_{i1}\right)dh_{i1}dh_{i2} \nonumber\\
&&\ge
\frac{1}{\pi}\int_{0}^{\frac{\pi}{2}}e^{-\frac{\rho\tau^2} {4\sin^2\theta}}d\theta \prod_{i=1}^2\int_{\frac{1}{2}\tilde{h}_i}^{\tilde{h}_i-\tau}f_{H_{i2}}\left(h_{i2}\right)\tau f_{H_{i1}}\left(h_{i2}\right)dh_{i1} \nonumber\\
&&\ge
\tau^2\prod_{i=1}^2\left(\frac{1}{2}\tilde{h}_i-\tau\right)\prod_{j=1}^2 f_{H_{ij}}\left(\frac{1}{2}\tilde{h}_i\right) \frac{1}{\pi}\int_{0}^{\frac{\pi}{2}}e^{-\frac{\rho\tau^2} {4\sin^2\theta}}d\theta
\end{eqnarray*}
Let $\tau=\frac{1}{\sqrt{\rho}}$. Then,
\begin{eqnarray*}
P\left(\mathbf{s}\rightarrow\mathbf{\hat{s}}\right)
&\ge&Q\left(\frac{1} {\sqrt{2}}\right)\rho^{-1}\prod_{i=1}^2\left(\frac{1}{2}\tilde{h}_i-\rho^{-\frac{1}{2}}\right)
\nonumber\\
&\times&\prod_{i=1}^2\prod_{j=1}^2f_{H_{ij}}\left(\frac{1}{2}\tilde{h}_i\right)\nonumber\\
&=&C_0\rho^{-1}+\mathcal{O}\left(\rho^{-1}\right)
\end{eqnarray*}
where
\begin{eqnarray*}
C_0=\frac{Q\left({1} /{\sqrt{2}}\right)}{4}\prod_{i=1}^2\tilde{h}_i\times\prod_{i=1}^2\prod_{j=1}^2f_{H_{ij}}\left(\frac{1}{2}\tilde{h}_i\right)
\end{eqnarray*}
is independent of $\rho$, giving us the desired. This completes the proof of Property~\ref{theorem:degenerate_code}.
~\hfill $\Box$

\subsection{Proof of Theorem~\ref{theorem:worst_case_pep}}\label{app:optimal_linear_stbc}
For presentation convenience,  let us denote the minimum distance of the constellation $\mathcal{P}\subseteq\mathbb{R}_+$ by $d_{\min}\left(\mathcal{P}\right)=\min_{p\neq\tilde{p},p,\tilde{p}\in\mathcal{P}}|p-\tilde{p}|$. Furthermore, let  $\frac{1}{2^K}\sum_{p\in\mathcal{P}}p$ be denoted by $P_{\mathcal{P}}$. Then,
for any $\mathbf{p}\neq \tilde{\mathbf{p}}$, let $\mathbf{X}\left(\mathbf{p}-\tilde{\mathbf{p}}\right)$ be denoted by $\mathbf{X}\left(\mathbf{e}\right)$. Theorem~\ref{theorem:rank_one_cover_length} tells us that the $i$-th cover length, say, $c_i$, of full-cover $\mathbf{X}^T\left(\mathbf{e}\right)\mathbf{X}\left(\mathbf{e}\right)$ is lower-bounded by
\begin{eqnarray*}
c_i\ge\frac{1}{\sqrt{\left[\mathbf{X}^T\left(\mathbf{e}\right)\mathbf{X}\left(\mathbf{e}\right)\right]_{ii}}}
\end{eqnarray*}
where $\left[\sum_{l=1}^L\mathbf{A}_l\left(|e_l|\right)^T\sum_{l=1}^L\mathbf{A}_l\left(|e_l|\right)\right]_{ii}$ is the $i$-th diagonal entry of $\sum_{l=1}^L\mathbf{A}_l\left(|e_l|\right)^T\sum_{l=1}^L\mathbf{A}_l\left(|e_l|\right)$.
Since $\mathbf{X}^T\left(\mathbf{e}\right)\mathbf{X}\left(\mathbf{e}\right)$ has full-cover, $\left[\mathbf{X}^T\left(\mathbf{e}\right)\mathbf{X}\left(\mathbf{e}\right)\right]_{ii}$ has non-zero by Theorem~\ref{theorem:full_cover_algebraic}.  Then, $\max_{\mathbf{p}\neq\tilde{\mathbf{p}}}\prod_{i=1}^N c_i^{\Omega_i}$ is lower-bounded by
\begin{eqnarray*}
\max_{\mathbf{p}\neq\tilde{\mathbf{p}}}\prod_{i=1}^N c_i^{\Omega_i}\ge\max_{\mathbf{p}\neq\tilde{\mathbf{p}}}\prod_{i=1}^N\left(\left[\mathbf{X}^T\left(\mathbf{e}\right)\mathbf{X}\left(\mathbf{e}\right)\right]_{ii}\right)^{-\Omega_i/2}
\end{eqnarray*}
In addition, we have
\begin{eqnarray*}
&&\max_{\mathbf{p}\neq\tilde{\mathbf{p}}}\prod_{i=1}^N\left(\left[\mathbf{X}^T\left(\mathbf{e}\right)\mathbf{X}\left(\mathbf{e}\right)\right]_{ii}\right)^{-\Omega_i/2}
\nonumber\\&&\ge
\max_{e_\ell\neq0,e_\ell'=0,\ell'\neq \ell}\prod_{i=1}^N\left(\left[\mathbf{X}^T\left(\mathbf{e}\right)\mathbf{X}\left(\mathbf{e}\right)\right]_{ii}\right)^{-\Omega_i/2}
\nonumber\\&&\ge
\max_{e_\ell^2=d_{\min}^2\left(\mathcal{P}\right)}\prod_{i=1}^N\left(d_{\min}^2\left(\mathcal{P}\right)\left[\mathbf{A}_\ell^T\mathbf{A}_\ell\right]_{ii}\right)^{-\Omega_i/2}
\nonumber\\
&&\ge
\sqrt[L]{\prod_{\ell=1}^L\prod_{i=1}^N\left(d_{\min}^2\left(\mathcal{P}\right)\left[\mathbf{A}_\ell^T\mathbf{A}_\ell\right]_{ii}\right)^{-\Omega_i/2}}
\end{eqnarray*}
For notational simplicity, let notation $a_{ki}^{\left(\ell\right)}$ denote the entry of $\mathbf{A}_\ell$ on the $k$-th row and the $i$-th column. Then, by Lemma~\ref{lemma:product_power}, we have
\begin{eqnarray}\label{eqn:product_power}
&&\prod_{\ell=1}^L\prod_{i=1}^N\left(\left[\mathbf{A}_\ell^T\mathbf{A}_\ell\right]_{ii}\right)^{\Omega_i}\nonumber\\
&=&\prod_{\ell=1}^L\prod_{i=1}^N\left(\sum_{k=1}^L \left(a_{ki}^{\left(\ell\right)}\right)^2\right)^{\Omega_i}\nonumber\\
&\le&\prod_{\ell=1}^L\prod_{i=1}^N\left(\sum_{k=1}^L a_{ki}^{\left(\ell\right)}\right)^{2\Omega_i}\nonumber\\
&\le&\prod_{\ell=1}^L\prod_{i=1}^N\Omega_i^{2\Omega_i}\left(\frac{\sum_{i=1}^N\sum_{k=1}^L a_{ki}^{\left(\ell\right)}}{\Omega}\right)^{2\Omega}
\nonumber\\
&=&\prod_{i=1}^N\Omega_i^{2L\Omega_i}\left(\frac{\prod_{\ell=1}^L\mathbf{1}^T\mathbf{A}_\ell\mathbf{1}}{\Omega^L}\right)^{2\Omega}
\end{eqnarray}
Further, using the arithmetic-mean geometric-mean inequality leads us to the following.
\begin{eqnarray*}
\prod_{\ell=1}^L\mathbf{1}^T\mathbf{A}_\ell\mathbf{1}
&\le&\left(\frac{\sum_{\ell=1}^L \mathbf{1}^T\mathbf{A}_\ell\mathbf{1}}{L}\right)^{L}\nonumber\\
&=&\left(\frac{L}{LP_{\mathcal{P}}}\right)^{L}=\frac{1}{P_{\mathcal{P}}^L}
\end{eqnarray*}
Then, substituting the above inequality into~\eqref{eqn:product_power} allows us to
attain the lower-bound of $\max_{\mathbf{p}\neq\tilde{\mathbf{p}}}\prod_{i=1}^N c_i^{\Omega_i}$ as follows.
\begin{eqnarray}\label{eqn:lower_cover_volume}
\max_{\mathbf{p}\neq\tilde{\mathbf{p}}}\prod_{i=1}^N c_i^{\Omega_i}\ge\prod_{i=1}^N\Omega_i^{-\Omega_i}\left(\frac{d_{\min}\left(\mathcal{P}\right)}{P_{\mathcal{P}}\Omega}\right)^{-\Omega}
\end{eqnarray}
Now, we solve $\min_{d_{\min}\left(\mathcal{P}\right)=1} P_{\mathcal{P}}$ to maximize $\frac{d_{\min}\left(\mathcal{P}\right)}{P_{\mathcal{P}}}$. Without loss of generality, we assume that all the $2^K$ elements of $\mathcal{P}$ satisfy $0\le p_0<p_1~\cdots~<p_{2^K-1}$. Since $d_{\min}\left(\mathcal{P}\right)=1$, we can have $p_{i+1}-p_i\ge1$ for any $0\le i\le  2^{K}-2$. Then, $p_i\ge i+ip_0$ and thus,
\begin{eqnarray*}
\sum_{i=0}^{2^K-1}p_i&\ge& \sum_{i=1}^{2^K-1}i+p_0+p_0\sum_{i=1}^{2^K-1}i\nonumber\\
&\ge&\frac{2^{K}(2^K-1)}{2}
\end{eqnarray*}
Then, $\frac{d_{\min}\left(\mathcal{P}\right)}{P_{\mathcal{P}}}\le \frac{1}{\frac{2^{K}(2^K-1)}{2^{K+1}}}= \frac{2}{2^K-1}$, where the equality holds if and only if $\mathcal{P}=\{0,1,~\cdots,~2^K-1\}$. Therefore,~\eqref{eqn:lower_cover_volume} can be further lower-bounded by
\begin{eqnarray}\label{eqn:lower_bound_loss}
\max_{\mathbf{p}\neq\tilde{\mathbf{p}}}\prod_{i=1}^N c_i^{\Omega_i}&\ge&\prod_{i=1}^N\Omega_i^{-\Omega_i}\left(\frac{2}{\Omega\left(2^K-1\right)}\right)^{-\Omega}
\nonumber\\
&=&\left(\frac{2^K-1}{2}\right)^{\Omega}\prod_{i=1}^N\left(\frac{\Omega}{\Omega_i}\right)^{\Omega_i}
\end{eqnarray}

In the following, let us consider a specific linear STBC given by~\eqref{eqn:linear_optimal}. Then, $\mathbf{X}^T\left(\mathbf{e}\right)\mathbf{X}(\mathbf{e})$ is given by
\begin{eqnarray*}
  \mathbf{X}^T\left(\mathbf{e}\right)\mathbf{X}\left(\mathbf{e}\right)=\frac{4\sum_{i=1}^Le_i^2}{\Omega^2\left(2^K-1\right)^2}
  \left(\begin{array}{llll}
  \Omega_1^2& \ldots&\Omega_1\Omega_N\\
    \vdots&\ddots&\vdots\\
   \Omega_1\Omega_N&\ldots&\Omega_N^2
  \end{array}
  \right)_{N\times N}
    \end{eqnarray*}
Notice that for any nonzero $\sum_{i=1}^Le_i^2$, all the entries of $\mathbf{X}^T\left(\mathbf{e}\right)\mathbf{X}\left(\mathbf{e}\right)$ is positive and thus, full-cover by Statement 1) of Theorem~\ref{theorem:full_cover_algebraic}. Furthermore, by Theorem~\ref{theorem:rank_one_cover_length}, the $i$-th cover length $c_i$ of this positive matrix $\mathbf{X}^T\left(\mathbf{e}\right)\mathbf{X}\left(\mathbf{e}\right)$ can be determined by
\begin{eqnarray*}
c_i=\frac{1}{\sqrt{\left[\mathbf{X}^T\left(\mathbf{e}\right)\mathbf{X}\left(\mathbf{e}\right)\right]_{ii}}}=\frac{\Omega\left(2^K-1\right)}{2\Omega_i\sqrt{\sum_{i=1}^Le_i^2}}
\end{eqnarray*}
for $i=1,2,~\cdots,~N$. Then, by computations, we attain
\begin{eqnarray*}
\max_{\mathbf{p}\neq\tilde{\mathbf{p}}}\prod_{i=1}^N c_i^{\Omega_i}&=&\max_{\mathbf{p}\neq\tilde{\mathbf{p}}}\prod_{i=1}^N\Omega_i^{-\Omega_i}\left(\frac{2\sqrt{\sum_{i=1}^Le_i^2}}{\Omega\left(2^K-1\right)}\right)^{-\Omega}\nonumber\\
&=&\prod_{i=1}^N\Omega_i^{-\Omega_i}\left(\frac{2\sqrt{\min_{\mathbf{p}\neq\tilde{\mathbf{p}}}\sum_{i=1}^Le_i^2}}{\Omega\left(2^K-1\right)}\right)^{-\Omega}\nonumber\\
&=&\prod_{i=1}^N\Omega_i^{-\Omega_i}\left(\frac{2}{\Omega\left(2^K-1\right)}\right)^{-\Omega}\nonumber\\
&=&\left(\frac{2^K-1}{2}\right)^{\Omega}\prod_{i=1}^N\left(\frac{\Omega}{\Omega_i}\right)^{\Omega_i}
\end{eqnarray*}
where the last but one equality is assured by the fact that $\min_{\mathbf{p}\neq\tilde{\mathbf{p}}}\sqrt{\sum_{i=1}^Le_i^2}= d_{\min}\left(\mathcal{P}\right) =1$ holds if $\mathcal{P}=\{0,1,~\cdots,~2^K-1\}$. Thus, the linear STBC given by~\eqref{eqn:linear_optimal} indeed achieves the lower-bound in~\eqref{eqn:lower_bound_loss}, and hence, optimal. This concludes the proof of Theorem~\ref{theorem:worst_case_pep}.~\hfill $\Box$
\subsection{Proof of Theorem~\ref{theorem:optimal_structure} }\label{app:optimal_coding_structure}

The optimal solution to  Problem~\ref{prob:optimal_non_linear} is attained by  successively solving the following two subproblems.
\begin{subproblem}\label{subprob:optimal_structure}
 Find a structure for the matrix elements of $\mathcal{X}\subseteq\mathbb{R}_+^{L\times N} $ such that
$\max_{\mathbf{X}\neq\tilde{\mathbf{X}},\mathbf{X},\tilde{\mathbf{X}}\in\mathcal{X} }\prod_{i=1}^N c_i^{\Omega_i}$ is minimized subject to $\tilde{d}_{\min}(\mathcal{X})=1$.~\hfill\QED
\end{subproblem}

\begin{subproblem}\label{subprob:optimal_non_linear_stbc}
Design a constellation $\mathcal{X}\subseteq\mathbb{R}_+^{L\times N} $ with the  structure of its matrix elements being the solution to Subproblem~\ref{subprob:optimal_structure}  in order to
minimize $\sum_{\mathbf{X}\in\mathcal{X} }\mathbf{1}^T\mathbf{X}\mathbf{1}$ under constraint that $\tilde{d}_{\min}(\mathcal{X})=1$.~\hfill\QED
\end{subproblem}

\underline{\textbf{Solution to Subproblem~\ref{subprob:optimal_structure})}}: By Theorem~\ref{theorem:rank_one_cover_length},  the $i$-th cover length, $c_i$, of $\Delta\mathbf{X}^T\Delta\mathbf{X}$ is lower-bounded
\begin{eqnarray*}
c_i\ge\frac{1}{\sqrt{\left[\Delta\mathbf{X}^T\Delta\mathbf{X}\right]_{ii}}}=\frac{1}{\|\mathbf{x}_{i}-\tilde{\mathbf{x}}_i\|_2}
\end{eqnarray*}
where $\left[\Delta\mathbf{X}^T\Delta\mathbf{X}\right]_{ii}$ is non-zero assured by full cover condition. This inequality allows us to lower-bound $\max_{\mathbf{X}\neq\tilde{\mathbf{X}}}\prod_{i=1}^N c_i^{\Omega_i}$ by
\begin{eqnarray*}
\max_{\mathbf{X}\neq\tilde{\mathbf{X}}}\prod_{i=1}^N c_i^{\Omega_i}\ge\max_{\mathbf{X}\neq\tilde{\mathbf{X}}}\prod_{i=1}^N\left(\|\mathbf{x}_{i}-\tilde{\mathbf{x}}_i\|_2\right)^{-\Omega_i}
\end{eqnarray*}
Notice that there exist two elements $\mathbf{X}^{(0)}$ and $\tilde{\mathbf{X}}^{(0)}$ in $\mathcal{X}$ such that $\sum_{i=1}^N\|\mathbf{x}_{i}^{(0)}-\tilde{\mathbf{x}}_{i}^{(0)}\|_{2}=\tilde{d}_{\min}(\mathcal{X})$, where $\mathbf{x}_{i}^{(0)}$ is the $i$-th column of $\mathbf{X}^{(0)}$. It follows that
\begin{eqnarray*}
\max_{\mathbf{X}\neq\tilde{\mathbf{X}}}\prod_{i=1}^N\left(\|\mathbf{x}_{i}-\tilde{\mathbf{x}}_i\|_2\right)^{-\Omega_i} \ge\prod_{i=1}^N\left(\|\mathbf{x}_{i}^{(0)}-\tilde{\mathbf{x}}_{i}^{(0)}\|_{2}\right)^{-\Omega_i}
\end{eqnarray*}
In addition, using Lemma~\ref{lemma:product_power} leads us to
\begin{eqnarray*}
&&\prod_{i=1}^N\left(\|\mathbf{x}_{i}^{(0)}-\tilde{\mathbf{x}}_{i}^{(0)}\|_{2}\right)^{\Omega_i}
\nonumber \\
&&\le\prod_{i=1}^N\Omega_i^{\Omega_i}\left(\frac{\sum_{i=1}^N\|\mathbf{x}_{i}^{(0)}-\tilde{\mathbf{x}}_{i}^{(0)}\|_{2}}{\Omega}\right)^{\Omega}
\nonumber\\
&&=\prod_{i=1}^N\Omega_i^{\Omega_i}\left(\frac{d_{\min}(\mathcal{X})}{\Omega}\right)^{\Omega}\nonumber\\
&&=\frac{1}{\Omega^{\Omega}}\prod_{i=1}^N\Omega_i^{\Omega_i}
\end{eqnarray*}
Then, we attain the lower-bound of $\max_{\mathbf{X}\neq\tilde{\mathbf{X}}}\prod_{i=1}^N c_i^{\Omega_i}$ by
\begin{eqnarray*}
\max_{\mathbf{X}\neq\tilde{\mathbf{X}}}\prod_{i=1}^N c_i^{\Omega_i}
&\ge&\Omega^{\Omega}\prod_{i=1}^N\Omega_i^{-\Omega_i}\nonumber\\
&=&\prod_{i=1}^N\left(\frac{\Omega}{\Omega_i}\right)^{\Omega_i}
\end{eqnarray*}

Let us consider a specific constellation $\breve{\mathcal{X}}$ satisfying $d_{\min}\left(\breve{\mathcal{X}}\right)=1$ and its element is given in the following form
\begin{eqnarray}\label{eqn:structure_space}
\breve{\mathbf{X}}=\left(
\begin{array}{lllll}
\Omega_1s_1&\Omega_2s_1&\cdots&\Omega_N s_1\\
\Omega_1s_2&\Omega_2s_2&\cdots&\Omega_Ns_2\\
\vdots&\vdots&\ddots&\vdots\\
\Omega_1s_L&\Omega_2s_L&\cdots&\Omega_Ns_L\\
\end{array}
\right)_{L\times N}
\end{eqnarray}
Then, the coding matrix is given by what follows.
\begin{eqnarray*}
  \Delta\breve{\mathbf{X}}^T\Delta\breve{\mathbf{X}}=\sum_{i=1}^Le_i^2
  \left(\begin{array}{llll}
  \Omega_1^2& \Omega_1\Omega_2&\ldots&\Omega_1\Omega_N\\
  \Omega_1\Omega_2& \Omega_2^2&\ldots&\Omega_2\Omega_N\\
  \vdots&\vdots&\ddots&\vdots\\
   \Omega_1\Omega_N&\Omega_2\Omega_N&\ldots&\Omega_N^2
  \end{array}
  \right)_{N\times N}
    \end{eqnarray*}
 Notice that all the entries of $\Delta\breve{\mathbf{X}}^T\Delta\breve{\mathbf{X}}$ are positive for any nonzero $\sum_{i=1}^Le_i^2$. By Theorem~\ref{theorem:rank_one_cover_length}, the cover length of this coding matrix is attained by
$\breve{c}_i=\frac{1}{\Omega_i\sqrt{\sum_{i=1}^Le_i^2}}$. Thus,
    \begin{eqnarray}\label{eqn:structure}
\max_{\mathbf{X}\neq\tilde{\mathbf{X}}}\prod_{i=1}^N \breve{c}_i^{\Omega_i}
&=&\max_{\mathbf{X}\neq\tilde{\mathbf{X}}}\prod_{i=1}^N\left(\Omega_i\sqrt{\sum_{i=1}^Le_i^2}\right)^{-\Omega_i}\nonumber\\
&=&\prod_{i=1}^N\left(\frac{1}{\Omega_i}\right)^{\Omega_i}\left(\min\sqrt{\sum_{i=1}^Le_i^2}\right)^{-\Omega}
\end{eqnarray}
In addition, our assumption that
 \begin{eqnarray*}
 \tilde{d}_{\min}\left(\breve{\mathcal{X}}\right)=\min\Omega\sqrt{\sum_{i=1}^Le_i^2}=1
 \end{eqnarray*}
 leads us to $\min \sqrt{\sum_{i=1}^Le_i^2}=\frac{1}{\Omega}$. Then, substituting this result into~\eqref{eqn:structure} produces \begin{eqnarray*}
\max_{\mathbf{X}\neq\tilde{\mathbf{X}}}\prod_{i=1}^N \breve{c}_i^{\Omega_i}
=\prod_{i=1}^N\left(\frac{\Omega}{\Omega_i}\right)^{\Omega_i}
\end{eqnarray*}
Thus, the coding structure given in~\eqref{eqn:structure_space} is the optimal solution to Subproblem~\ref{subprob:optimal_structure} in the sense of minimizing the worst-case small-scale diversity loss.

 \underline{\textbf{Solution to Subproblem~\ref{subprob:optimal_non_linear_stbc})}}:
If the structure of the matrix elements of  $\mathcal{X}$ is given by~\eqref{eqn:structure_space}, then, Subproblem~\ref{subprob:optimal_non_linear_stbc} is equivalent to minimizing $\sum_{\mathbf{X}\in\mathcal{X}}\mathbf{1}^T\mathbf{X}\mathbf{1}$ subject to $\min\Omega\sqrt{\sum_{i=1}^Ne_i^2}=1$. This problem can be equivalently formulated into Subproblem~\ref{problem:multi_dimension_constellation}. Then, the proof of Theorem~\ref{theorem:optimal_structure} is complete.
 ~\hfill $\Box$

\subsection{Proof of Theorem~\ref{theorem:diophantine_conste}}\label{app:diopantine_constellation}
To prove Theorem~\ref{theorem:diophantine_conste}, we consider the following cases.
\begin{enumerate}
  \item \underline{$1\le L<4$}.
  When $L=1$, $\cup_{q=0}^{\infty}\mathcal{S}^{(L)}_{q}=\mathbb{N}$. By computations, we attain $\min_{s\neq\tilde{s},s,\tilde{s}\in\cup_{q=0}^{\infty}\mathcal{S}^{(L)}_{q}}|s-\tilde{s}|=1$ for $L=1$. If $L=2,3$, then,  \begin{eqnarray*}
\mathcal{S}_{q}^{(L)}=\left\{\mathbf{x}:\mathbf{1}^T\mathbf{x}\mathbf{1}=q,\mathbf{x}\in\mathbb{N}^L\right\}.
\end{eqnarray*}
We notice that when $L=2,3$, all the entries of the elements of $\cup_{q=0}^{\infty}\mathcal{S}^{(L)}_{q}$ are integers. This observation tells us that for any $\mathbf{s},\tilde{\mathbf{s}}\in\cup_{q=0}^{\infty}\mathcal{S}^{(L)}_{q}$ with $\mathbf{s}\neq\tilde{\mathbf{s}}$,  $\|\mathbf{s}-\tilde{\mathbf{s}}\|_2\ge1$. Thus, when $L=2,3$,
$\min_{\mathbf{s}\neq\tilde{\mathbf{s}},\mathbf{s},\tilde{\mathbf{s}}\in\cup_{q=0}^{\infty}\mathcal{S}^{(L)}_{q}}\|\mathbf{s}-\tilde{\mathbf{s}}\|_2\ge1$ for any given positive integer $K$. Therefore, if $1\le L<4$, then, it indeed holds that
  \begin{eqnarray*}
\min_{\mathbf{s}\neq\tilde{\mathbf{s}},\mathbf{s},\tilde{\mathbf{s}}\in\cup_{q=0}^{\infty}\mathcal{S}^{(L)}_{q}}\|\mathbf{s}-\tilde{\mathbf{s}}\|_2\ge1.
  \end{eqnarray*}
\item \underline{$L\ge4$}. In this case, we rewrite $\cup_{q=0}^{\infty}\mathcal{S}^{(L)}_{q}$  into
\begin{eqnarray*}
\cup_{q=0}^{\infty}\mathcal{S}^{(L)}_{q}=\cup_{n=1}^{\lfloor\sqrt{L}\rfloor-1}\hat{\mathcal{S}}^{(n)}\cup\tilde{\mathcal{S}}
\end{eqnarray*}
where
 \begin{eqnarray*}
\left\{
\begin{array}{llll}
 \hat{\mathcal{S}}^{(n)}=\cup_{q=0}^{\infty}\left\{\frac{n\mathbf{1}_{L\times1}}{\lfloor \sqrt{L}\rfloor}+\mathbf{x}:\mathbf{1}^T\mathbf{x}\mathbf{1}=q,\mathbf{x}\in\mathbb{N}^L\right\},\\
\tilde{\mathcal{S}}=\cup_{q=0}^{\infty}\left\{\mathbf{x}:\mathbf{1}^T\mathbf{x}\mathbf{1}=q,\mathbf{x}\in\mathbb{N}^L\right\}. \end{array}
\right.
 \end{eqnarray*}
 Without much difficulty, we can attain that $\hat{\mathcal{S}}^{(n)}\cap\tilde{\mathcal{S}}=\emptyset$ for any given $n$, and $\hat{\mathcal{S}}^{(n_1)}\cap\hat{\mathcal{S}}^{(n_2)}=\emptyset$ for any $n_1\neq n_2$. Then, we consider the following four sub-possibilities.
\begin{enumerate}
  \item \underline{$\mathbf{s},\tilde{\mathbf{s}}\in\tilde{\mathcal{S}}$ with $\mathbf{s}\neq\tilde{\mathbf{s}}$}. Following the argument similar to the cases with  $L=1,2,3$ leads us to the fact that $\min_{\mathbf{s}\neq\tilde{\mathbf{s}},\mathbf{s},\tilde{\mathbf{s}}\in\tilde{\mathcal{S}}}\|\mathbf{s}-\tilde{\mathbf{s}}\|_2\ge1$.
  \item \underline{$\hat{\mathbf{s}}\in\hat{\mathcal{S}}^{(n)}$ and $\tilde{\mathbf{s}}\in\tilde{\mathcal{S}}$}.  In this case, $\hat{\mathbf{s}}=\frac{n}{\lfloor \sqrt{L}\rfloor}\mathbf{1}_{L\times1}+\hat{\mathbf{x}}$ and $\tilde{\mathbf{s}}=\tilde{\mathbf{x}}$, where $\hat{\mathbf{x}},\tilde{\mathbf{x}}\in\mathbb{N}^{L}$. Then, we attain
\begin{eqnarray*}
&&\|\hat{\mathbf{s}}-\tilde{\mathbf{s}}\|_2=\sqrt{\sum_{i=1}^L\left(\frac{n}{\lfloor \sqrt{L}\rfloor}+\hat{x}_i-\tilde{x}_i\right)^2}
\nonumber\\
&&=\frac{1}{\lfloor \sqrt{L}\rfloor}\sqrt{\sum_{i=1}^L\left(n+\lfloor \sqrt{L}\rfloor\left(\hat{x}_i-\tilde{x}_i\right)\right)^2}
\end{eqnarray*}
Since $n\le \lfloor \sqrt{L}\rfloor-1$ and  $(\hat{x}_{i}-\tilde{x}_i)$ is integer-valued,  we attain that $\left(n+\lfloor \sqrt{L}\rfloor\left(\hat{x}_i-\tilde{x}_i\right)\right)\neq0$ and thus,  $\left(n+\lfloor \sqrt{L}\rfloor\left(\hat{x}_i-\tilde{x}_i\right)\right)^2\ge1$. This observation leads us to
\begin{eqnarray*}
\|\hat{\mathbf{s}}-\tilde{\mathbf{s}}\|_2\ge\frac{1}{\lfloor \sqrt{L}\rfloor}\sqrt{\sum_{i=1}^L1}=\frac{\sqrt{L}}{\lfloor \sqrt{L}\rfloor}\ge1
\end{eqnarray*}

\item \underline{$\mathbf{s}\neq\tilde{\mathbf{s}},\mathbf{s},\tilde{\mathbf{s}}\in\hat{\mathcal{S}}^{(n)}$ with $n\neq0$}. In this situation, we assume $\mathbf{s}=\frac{n}{\lfloor \sqrt{L}\rfloor}\mathbf{1}_{L\times1}+\mathbf{x}$ and $\tilde{\mathbf{s}}=\frac{n}{\lfloor \sqrt{L}\rfloor}\mathbf{1}_{L\times1}+\tilde{\mathbf{x}}$. Then,
\begin{eqnarray*}
\|\mathbf{s}-\tilde{\mathbf{s}}\|_2=\|\mathbf{x}-\tilde{\mathbf{x}}\|_2
\end{eqnarray*}
Notice that if $\mathbf{s},\tilde{\mathbf{s}}\in\hat{\mathcal{S}}^{(n)}$, then, $\mathbf{s}\neq\tilde{\mathbf{s}}$ is equivalent to $\mathbf{x}\neq\tilde{\mathbf{x}}$. Therefore, for any $\mathbf{s}\neq\tilde{\mathbf{s}}$, it is true that $\|\mathbf{s}-\tilde{\mathbf{s}}\|_2\ge1$.
\item \underline{$\hat{\mathbf{s}}\in\hat{\mathcal{S}}^{(n_1)}$ and $\tilde{\mathbf{s}}\in\hat{\mathcal{S}}^{(n_2)}$ with $n_1\neq n_2$}. To assure $\lfloor \sqrt{L}\rfloor\ge3$, this case is possible only if  $L\ge9$. In this case, $\hat{\mathbf{s}}=\frac{n_1}{\lfloor \sqrt{L}\rfloor}\mathbf{1}_{L\times1}+\hat{\mathbf{x}}$ and $\tilde{\mathbf{s}}=\frac{n_2}{\lfloor \sqrt{L}\rfloor}\mathbf{1}_{L\times1}+\tilde{\mathbf{x}}$. Without loss of generality, we assume that $n_2>n_1$, implying $n_2\ge n_1+1$. By following the argument similar to  that in Item b), we have
\begin{eqnarray*}
&&\|\hat{\mathbf{s}}-\tilde{\mathbf{s}}\|_2=\sqrt{\sum_{i=1}^L\left(\frac{n_1-n_2}{\lfloor \sqrt{L}\rfloor}+\hat{x}_i-\tilde{x}_i\right)^2}\nonumber\\
&&\ge\frac{1}{\lfloor \sqrt{L}\rfloor}\sqrt{\sum_{i=1}^L1}=\frac{\sqrt{L}}{\lfloor \sqrt{L}\rfloor}\ge1
\end{eqnarray*}
\end{enumerate}
\end{enumerate}
To sum up, we can conclude that for any positive integers $K$ and $L$ satisfying $L\ge1$, it holds that
\begin{eqnarray*}
\min_{\mathbf{s}\neq\tilde{\mathbf{s}},\mathbf{s},\tilde{\mathbf{s}}\in\cup_{q=0}^{\infty}\mathcal{S}^{(L)}_{q}}\|\mathbf{s}-\tilde{\mathbf{s}}\|_2\ge1
\end{eqnarray*}
where the equality holds when $\mathbf{s}=\mathbf{0}_{L\times1}\in\cup_{q=0}^{\infty}\mathcal{S}^{(L)}_{q}$ and $\tilde{\mathbf{s}}=[1,\mathbf{0}_{1\times(L-1)}]^T\in\cup_{q=0}^{\infty}\mathcal{S}^{(L)}_{q}$. Therefore, this completes the proof of Theorem~\ref{theorem:diophantine_conste}.
~\hfill$\Box$

\subsection{Proof of Theorem~\ref{theorem:golden_code_optical}}\label{app:proof_golden_optical}

\subsubsection{\textbf{Equivalent Simplification}}
From the constraint condition in~\eqref{eqn:design_criterion_variance}, we know
\begin{eqnarray*}
&&\forall i\neq j,\left(f_{i1}e_1+f_{i2}e_2\right)\left(f_{j1}e_1+f_{j2}e_2\right)>0,\nonumber\\
&&\left(g_{i1}e_1+g_{i2}e_2\right)\left(g_{j1}e_1+g_{j2}e_2\right)>0
\end{eqnarray*}
With this condition, we use Lemma~\ref{lemma:product_power} and then, arrive at the following inequaltiy
\begin{eqnarray*}
&&\prod_{i=1}^N\left(f_{i1}e_1+f_{i2}e_2\right)^{2\Omega_i}\nonumber\\
&&\le \left(\frac{\sum_{i=1}^N(f_{i1}e_1+f_{i2}e_2)}{\Omega}\right)^{2\Omega}\prod_{i=1}^{N}\Omega_i^{2\Omega_i}
\nonumber\\
&&= \left(\frac{\sum_{i=1}^Nf_{i1}e_1+\sum_{i=1}^Nf_{i2}e_2}{\Omega}\right)^{2\Omega}\prod_{i=1}^{N} \Omega_i^{2\Omega_i}
\end{eqnarray*}
where the equality holds if and only if
\begin{eqnarray}\label{eqn:equality_f}
\forall i\neq j,\frac{f_{i1}e_1+f_{i2}e_2}{f_{j1}e_1+f_{j2}e_2}=\frac{\Omega_i}{\Omega_j}
\end{eqnarray}
In the same token, we have
\begin{eqnarray*}
&&\prod_{i=1}^N\left(g_{i1}e_1+g_{i2}e_2\right)^{2\Omega_i}\nonumber\\
&&\le \left(\frac{\sum_{i=1}^Ng_{i1}e_1+\sum_{i=1}^Ng_{i2}e_2}{\sum_{i=1}^N\Omega_i}\right)^{2\Omega}\prod_{i=1}^{N}\Omega_i^{2\Omega_i}
\end{eqnarray*}
where the equality holds if and only if
\begin{eqnarray}\label{eqn:equality_g}
\forall i\neq j,\frac{g_{i1}e_1+g_{i2}e_2}{g_{j1}e_1+g_{j2}e_2}=\frac{\Omega_i}{\Omega_j}
\end{eqnarray}
Therefore, $\prod_{i=1}^N\left(f_{i1}e_1+f_{i2}e_2\right)^{2\Omega_i}\left(g_{i1}e_1+g_{i2}e_2\right)^{2\Omega_i}$ can be upper-bounded by
\begin{eqnarray}\label{eqn:upper_bound}
&& \prod_{i=1}^N\left(f_{i1}e_1+f_{i2}e_2\right)^{2\Omega_i}\left(g_{i1}e_1+g_{i2}e_2\right)^{2\Omega_i} \nonumber\\
&&
\le
\left(\sum_{i=1}^Nf_{i1}e_1+\sum_{i=1}^Nf_{i2}e_2\right)^{2\Omega}
\nonumber\\
&&\times\left(\sum_{i=1}^Ng_{i1}e_1+\sum_{i=1}^Ng_{i2}e_2\right)^{2\Omega}\frac{1}{\Omega^{4\Omega}}
\prod_{i=1}^{N}\Omega_i^{4\Omega_i}
\end{eqnarray}
For notational simplicity, let us denote $\tilde{f}_{1}=\sum_{i=1}^Nf_{i1}$, $\tilde{f}_{2}=\sum_{i=1}^Nf_{i2}$, $\tilde{g}_{1}=\sum_{i=1}^Ng_{i1}$ and $\tilde{g}_{2}=\sum_{i=1}^Ng_{i2}$.
Then, by~\eqref{eqn:equality_f} and~\eqref{eqn:equality_g}, the equality in~\eqref{eqn:upper_bound} holds if and only if
\begin{eqnarray}\label{eqn:equality_fg}
 f_{i1}=\frac{\Omega_i\tilde{f}_{1}}{\Omega}, f_{i2}=\frac{\Omega_i\tilde{f}_{2}}{\Omega},g_{i1}=\frac{\Omega_i\tilde{g}_{1}}{\Omega}, g_{i2}=\frac{\Omega_i\tilde{g}_{2}}{\Omega}
\end{eqnarray}
The upper-bound in~\eqref{eqn:upper_bound} allows us to equivalently transform~\eqref{eqn:design_criterion_variance} into the following problem.
\begin{eqnarray}\label{eqn:design_criterion}
&&\max_{\tilde{f}_{1},\tilde{f}_{2},\tilde{g}_{1},\tilde{g}_{2}}\min_{e_1,e_2,e_1^2+e_2^2\neq0} \left(\tilde{f}_{1}e_1+\tilde{f}_{2}e_2\right)^{2}\left(\tilde{g}_{1}e_1+\tilde{g}_{2}e_2\right)^{2}
\nonumber\\
&&s.t.
\left\{\begin{array}{ll}
 \tilde{f}_{1},\tilde{f}_{2},\tilde{g}_{1},\tilde{g}_{2}>0,\tilde{f}_{1}+\tilde{f}_{2}+\tilde{g}_{1}+\tilde{g}_{2}=1,\\
e_1\in\left\{0,\pm1,\ldots,\pm\left(2^{K_1}-1\right)\right\},\\
e_2\in\left\{0,\pm1,\ldots,\pm\left(2^{K_2}-1\right)\right\},\\
 \left(\tilde{f}_{1}e_1+\tilde{f}_{2}e_2\right)\left(\tilde{g}_{1}e_1+\tilde{g}_{2}e_2\right)\neq0.
 \end{array}
\right.
\end{eqnarray}

\subsubsection{\textbf{Solution for $K_1=K_2=1$}}
We begin by solving the following optimization problem for $K_1=K_2=1$,
\begin{eqnarray}\label{eqn:ook}
&&\max_{\tilde{f}_{1},\tilde{f}_{2},\tilde{g}_{1},\tilde{g}_{2}}\min \left(\tilde{f}_{1}^2\tilde{g}_{1}^2,\tilde{f}_{2}^2\tilde{g}_{2}^2,\left(\tilde{f}_{1}-\tilde{f}_{2}\right)^{2}\left(\tilde{g}_{1}-\tilde{g}_{2}\right)^{2}\right) \nonumber\\
&&s.t.
\left\{\begin{array}{ll}
\tilde{f}_{1}\neq \tilde{f}_{2},\tilde{g}_{1}\neq \tilde{g}_{2}, \\
 \tilde{f}_{1}+\tilde{f}_{2}+\tilde{g}_{1}+\tilde{g}_{2}=1,\\
\tilde{f}_{1},\tilde{f}_{2},\tilde{g}_{1},\tilde{g}_{2}>0
 \end{array}
\right.
\end{eqnarray}

Let us define $r_1=\frac{\tilde{f}_{1}}{\tilde{f}_{2}}$ and $r_2=\frac{\tilde{g}_{1}}{\tilde{g}_{2}}$, where $r_1,r_2\in(0,1)\cup(1,\infty)$. Based on $r_1$ and $r_2$, our discussions for~\eqref{eqn:ook} fall into the following four possibilities.
\begin{enumerate}
  \item \underline{$r_1>1$ and $r_2>1$}.  For this situation, we consider the two sub-possibilities: $\left(r_{1}-1\right)\left(r_{2}-1\right)\le1$ and $\left(r_{1}-1\right)\left(r_{2}-1\right)\ge1$~\footnote{Although these two sub-possibilities are not disjoint, we still abuse these conditions since we will prove that the local optimality happens when $\left(r_{1}-1\right)\left(r_{2}-1\right)=1$.}.
\begin{enumerate}
  \item \underline{$\left(r_{1}-1\right)\left(r_{2}-1\right)\le1$}. In this case, it holds that
      \begin{eqnarray*}
      &&\min \left(\tilde{f}_{1}^2\tilde{g}_{1}^2,\tilde{f}_{2}^2\tilde{g}_{2}^2,\left(\tilde{f}_{1}-\tilde{f}_{2}\right)^{2}\left(\tilde{g}_{1}-\tilde{g}_{2}\right)^{2}\right)
      \nonumber\\
      &&= \left(\tilde{f}_{1}-\tilde{f}_{2}\right)^2\left(\tilde{g}_{1}-\tilde{g}_{2}\right)^2
      \end{eqnarray*}
      Thus, our task for this situation is to solve the following optimization problem

 \begin{eqnarray*}
&&\max_{\mathbf{F}}\left(\tilde{f}_{1}-\tilde{f}_{2}\right)\left(\tilde{g}_{1}-\tilde{g}_{2}\right) \nonumber\\
&&s.t.
\left\{\begin{array}{ll}
\tilde{f}_{1}> \tilde{f}_{2}>0,\tilde{g}_{1}> \tilde{g}_{2}>0, \\
 \tilde{f}_{1}+\tilde{f}_{2}+\tilde{g}_{1}+\tilde{g}_{2}=1,\\
  \left(r_{1}-1\right)\left(r_{2}-1\right)\le1,\\
r_1>1,r_2>1.
 \end{array}
\right.
\end{eqnarray*}
 We observe that
  \begin{eqnarray}\label{eqn:case1_upper_bound}
&&\left(\tilde{f}_{1}-\tilde{f}_{2}\right)\left(\tilde{g}_{1}-\tilde{g}_{2}\right)\nonumber\\
&&=\left(\tilde{f}_{1}+\tilde{f}_{2}\right)\left(\tilde{g}_{1}+\tilde{g}_{2}\right)\frac{\left(r_{1}-1\right)\left(r_{2}-1\right)}{\left(r_{1}+1\right)\left(r_{2}+1\right)} \nonumber\\
&&\le\frac{\left(\tilde{f}_{1}+\tilde{f}_{2}+\tilde{g}_{1}+\tilde{g}_{2}\right)^2\left(r_{1}-1\right)\left(r_{2}-1\right)}{4\left(r_{1}+1\right)\left(r_{2}+1\right)} \nonumber\\
&&=\frac{\left(r_{1}-1\right)\left(r_{2}-1\right)}{4\left(r_{1}+1\right)\left(r_{2}+1\right)}
  \end{eqnarray}
 where the equality holds if and only if $\tilde{f}_{1}+\tilde{f}_{2}=\tilde{g}_{1}+\tilde{g}_{2}=\frac{1}{2}$.   Then, we maximize $\frac{\left(r_{1}-1\right)\left(r_{2}-1\right)}{\left(r_{1}+1\right)\left(r_{2}+1\right)}$ under the conditions that  $\left(r_{1}-1\right)\left(r_{2}-1\right)\le1$ and $r_1>1,r_2>1$.
    Since
    \begin{eqnarray*}
   \frac{\left(r_{1}-1\right)\left(r_{2}-1\right)}{\left(r_{1}+1\right)\left(r_{2}+1\right)}=\frac{r_{2}-1}{r_{2}+1}\left(1-\frac{2}{r_{1}+1}\right),
    \end{eqnarray*}
we conclude that for any given $r_2>1$, $\frac{\left(r_{1}-1\right)\left(r_{2}-1\right)}{\left(r_{1}+1\right)\left(r_{2}+1\right)}$ is monotonically increasing with respect to $r_1$. In addition, $\left(r_{1}-1\right)\left(r_{2}-1\right)\le1$ and $r_1>1,r_2>1$ give us that $r_{1}\le\frac{r_2}{r_2-1}$. Thus, we can upper-bound $\frac{\left(r_{1}-1\right)\left(r_{2}-1\right)}{\left(r_{1}+1\right)\left(r_{2}+1\right)}$ by
   \begin{eqnarray*}
\frac{\left(r_{1}-1\right)\left(r_{2}-1\right)}{\left(r_{1}+1\right)\left(r_{2}+1\right)}&\le & \frac{\left(\frac{r_2}{r_2-1}-1\right)\left(r_{2}-1\right)}{\left(\frac{r_2}{r_2-1}+1\right)\left(r_{2}+1\right)} \nonumber\\
&=&\frac{r_{2}-1}{\left(2r_2-1\right)\left(r_{2}+1\right)}.
   \end{eqnarray*}
For notational simplicity, we denote the above upper-bound  by $F_{1}\left(r_2\right)$. Now, we give the first-order derivative of $F_{1}\left(r_2\right)$ by
\begin{eqnarray*}
\frac{\partial F_{1}\left(r_2\right)}{\partial r_2}=\frac{-2r_2\left(r_2-2\right)}{\left(2r_2-1\right)^2\left(r_{2}+1\right)^2}
\end{eqnarray*}
Notice that $\frac{\partial F_{1}\left(r_2\right)}{\partial r_2}>0$ with $r_2<2$ and $\frac{\partial F_{1}\left(r_2\right)}{\partial r_2}<0$ with $r_2>2$. This observation indicates that $F_{1}\left(r_2\right)$ is maximized if and only if $r_2=2$. Combining $r_2=2$ with $r_{1}=\frac{r_2}{r_2-1}$ and $\tilde{f}_{1}+\tilde{f}_{2}=\tilde{g}_{1}+\tilde{g}_{2}=\frac{1}{2}$ produces that
\begin{eqnarray*}
\tilde{f}_{1}=\tilde{g}_{1}=\frac{1}{3},\tilde{f}_{2}=\tilde{g}_{2}=\frac{1}{6}
\end{eqnarray*}
  \item \underline{$\left(r_{1}-1\right)\left(r_{2}-1\right)\ge1$}.
In this case, it holds that
\begin{eqnarray*}
\min \left(\tilde{f}_{1}^2\tilde{g}_{1}^2,\tilde{f}_{2}^2\tilde{g}_{2}^2,\left(\tilde{f}_{1}-\tilde{f}_{2}\right)^{2}\left(\tilde{g}_{1}-\tilde{g}_{2}\right)^{2}\right)= \tilde{f}_{2}^2\tilde{g}_{2}^2
\end{eqnarray*}
Thus, our task for this situation is to solve the following optimization problem
      \begin{eqnarray}
&&\max_{\mathbf{F}}\tilde{f}_{2}\tilde{g}_{2} \nonumber\\
&&s.t.
\left\{\begin{array}{ll}
\tilde{f}_{1}> \tilde{f}_{2}>0,\tilde{g}_{1}> \tilde{g}_{2}>0, \\
 \tilde{f}_{1}+\tilde{f}_{2}+\tilde{g}_{1}+\tilde{g}_{2}=1,\\
 \left(r_{1}-1\right)\left(r_{2}-1\right)\ge1.
 \end{array}
\right.
\end{eqnarray}
Using arithmetic mean geometric mean inequality enables us to upper-bound $\tilde{f}_{2}\tilde{g}_{2}$ by
  \begin{eqnarray}\label{eqn:case1_supreme}
 \tilde{f}_{2}\tilde{g}_{2}&=&\frac{\left(\tilde{f}_{1}+\tilde{f}_{2}\right)\left(\tilde{g}_{1}+\tilde{g}_{2}\right)}{\left(r_{1}+1\right)\left(r_{2}+1\right)} \nonumber\\
  &\le&\frac{\left(\tilde{f}_{1}+\tilde{f}_{2}+\tilde{g}_{1}+\tilde{g}_{2}\right)^2}{4\left(r_{1}+1\right)\left(r_{2}+1\right)}  \nonumber\\ & =&\frac{1}{4\left(r_{1}+1\right)\left(r_{2}+1\right)}
  \end{eqnarray}
  where the equality holds if and only if $\tilde{f}_{1}+\tilde{f}_{2}=\tilde{g}_{1}+\tilde{g}_{2}=\frac{1}{2}$.

  Since $\left(r_{1}-1\right)\left(r_{2}-1\right)\ge1$ gives $r_{1}\ge\frac{r_2}{r_2-1}$, we can upper-bound the righthand side of~\eqref{eqn:case1_supreme} by
 \begin{eqnarray}\label{eqn:upper_bound_tem}
\frac{1}{4\left(r_{1}+1\right)\left(r_{2}+1\right)}\le\frac{r_2-1}{4\left(2r_{2}-1\right)\left(r_{2}+1\right)} \end{eqnarray}
 By using the same techniques  as $F_{1}\left(r_{2}\right)$ for the above sub-possibilities, we can find that the upper-bound in~\eqref{eqn:upper_bound_tem} can be maximized if and only if $r_1=r_2=2$. That being said, the optimal solution is given by
   \begin{eqnarray*}
  \tilde{f}_{1}=\tilde{g}_{1}=\frac{1}{3},\tilde{f}_{2}=\tilde{g}_{2}=\frac{1}{6},
   \end{eqnarray*}
\end{enumerate}
Now, we can conclude that when $r_{1}>1$ and $r_{2}>1$, the maximum value of the objective function of~\eqref{eqn:ook} is given by
\begin{eqnarray*}
\max \min \left(\tilde{f}_{1}^2\tilde{g}_{1}^2,\tilde{f}_{2}^2\tilde{g}_{2}^2,\left(\tilde{f}_{1}-\tilde{f}_{2}\right)^{2}\left(\tilde{g}_{1}-\tilde{g}_{2}\right)^{2}\right)=\frac{1}{1296}.
\end{eqnarray*}
  \item \underline{$r_1>1$ and $r_{2}<1$}. In this situation, four sub-cases are  considered~\footnote{For the same reason as the case with $r_1>1$ and $r_{2}<1$ , these four possibilities are not disjoint and in fact, we will verify that the local optimality occurs when $r_1r_2=1$ and $\left(r_1-1\right)\left(1-r_2\right)=1$. }:
\begin{enumerate}
  \item $r_1r_2\ge1$ and $\left(r_1-1\right)\left(1-r_2\right)\ge1$.
  \item $r_1r_2\ge1$ and $\left(r_1-1\right)\left(1-r_2\right)\le1$.
  \item $r_1r_2\le1$ and $\left(r_1-1\right)\left(1-r_2\right)\ge1$.
  \item $r_1r_2\le1$ and $\left(r_1-1\right)\left(1-r_2\right)\le1$.
\end{enumerate}
Discussions for these four possibilities are given as follows.
\begin{enumerate}
  \item \underline{$r_1r_2\ge1$ and $\left(r_1-1\right)\left(1-r_2\right)\ge1$}. In this case, we can have
      \begin{eqnarray*}
      \min \left(\tilde{f}_{1}^2\tilde{g}_{1}^2,\tilde{f}_{2}^2\tilde{g}_{2}^2,\left(\tilde{f}_{1}-\tilde{f}_{2}\right)^{2}\left(\tilde{g}_{1}-\tilde{g}_{2}\right)^{2}\right)= \tilde{f}_{2}^2\tilde{g}_{2}^2
      \end{eqnarray*}
      Accordingly, the local optimization problem can be formulated into
  \begin{eqnarray}\label{eqn:case1}
&&\max\tilde{f}_{2}\tilde{g}_{2} \nonumber\\
&& s.t.
\left\{\begin{array}{ll}
 \tilde{f}_{2}\left(1+r_1\right)+\tilde{g}_{2}\left(1+r_2\right)=1,\\
r_1>1,r_2<1,r_1r_2\ge1,\\
\left(r_1-1\right)\left(1-r_2\right)\ge1
 \end{array}
\right.
\end{eqnarray}
Notice that
\begin{eqnarray*}
\tilde{f}_{2}\tilde{g}_{2}&=&\frac{\tilde{f}_{2}\left(1+r_1\right)\tilde{g}_{2}\left(1+r_2\right)}{\left(1+r_1\right)\left(1+r_2\right)} \nonumber\\
&\le&\frac{\left(\tilde{f}_{2}\left(1+r_1\right)+\tilde{g}_{2}\left(1+r_2\right)\right)^2}{4\left(1+r_1\right)\left(1+r_2\right)}
\nonumber\\&=&\frac{1}{4\left(1+r_1\right)\left(1+r_2\right)}
\end{eqnarray*}
where the equality holds if and only if $\tilde{f}_{2}\left(1+r_1\right)=\tilde{g}_{2}\left(1+r_2\right)=\frac{1}{2}$.

Now, we are in a position to minimize $\left(1+r_1\right)\left(1+r_2\right)$.
By $\left(r_1-1\right)\left(1-r_2\right)\ge1$, we attain $r_1+r_2\ge r_1r_2+2$. Combining this inequality with $r_1r_2\ge1$ produces
\begin{eqnarray*}
\left(1+r_1\right)\left(1+r_2\right)&=&r_1r_2+r_1+r_2+1\nonumber\\
&\ge& 2r_1r_2+3\ge5
\end{eqnarray*}
 where the equality holds if and only if $r_1r_2=1$ and $\left(r_1-1\right)\left(1-r_2\right)=1$. The solution to these two equations is determined by $r_1=\frac{3+\sqrt{5}}{2}$ and $r_2=\frac{3-\sqrt{5}}{2}$. Hence, in this case, combining $\tilde{f}_{2}\left(1+r_1\right)=\tilde{g}_{2}\left(1+r_2\right)=\frac{1}{2}$ and $r_1=\frac{3+\sqrt{5}}{2},r_2=\frac{3-\sqrt{5}}{2}$  gives us  the optimal solution to~\eqref{eqn:case1} by
\begin{eqnarray}\label{eqn:local_optimal}
  \tilde{f}_{1}=\tilde{g}_{2}=\frac{5+\sqrt{5}}{20},\tilde{f}_{2}=\tilde{g}_{1}=\frac{5-\sqrt{5}}{20}.
   \end{eqnarray}

  \item \underline{$r_1r_2\ge1$ and $\left(r_1-1\right)\left(1-r_2\right)\le1$}. In this case, we have
      \begin{eqnarray*}
      &&\min \left(\tilde{f}_{1}^2\tilde{g}_{1}^2,\tilde{f}_{2}^2\tilde{g}_{2}^2,\left(\tilde{f}_{1}-\tilde{f}_{2}\right)^{2}\left(\tilde{g}_{1}-\tilde{g}_{2}\right)^{2}\right)
      \nonumber\\
      &&= \left(\tilde{f}_{1}-\tilde{f}_{2}\right)^{2}\left(\tilde{g}_{1}-\tilde{g}_{2}\right)^{2}
      \end{eqnarray*}
      The corresponding local optimization problem is given by
  \begin{eqnarray}\label{eqn:third_term}
&&\max\tilde{f}_{2}\tilde{g}_{2}\left(r_1-1\right)\left(1-r_2\right) \nonumber\\
&& s.t.\left\{\begin{array}{lll}
\tilde{f}_{2}\left(1+r_1\right)+\tilde{g}_{2}\left(1+r_2\right)=1,\\
r_1>1,r_2<1,r_1r_2\ge1,\\
\left(r_1-1\right)\left(1-r_2\right)<1.
\end{array}
\right.
\end{eqnarray}
Following the techniques similar to those for~\eqref{eqn:case1_upper_bound}, we have
\begin{eqnarray*}
\tilde{f}_{2}\tilde{g}_{2}\left(r_1-1\right)\left(1-r_2\right)\le\frac{\left(r_1-1\right)\left(1-r_2\right)}{4\left(1+r_1\right)\left(1+r_2\right)}
\end{eqnarray*}
 where the equality holds if and only if $\tilde{f}_{2}\left(1+r_1\right)=\tilde{g}_{2}\left(1+r_2\right)=\frac{1}{2}$.
Since
\begin{eqnarray*}
\frac{\left(r_1-1\right)\left(1-r_2\right)}{\left(1+r_1\right)\left(1+r_2\right)}=1-2\frac{1}{\frac{r_1+r_2}{r_1r_2+1}+1}
\end{eqnarray*}
we can find that  in this situation, $\max\tilde{f}_{2}\tilde{g}_{2}\left(r_1-1\right)\left(1-r_2\right) $ is equivalent to $\max\frac{r_1+r_2}{r_1r_2+1}$.
In addition, $\left(r_1-1\right)\left(1-r_2\right)\le1$ implies $r_1+r_2\le r_1r_2+2$. Then,
\begin{eqnarray*}
\frac{r_1+r_2}{ r_1r_2+1}&\le& \frac{r_1r_2+2}{ r_1r_2+1}\nonumber\\
&=&1+\frac{1}{ r_1r_2+1}\le 1+\frac{1}{ 2}.
\end{eqnarray*}
As a result, when $r_1r_2\ge1$ and $\left(r_1-1\right)\left(1-r_2\right)\le1$, it follows that $\tilde{f}_{2}\tilde{g}_{2}\left(r_1-1\right)\left(1-r_2\right)\le\frac{1}{400}$, where the equality holds if and only if $r_1r_2=1$ and $\left(r_1-1\right)\left(1-r_2\right)=1$.  Then,  the optimal solution, in this situation, is given by~\eqref{eqn:local_optimal}.
  \item \underline{$r_1r_2\le1$ and $\left(r_1-1\right)\left(1-r_2\right)\ge r_1r_2$}.  In this case, it is true that
      \begin{eqnarray*}
      \min \left(\tilde{f}_{1}^2\tilde{g}_{1}^2,\tilde{f}_{2}^2\tilde{g}_{2}^2,\left(\tilde{f}_{1}-\tilde{f}_{2}\right)^{2}\left(\tilde{g}_{1}-\tilde{g}_{2}\right)^{2}\right)= \tilde{f}_{1}^2\tilde{g}_{1}^2
      \end{eqnarray*}
      Thus, the local optimization problem can be formulated into
  \begin{eqnarray}\label{eqn:case3}
&&\max_{\mathbf{F},r_1>1,r_2<1}\tilde{f}_{1}\tilde{g}_{1} \nonumber\\
&& s.t.
\left\{\begin{array}{ll}
\tilde{f}_{1}\left(1+\frac{1}{r_1}\right)+\tilde{g}_{1}\left(1+\frac{1}{r_2}\right)=1,\\
r_1r_2\le1,\left(1-\frac{1}{r_1}\right)\left(\frac{1}{r_2}-1\right)\ge1.
\end{array}
\right.
\end{eqnarray}
In the same token, we can attain that
 \begin{eqnarray*}
\tilde{f}_{1}\tilde{g}_{1}&=&\frac{\tilde{f}_{1}\tilde{g}_{1}\left(1+\frac{1}{r_1}\right)\left(1+\frac{1}{r_2}\right)}{\left(1+\frac{1}{r_1}\right)\left(1+\frac{1}{r_2}\right)} \nonumber\\ &\le&\frac{\left(\tilde{g}_{1}\left(1+\frac{1}{r_1}\right)+\tilde{f}_{1}\left(1+\frac{1}{r_2}\right)\right)^2}{4\left(1+\frac{1}{r_1}\right)\left(1+\frac{1}{r_2}\right)}\nonumber\\
&=&\frac{1}{4\left(1+\frac{1}{r_1}\right)\left(1+\frac{1}{r_2}\right)}
 \end{eqnarray*}
 where the equality holds if and only if $\tilde{f}_{1}\left(1+\frac{1}{r_1}\right)=\tilde{g}_{1}\left(1+\frac{1}{r_2}\right)=\frac{1}{2}$.

We next attack the optimization problem $\min \left(1+\frac{1}{r_1}\right)\left(1+\frac{1}{r_2}\right)$. By the constraints $r_1r_2\le1$ and $\left(1-\frac{1}{r_1}\right)\left(\frac{1}{r_2}-1\right)\ge1$, we have
\begin{eqnarray*}
\left(1+\frac{1}{r_1}\right)\left(1+\frac{1}{r_2}\right)=\frac{1}{r_1}+\frac{1}{r_2}+\frac{1}{r_1r_2}+1\ge5
\end{eqnarray*}
Putting things together yields that
\begin{eqnarray*}
\min \left(1+\frac{1}{r_1}\right)\left(1+\frac{1}{r_2}\right)\le5
\end{eqnarray*}
It should be noted that this equality holds if and only if $r_1r_2=1$ and $\left(1-\frac{1}{r_1}\right)\left(\frac{1}{r_2}-1\right)=1$.  Together with $\tilde{f}_{1}\left(1+\frac{1}{r_1}\right)=\tilde{g}_{1}\left(1+\frac{1}{r_2}\right)=\frac{1}{2}$, we can  attain the solution to~\eqref{eqn:case3} given by~\eqref{eqn:local_optimal}.

\item \underline{$r_1r_2<1$ and $\left(r_1-1\right)\left(1-r_2\right)\le r_1r_2$}. In this case, we can have
    \begin{eqnarray*}
    &&\min \left(\tilde{f}_{1}^2\tilde{g}_{1}^2,\tilde{f}_{2}^2\tilde{g}_{2}^2,\left(\tilde{f}_{1}-\tilde{f}_{2}\right)^{2}\left(\tilde{g}_{1}-\tilde{g}_{2}\right)^{2}\right)
    \nonumber\\
    &&= \left(\tilde{f}_{1}-\tilde{f}_{2}\right)^{2}\left(\tilde{g}_{1}-\tilde{g}_{2}\right)^{2}
    \end{eqnarray*}
     The corresponding local optimization problem is given by
  \begin{eqnarray*}
&&\max_{\mathbf{F}}\tilde{f}_{2}\tilde{g}_{2}\left(r_1-1\right)\left(1-r_2\right) \nonumber\\
&& s.t.
\left\{\begin{array}{ll}
\tilde{f}_{1}\left(1+\frac{1}{r_1}\right)+\tilde{g}_{1}\left(1+\frac{1}{r_2}\right)=1,\\
r_1r_2\le1,\left(1-\frac{1}{r_1}\right)\left(\frac{1}{r_2}-1\right)\le1.
\end{array}
\right.
\end{eqnarray*}
By~\eqref{eqn:case1_upper_bound}, we have
\begin{eqnarray*}
\tilde{f}_{2}\tilde{g}_{2}\left(r_1-1\right)\left(1-r_2\right)
\le \frac{\left(r_1-1\right)\left(1-r_2\right)}{4\left(1+r_1\right)\left(1+r_2\right)}
\end{eqnarray*}
where the equality holds if and only if $\tilde{f}_{2}\left(1+r_1\right)=\tilde{g}_{2}\left(1+r_2\right)=\frac{1}{2}$.
Using the same strategy for~\eqref{eqn:third_term}, we can attain the optimal solution given by~\eqref{eqn:local_optimal}.
\end{enumerate}

Thus far, the discussions for the case with $r_1>1$ and $r_2<1$ have been complete. The optimal solution is given by
\begin{eqnarray*}
\tilde{f}_{1}=\tilde{g}_{2}=\frac{5+\sqrt{5}}{20},\tilde{f}_{2}=\tilde{g}_{1}=\frac{5-\sqrt{5}}{20}
\end{eqnarray*}
Thus, when  $r_1>1$ and $r_2<1$, we have
\begin{eqnarray*}
\max \min \left(\tilde{f}_{1}^2\tilde{g}_{1}^2,\tilde{f}_{2}^2\tilde{g}_{2}^2,\left(\tilde{f}_{1}-\tilde{f}_{2}\right)^{2}\left(\tilde{g}_{1}-\tilde{g}_{2}\right)^{2}\right)=\frac{1}{400}.
\end{eqnarray*}

  \item \underline{$r_{1}<1,r_{2}>1$}.
The case with $\tilde{f}_{1}<\tilde{f}_{2}$ and $\tilde{g}_{1}>\tilde{g}_{2}$ is parallel to that with $\tilde{f}_{1}>\tilde{f}_{2},\tilde{g}_{1}<\tilde{g}_{2}$. Thus, the local maximum value of the objective function is equal to $\frac{1}{400}$ with
$\tilde{f}_{1}=\tilde{g}_{2}=\frac{5-\sqrt{5}}{20}$ and $\tilde{f}_{2}=\tilde{g}_{1}=\frac{5+\sqrt{5}}{20}$.

  \item \underline{$r_{1}<1,r_{2}<1$}. In the same token, the case with $\tilde{f}_{1}<\tilde{f}_{2},\tilde{g}_{1}<\tilde{g}_{2}$ is parallel to that of $\tilde{f}_{1}>\tilde{f}_{2},\tilde{g}_{1}>\tilde{g}_{2}$ and thus, the maximum value of the objective function is equal to $\frac{1}{1296}$ with $\tilde{f}_{1}=\tilde{g}_{1}=\frac{1}{6}$ and $\tilde{f}_{2}=\tilde{g}_{2}=\frac{1}{3}$.

\end{enumerate}
By comparing the local solution for the above four cases, we attain the globally optimal solution to~\eqref{eqn:ook}, which is given by
\begin{subequations}\label{eqn:golen_ook}
\begin{eqnarray}
  \tilde{f}_{1}=\tilde{g}_{2}=\frac{5-\sqrt{5}}{20},\tilde{f}_{2}=\tilde{g}_{1}=\frac{5+\sqrt{5}}{20}
   \end{eqnarray}
   or
   \begin{eqnarray}
  \tilde{f}_{1}=\tilde{g}_{2}=\frac{5+\sqrt{5}}{20},\tilde{f}_{2}=\tilde{g}_{1}=\frac{5-\sqrt{5}}{20}
   \end{eqnarray}
\end{subequations}
\subsubsection{\textbf{Solution for generic $K_1$ and $K_2$}}
   In the following, we prove that~\eqref{eqn:golen_ook} is exactly the solution to~\eqref{eqn:design_criterion}. With $e_1\in\left\{0,\pm1,\ldots,\pm\left(2^{K_1}-1\right)\right\},
e_2\in\left\{0,\pm1,\ldots,\pm\left(2^{K_1}-1\right)\right\}$ and $\tilde{f}_{1}+\tilde{f}_{2}+\tilde{g}_{1}+\tilde{g}_{2}=1$, we have
    \begin{eqnarray*}
&&\min_{e_1,e_2,e_1^2+e_2^2\neq0} \left(\tilde{f}_{1}e_1+\tilde{f}_{2}e_2\right)^{2}\left(\tilde{g}_{1}e_1+\tilde{g}_{2}e_2\right)^{2}
\nonumber\\
&&\le\min \left(\tilde{f}_{1}^2\tilde{g}_{1}^2,\tilde{f}_{2}^2\tilde{g}_{2}^2,\left(\tilde{f}_{1}-\tilde{f}_{2}\right)^{2}\left(\tilde{g}_{1}-\tilde{g}_{2}\right)^{2}\right)  \end{eqnarray*}
Then, we arrive at the following inequality
  \begin{eqnarray*}
 &&\max_{\tilde{f}_{1},\tilde{f}_{2},\tilde{g}_{1},\tilde{g}_{2}}\min_{e_1,e_2,e_1^2+e_2^2\neq0} \left(\tilde{f}_{1}e_1+\tilde{f}_{2}e_2\right)^{2}\left(\tilde{g}_{1}e_1+\tilde{g}_{2}e_2\right)^{2}\nonumber\\
 &&\le\max_{\tilde{f}_{1},\tilde{f}_{2},\tilde{g}_{1},\tilde{g}_{2}}\min \left(\tilde{f}_{1}^2\tilde{g}_{1}^2,\tilde{f}_{2}^2\tilde{g}_{2}^2,\left(\tilde{f}_{1}-\tilde{f}_{2}\right)^{2}\left(\tilde{g}_{1}-\tilde{g}_{2}\right)^{2}\right)  \end{eqnarray*}
 Hence, we attain
  \begin{eqnarray}\label{eqn:small_inequality}
    \max_{\tilde{f}_{1},\tilde{f}_{2},\tilde{g}_{1},\tilde{g}_{2}}\min_{e_1^2+e_2^2\neq0} \left(\tilde{f}_{1}e_1+\tilde{f}_{2}e_2\right)^{2}\left(\tilde{g}_{1}e_1+\tilde{g}_{2}e_2\right)^{2}
\le\frac{1}{400}
  \end{eqnarray}
  In the following, we prove that the upper-bound in~\eqref{eqn:small_inequality} can be achieved by~\eqref{eqn:golen_ook}. Substituting~\eqref{eqn:golen_ook} into $\left(\tilde{f}_{1}e_1+\tilde{f}_{2}e_2\right)^{2}\left(\tilde{g}_{1}e_1+\tilde{g}_{2}e_2\right)^{2}$ and performing some manipulations yield what follows.
  \begin{eqnarray*}
  \left(\tilde{f}_{1}e_1+\tilde{f}_{2}e_2\right)^{2}\left(\tilde{g}_{1}e_1+\tilde{g}_{2}e_2\right)^{2}=\frac{1}{400}\left(3e_1e_2+\left(e_1^2+e_2^2\right)\right)^2
  \end{eqnarray*}
   Over the feasible sets of $e_1$ and $e_2$,
  \begin{eqnarray}\label{eqn:farey_sequency}
&&e_1^2+e_2^2\neq 0, e_1\in\left\{0,\pm1,\ldots,\pm\left(2^{K_1}-1\right)\right\},\nonumber\\
&&e_2\in\left\{0,\pm1,\ldots,\pm\left(2^{K_2}-1\right)\right\},
  \end{eqnarray}
  the following two possibilities are considered.
  \begin{enumerate}
    \item $e_1e_2=0$. In this case,
    \begin{eqnarray*}
      \frac{1}{400}\left(3e_1e_2+\left(e_1^2+e_2^2\right)\right)^2=\frac{1}{400}\left(e_1^2+e_2^2\right)^2
    \end{eqnarray*}
    By~\eqref{eqn:farey_sequency}, we notice that  $\left(e_1^2+e_2^2\right)$ is integer-valued and  $e_1^2+e_2^2\neq0$. Then, when $e_1e_2=0$ and $e_1^2+e_2^2\neq0$ , we arrive at the following
    \begin{eqnarray*}
    \frac{1}{400}\left(3e_1e_2+\left(e_1^2+e_2^2\right)\right)^2\ge\frac{1}{400},
    \end{eqnarray*}
    where the equality holds if and only if
    \begin{eqnarray*}
    \left(e_1,e_2\right)^2\in\left\{\pm\left(0,1\right)^T,\pm\left(1,0\right)^T\right\}.
    \end{eqnarray*}
    \item $e_1e_2\neq0$. Letting $3e_1e_2+\left(e_1^2+e_2^2\right)=0$ gives us that $\frac{e_1}{e_2}=-\frac{3\pm\sqrt{5}}{2}$, which is impossible over the feasible sets of $e_1$ and $e_2$ given by~\eqref{eqn:farey_sequency}.   Then, by~\eqref{eqn:farey_sequency}, $\left(3e_1e_2+\left(e_1^2+e_2^2\right)\right)^2$ is non-zero and integer-valued. Consequently, when $e_1e_2\neq0$, we have
        \begin{eqnarray*}
    \frac{1}{400}\left(3e_1e_2+\left(e_1^2+e_2^2\right)\right)^2\ge\frac{1}{400}
    \end{eqnarray*}
     where the equality holds if $\left(e_1,e_2\right)^T=\left(-1,1\right)^T$.
    \end{enumerate}
  According to the above discussions, when $\tilde{f}_{1},\tilde{f}_{2},\tilde{g}_{1}$ and $\tilde{g}_{2}$ are given by~\eqref{eqn:golen_ook}, we can conclude that the upper-bound in~\eqref{eqn:small_inequality} can be achieved by~\eqref{eqn:golen_ook}, giving us that~\eqref{eqn:golen_ook} is indeed the optimal solution to~\eqref{eqn:design_criterion}. Further, combining~\eqref{eqn:golen_ook} with~\eqref{eqn:equality_fg} produces~\eqref{eqn:goledn_optical} and~\eqref{eqn:goledn_optical_equivalent}, which are the solutions to~\eqref{eqn:design_criterion_variance}. Then, the proof of Theorem~\ref{theorem:golden_code_optical} is complete.
~\hfill $\Box$

\bibliographystyle{ieeetr}
\bibliography{tzzt}
\end{document}